\documentclass[useAMS,usenatbib]{mn2e}

\usepackage{aecompl}
\usepackage[toc,page]{appendix}

\usepackage{graphicx}
\usepackage{amssymb}
\usepackage{amsmath}
\usepackage{aas_macros}
\usepackage{deluxetable}
\usepackage{lscape}
\usepackage{rotating}
\usepackage[symbol*]{footmisc}
\usepackage{gensymb}



\title[The problems of mapping from PPV to PPP]{Synthetic C$^{18}$O observations of fibrous filaments: the problems of mapping from PPV to PPP}
\author[S. D. Clarke et al.]{S. D. Clarke$^{1}$\thanks{E-mail: clark@ph1.uni-koeln.de }, A. P. Whitworth$^{2}$, R. L. Spowage$^{2}$, A. Duarte-Cabral$^{2}$, S. T. Suri$^{1}$,\newauthor S. E. Jaffa$^2$, S. Walch$^{1} $ and P. C. Clark$^{2}$\\$^{1}$I. Physikalisches Institut, Universit{\"a}t zu K{\"o}ln, Z{\"u}lpicher Str. 77, D-50937 K{\"o}ln, Germany\\$^{2}$School of Physics and Astronomy, Cardiff University, Cardiff, CF24 3AA, UK}

\begin{document}

\date{}

\pagerange{\pageref{firstpage}--\pageref{lastpage}} \pubyear{2002}

\maketitle

\label{firstpage}

\begin{abstract}
Molecular-line observations of filaments in star-forming regions have revealed the existence of elongated coherent features within the filaments; these features are termed {\it fibres}. Here we caution that, since fibres are traced in PPV space, there is no guarantee that they represent coherent features in PPP space. We illustrate this contention using simulations of the growth of a filament from a turbulent medium. Synthetic C$^{18}$O observations of the simulated filaments reveal the existence of fibres very similar to the observed ones, i.e. elongated coherent features in the resulting PPV data-cubes. Analysis of the PPP data-cubes (i.e. 3D density fields) also reveals elongated coherent features, which we term {\it sub-filaments}. Unfortunately there is very poor correspondence between the fibres and the sub-filaments in the simulations. Both fibres and sub-filaments derive from inhomogeneities in the turbulent accretion flow onto the main filament. As a consequence, fibres are often affected by line-of-sight confusion. Similarly, sub-filaments are often affected by large velocity gradients, and even velocity discontinuities. These results suggest that extreme care should be taken when using velocity coherent features to constrain the underlying substructure within a filament.          
\end{abstract}


\section{Introduction}%

Filaments have long been known to play an important role in the formation of stars, harbouring significant amounts of high-density molecular gas and acting as sites of core formation \citep{Bar1907,SchElm79}. Recent observations by the Herschel Space Observatory have revealed just how important this role is \citep{And10, Arz13, Kon15, Mar16}. As a result, filaments have been the focus of numerous theoretical and numerical studies \citep{FisMar12,Hei13,Hen13, HenAnd13, Smi14, Fre14,ClaWhi15,SeiWal15, Cla16, Smi16, Cla17}. 

Observations of molecular line emission give information about a filament's gas-phase chemical composition and its internal kinematics. Filaments are found to be kinematically complex, exhibiting multiple velocity components and velocity-coherent features, which have been termed fibres \citep[][Suri et al. in prep.]{Hac13,TafHac15,Hac17,Dha18}. 

Models of filament fragmentation show that, due to their geometry, {\it equilibrium} filaments are prone to fragment into cores, but not sub-filaments \citep{InuMiy92,InuMiy97,Pon11}. \citet{Cla16} show that the {\it non-equilibrium} evolution of an accreting cylindrically symmetric filament changes the spacing of the resulting cores but not the general manner of fragmentation. 

Some recent numerical studies have been able to produce sub-filaments \citep{Smi16,Cla17}. \citet{Smi16} simulate a turbulent self-gravitating cloud, in which small filaments form due to the turbulent fragmentation, and are then swept up into a larger main filament by large-scale motions. This scenario is described as `fray and gather', and the swept-up small filaments are identified as fibres. 

\citet{Cla17} present simulations of single filaments forming in, and accreting from, a turbulent medium. In simulations in which the turbulent energy is comparable to the gravitational energy, sub-filaments form within the main filament, due to the turbulent internal velocity of the main filament; the turbulence is driven by accretion. This scenario conforms to the `fray and fragment' scenario proposed by \citet{TafHac15}. 

Although these simulations produce extended coherent features in position-position-position (PPP) space, it is presently unclear how these features would appear in molecular-line observations, and how they compare with the observed fibres identified in position-position-velocity (PPV) space. It is also unclear whether the observed fibres correspond to coherent features in PPP space. 

In this paper, we present the results of moving-mesh simulations with the same initial setup used in \citet{Cla17}, an initially sub-critical filament which accretes from a supersonic turbulent medium. We use {\it sub-filaments} exclusively to mean extended coherent features in PPP space, and {\it fibres} to mean extended coherent features in PPV space. In Section \ref{SEC:NUM}, we detail the numerical setup, the initial conditions and the production of synthetic observations. In Section \ref{SEC:RES} we present the results of the simulations and the synthetic C$^{18}$O observations. In Section \ref{SEC:DIS}, we discuss the significance of the results and compare to previous work and observations. In Section \ref{SEC:CON}, we summarise our conclusions. 

\section{Numerical Setup}\label{SEC:NUM}%

\subsection{Simulations}%

The simulations presented in this paper\footnote{The preliminary results from these simulations were previously presented in the thesis of Seamus Clarke (http://orca.cf.ac.uk/102784/)} have been performed using the moving-mesh code \textsc{Arepo} \citep{Spr10}. The code uses self-gravitating hydrodynamics, with time-dependent coupled chemistry and thermodynamics. The boundary conditions are periodic for the hydrodynamics, but not for self-gravity. Ten simulations are performed with different random seeds (labelled {\sc Sim}01 to {\sc Sim}10).

The computational domain is defined by Cartesian co-ordinates $(x,y,z)$, with $|x|\!<\!3.0\,{\rm pc}$, $|y|\!<\!3.0\,{\rm pc}$ and $|z|\!<\!2.5\,{\rm pc}$. The initial density field is cylindrically symmetric about the $z$ axis, so we also introduce a radius variable $w\!=\!(x^2+y^2)^{1/2}$, and put
\begin{eqnarray}
\rho(w,z)\!\!\!&\!\!\!=\!\!\!&\!\!\!\left\{\!\!\begin{array}{ll}
15\,{\rm M}_{_\odot}\,{\rm pc}^{-3}\!\left(\!w/{\rm pc}\!\right)^{\!-1},&\!\!\!\!w\!<\!3.0\,{\rm pc},\;|z|\!<\!1.5\,{\rm pc};\\
0.015\,{\rm M}_{_\odot}\,{\rm pc}^{-3},&\!\!\!\!{\rm elsewhere}.\\
\end{array}\right.
\end{eqnarray}
Here the low-density gas is simply a filler in the outer regions of the computational domain, and plays no significant role in the evolution of the filament; it constitutes $\sim 0.01\%$ of the total mass. The initial density field is set up and settled with $\sim 10^6$ cells all having approximately the same mass.

The initial velocity field is given by
\begin{eqnarray}
v(w,z)\!\!\!&\!\!\!=\!\!\!&\!\!\!\left\{\!\!\begin{array}{ll}
-0.75\,\hat{\bf r}\,{\rm km}\,{\rm s}^{-1}\!+\!{\bf v}_{\rm turb},&\!\!\!\!w\!<\!3.0\,{\rm pc},\;|z|\!<\!1.5\,{\rm pc};\\
0\,{\rm km}\,{\rm s}^{-1},&\!\!\!\!{\rm elsewhere}.\\
\end{array}\right.
\end{eqnarray}
Without the turbulent component, ${\bf v}_{\rm turb}$, this gives a cylindrically symmetric inflow of $70\,{\rm M}_{_\odot}\,{\rm Myr}^{-1}\,{\rm pc}^{-1}$ towards the $z$ axis. The dense gas initially at $w\!\sim\!3\,{\rm pc}$ takes $\sim\! 4\,{\rm Myr}$ to reach the $z$ axis. Since the simulations are only run for $\sim\!0.45\,{\rm Myr}$, the low-density filler gas outside $w\!\sim\!3\,{\rm pc}$ does not have time to influence the dynamics near the $z$ axis where the filament is accumulating. The turbulent velocity field, ${\bf v}_{\rm turb}$, is generated assuming a power spectrum $P_k\propto k^{-4}$, with $k_{\rm min}\!=\!4.2\,{\rm pc}^{-1}$, a thermal mix of compressive and solenoidal modes, and a mean velocity dispersion of $1\,{\rm km}\,{\rm s}^{-1}$.

The chemical network in the simulations is a combination of the hydrogen network introduced in \citet{GloMac07a,GloMac07b} and the CO network of \citet{NelLan97}; this combined network is introduced as NL97 in \citet{GloCla12}. We use the cosmic ray heating rate, and the radiative heating and cooling rates presented in \citet{GloMac07a,GloMac07b}; a standard dust-to-gas ratio of 0.01; and solar elemental abundances (relative to hydrogen the abundances by number of helium, carbon and oxygen are respectively $ \chi_{_{\rm He}}\!=\!0.1$, $\chi_{_{\rm C}}\!=\!1.41\times 10^{-4}$ and $\chi_{_{\rm O}}\!=\!3.16\times 10^{-4}$ \citep{GloCla12}). 

We combine the interstellar radiation field (ISRF) defined by \citet{Dra78} at ultraviolet wavelengths, with that defined by \citet{Bla94} at longer wavelengths; the ISRF is normalised to the local ISRF, $G_{_0}\!=\!1.7$ in \citet{Hab68} units. The ISRF is attenuated using the \textsc{TreeCol} algorithm presented in \citet{Cla12}, we direct the reader to that paper for more details. The cosmic ray ionization rate is $\zeta_{_{\rm CR}} = 10^{-17} \, \rm s^{-1}$, consistent with that measured in dense gas \citep{Cas98,Ber99}. 

The gas is initially fully atomic and at 40 K; runs with initially fully molecular gas, and runs at different initial temperatures, show no significant differences. This supports the results of \citet{GloCla12} and \citet{ClaGlo15}, who find that the initial chemical state of a cloud does not significantly alter the global dynamic evolution once the gas density is above $\sim\!100\,{\rm cm}^{-3}$.    

The resolution of the simulation is of order $r_{\rm cell} = (3V_{\rm cell} / 4 \pi)^{1/3}$, where $V_{\rm cell}$ is the volume of a cell. Cell refinement is used to ensure that the resolution always satisfies the Truelove criterion \citep{Tru97}, i.e. $r_{\rm cell}\!<\!\lambda_{\rm Jeans}/8$, where $\lambda_{\rm Jeans}$ is the local Jeans length. The majority of the gas in the filament has density in the range $\sim 10^{-21}$ to $\sim 10^{-18}\,{\rm g}\,{\rm cm}^{-3}$, giving spatial resolution in the range $\sim 3\times 10^{-3}$ to $\sim 3 \times 10^{-4}\,{\rm pc}$. Due to cell refinement the simulations end with $\sim\!10^7$ cells (having started with $\sim\!10^6$). 

\begin{figure*}
\centering
\includegraphics[width = 0.45\linewidth]{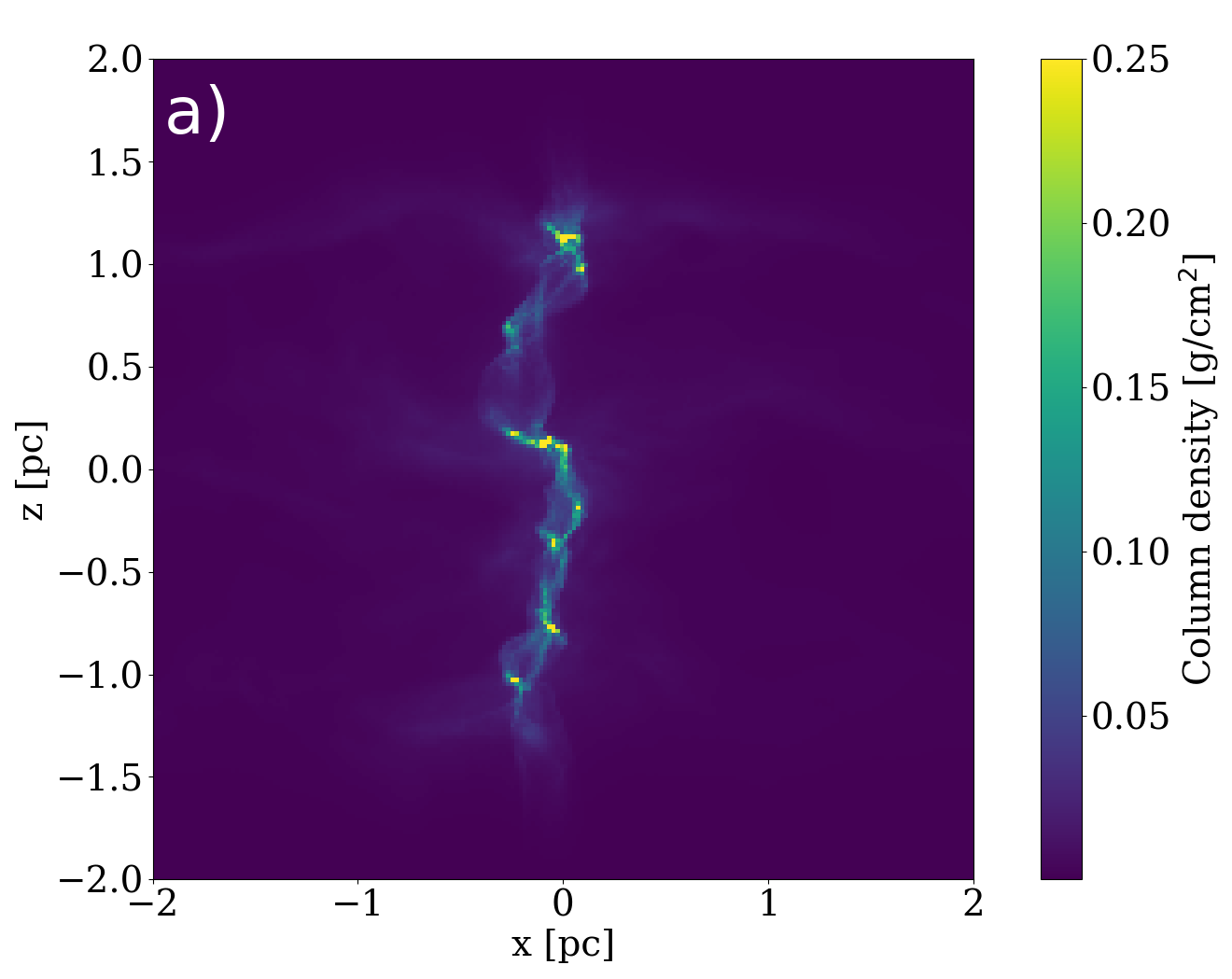}
\includegraphics[width = 0.45\linewidth]{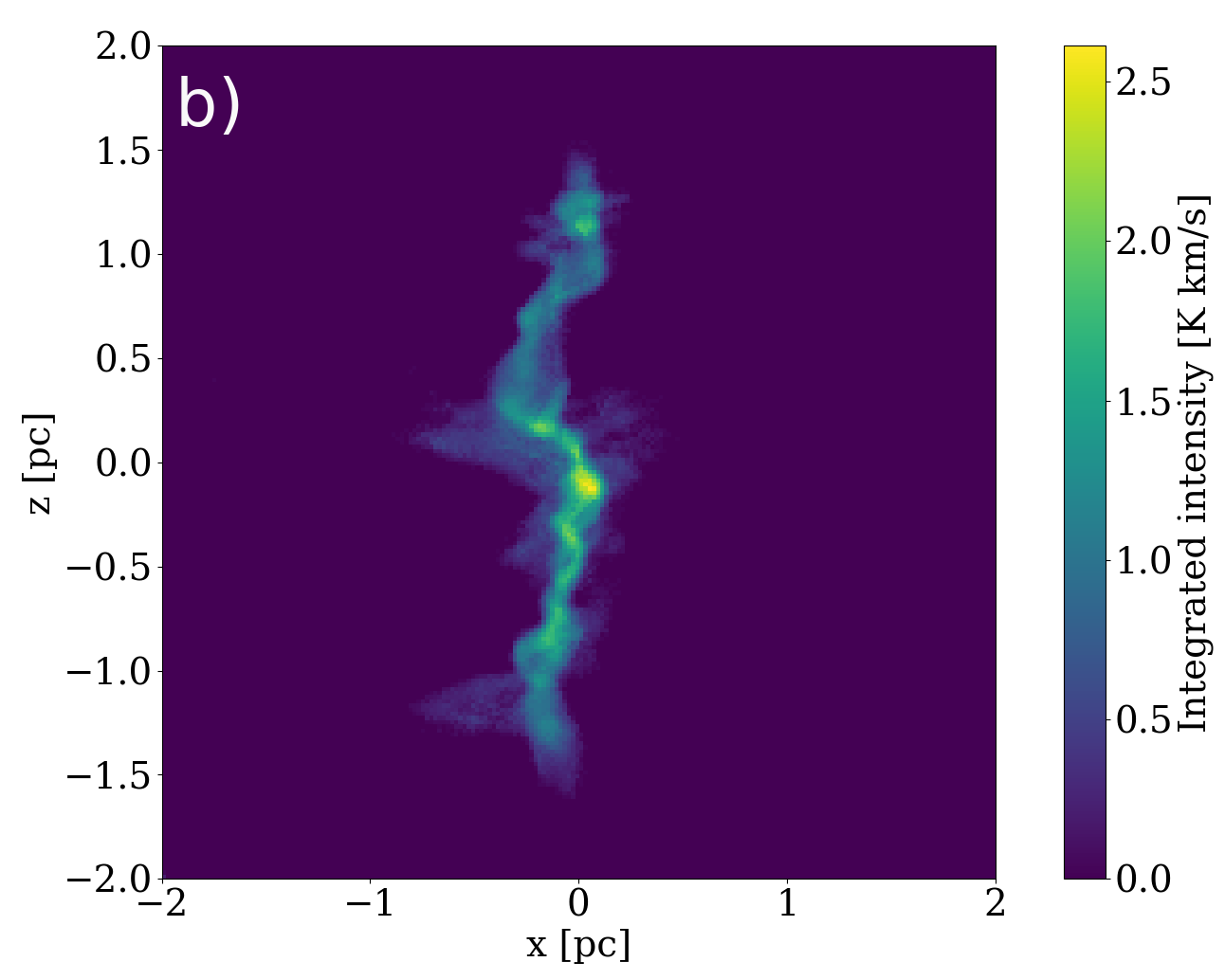}
\includegraphics[width = 0.45\linewidth]{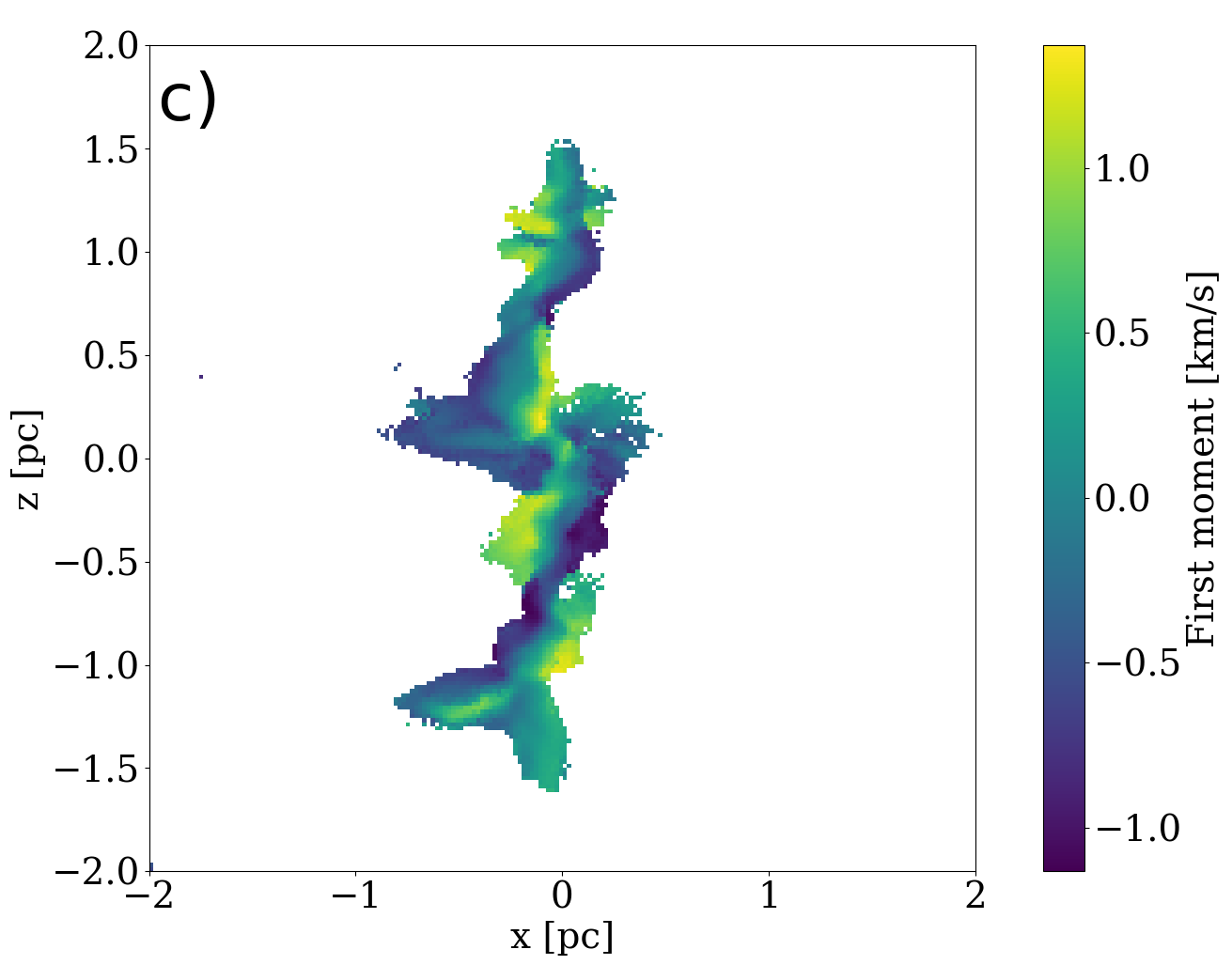}
\includegraphics[width = 0.45\linewidth]{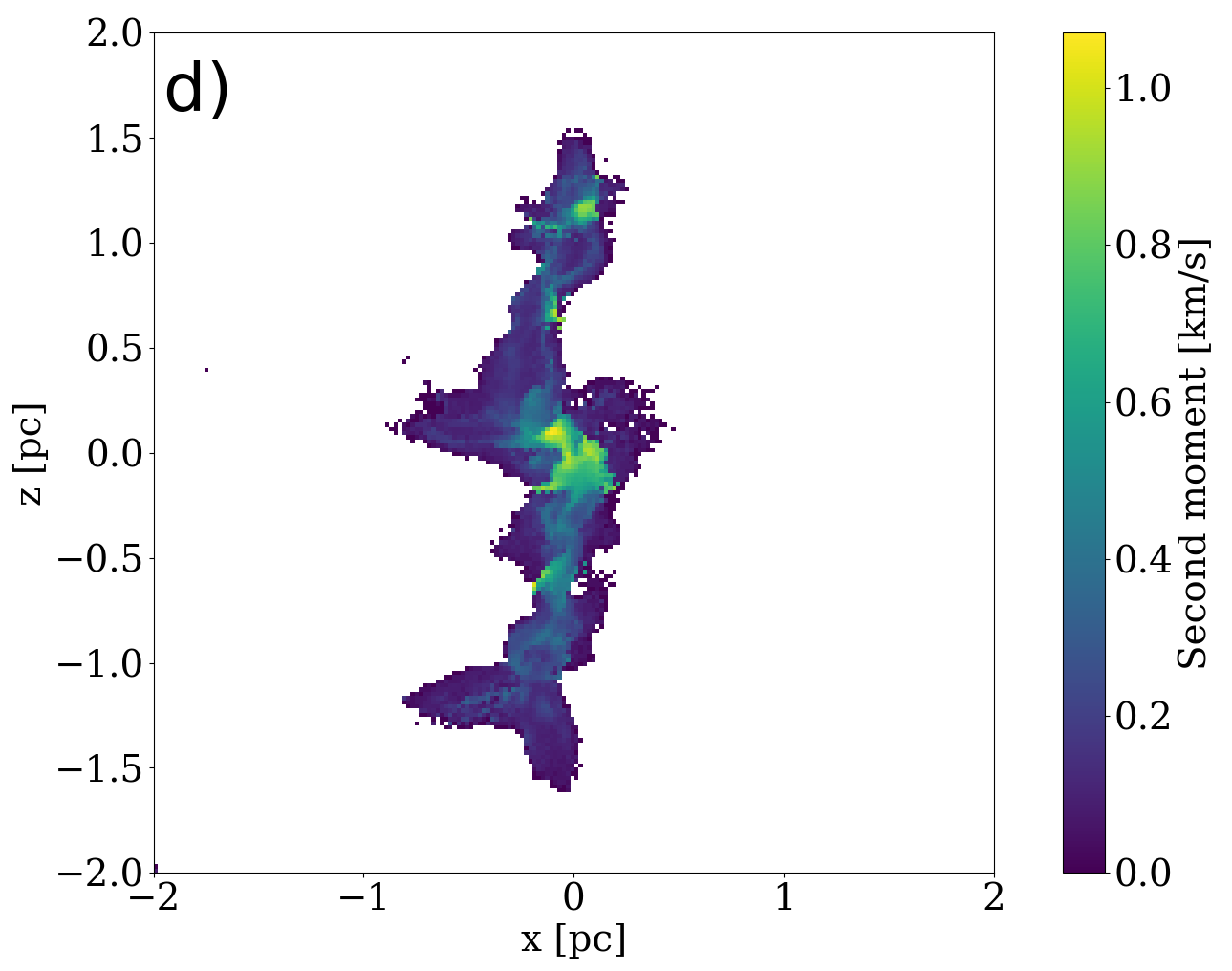}
\caption{Maps showing a) column density, b) integrated intensity, c) intensity-weighted velocity centroid, and d) intensity-weighted velocity dispersion, for the synthetic C$^{18}$O observations of a single representative frame from {\sc Sim}02. The column density map and the synthetic C$^{18}$O maps all have the same pixel size, $0.02\,{\rm pc}$; the column density map has not been convolved with a beam.}
\label{fig::column}
\end{figure*} 

\subsection{Synthetic observations}\label{SSEC:SYN}%

We generate maps of the C$^{18}$O$(J\!=\!1\!-\!0)$ monochromatic intensity, $I_{_{\!v}}^{\rm obs}$ from the simulations, using the post-processing radiative transfer code \textsc{RADMC-3D} \citep{Dul12}. The C$^{18}$O$(J\!=\!1\!-\!0)$ line is chosen because it was used in the first detection of fibres by \citet{Hac13}. 

To run \textsc{RADMC-3D} we use the in-built \textsc{Arepo} algorithm to map the Voronoi mesh onto a fixed Cartesian grid, with a resolution of 0.01 pc, and covering $-2.0\,{\rm pc}< x,y,z <+2.0\,{\rm pc}$; this domain includes the filament and the accretion flow, while omitting most of the low density `filler' gas.

\textsc{RADMC-3D} is run assuming non-local thermodynamic equilibrium and uses the large velocity gradient approximation \citep{Sob57}. To test the applicability of the large velocity gradient approximation the Sobolev length scale is calculated and compared to the grid size. The Sobolev length scale, $L$, is defined as
\begin{equation}
L = \frac{\sigma}{\mid \rm{d}v/\rm{d}r \mid},
\end{equation}
where $\sigma$ is the thermal width of C$^{18}$O and $\mid \rm{d}v/\rm{d}r \mid$ is the absolute velocity gradient in the line of sight. At 10 K the thermal width of C$^{18}$O is $\sim 0.05 \, {\rm km}\,{\rm s}^{-1}$. We find that the median value of the Sobolev length scale is 0.009 pc. As the grid spacing is 0.01 pc, the Sobolev length scale is comparable to the grid size confirming that the large velocity gradient approximation is applicable. We refer the reader to \citet{Oss02} for a discussion on the applicability of the large velocity gradient approximation in a turbulent medium and its accuracy.

The energy levels and excitation coefficients of C$^{18}$O are taken from the Leiden Atomic and Molecular Database \citep{Sch05} and uses the work of \citet{Yan10}. The number density of C$^{18}$O is given by $n_{_{\rm{C^{18}O}}} = n_{_{\rm{C^{16}O}}} / 500$ \citep{WilRoo94}; C$^{16}$O is the isotopologue traced in the modified NL97 chemical network. 

Micro-turbulence is usually included in synthetic observations, to account for the unresolved velocity dispersion on scales smaller than the grid resolution. We do not include micro-turbulence here, since the gas inside the filament has a velocity dispersion of only $\sim 0.05\,{\rm km}\,{\rm s}^{-1}$ on scales of $\sim 0.01\,{\rm pc}$. This is much less than the typical C$^{18}$O line-widths in the synthetic spectra, so the inclusion of micro-turbulence would have very little effect on the analysis.

\begin{figure*}
\centering
\includegraphics[width = 0.95\linewidth]{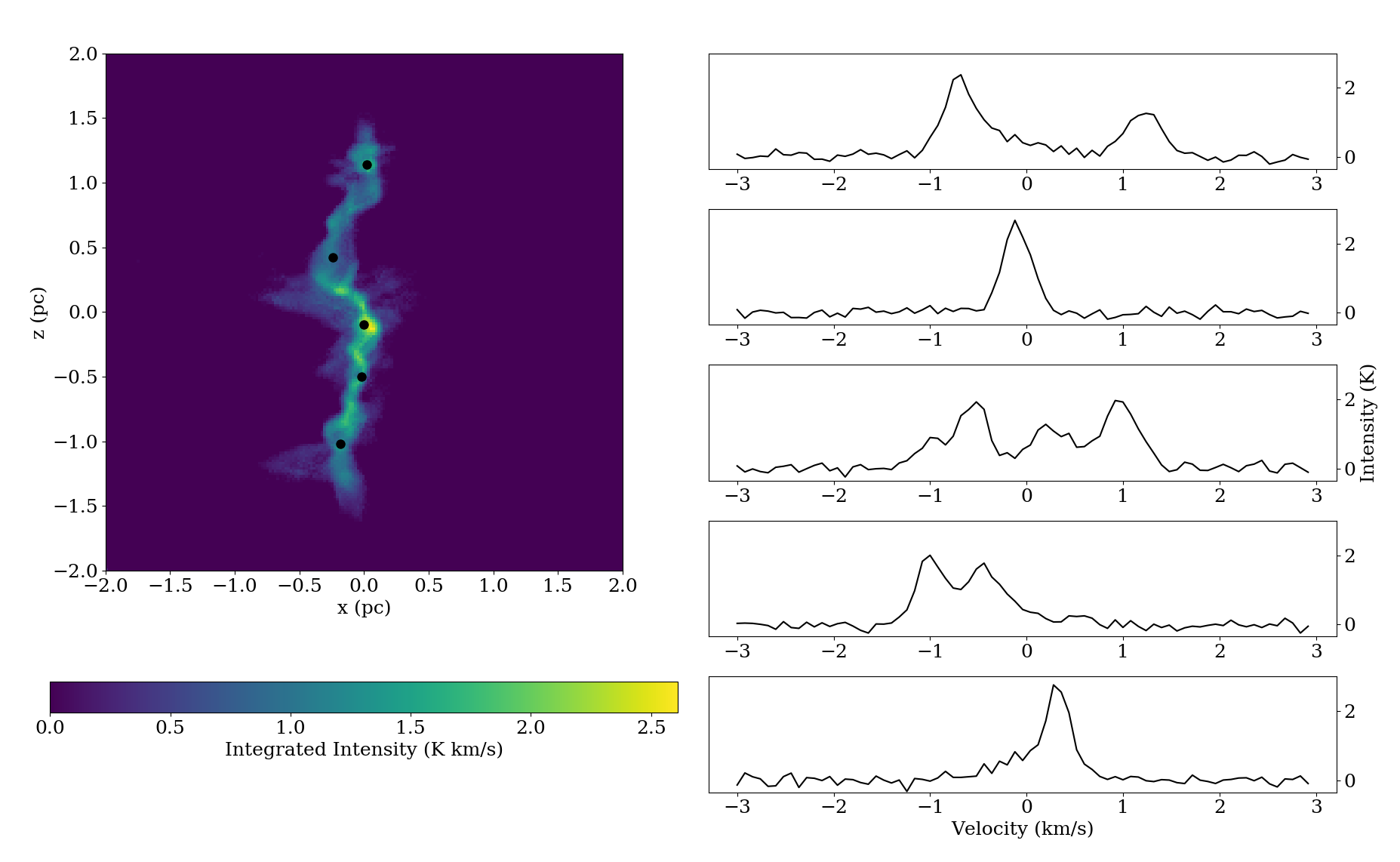}
\caption{On the left, the integrated intensity map of the C$^{18}$O emission, with black dots showing the location of the five spectra displayed on the right. Four of the five spectra show multiple velocity components (as shown in Section \ref{SSEC::FITTING} and Fig. \ref{fig::fitting}).}
\label{fig::multvel}
\end{figure*} 

At high densities, CO is expected to freeze-out on to dust grains \citep{Cas99,Red02,Sav03,Chr12,Gia16}. Since CO freeze-out is not included in the modified NL97 chemical network, we use the freeze-out approximation introduced in \citet{Hol09}, which estimates the instantaneous equilibrium between freeze-out and desorption. We note that this method probably overestimates the degree of freeze-out, since it yields $>50\%$ freeze-out at number densities $n_{\rm H_2}\!>\!3\times 10^{4}\,{\rm cm}^{-3}$, whereas observations suggest that this only occurs  for $n_{\rm H_2}\!>\!10^{5}\,{\rm cm}^{-3}$ \citep{Lip13}. 

\citet{Hac13} used the 14m FCRAO telescope to produce their C$^{18}$O map of L1495 in Taurus. This gave a velocity resolution of $0.07\,{\rm km}\,{\rm s}^{-1}$; a spatial resolution of $60"$, corresponding to $\sim 0.04\,{\rm pc}$ at the distance of Taurus ($140\,{\rm pc}$; \citet{Eli78}); and a noise level of $\sim\!0.1\,{\rm K}$ per velocity channel. The synthetic image from \textsc{RADMC-3D} is noiseless and has better spatial and velocity resolution than the \citet{Hac13} data, {\it viz.} $0.01\,{\rm pc}$ and $0.02\,{\rm km\,s}^{-1}$, respectively. Therefore we degrade the synthetic images by rebinning the velocity channels, convolving each velocity channel map with a 2D-Gaussian having full width half maximum (FWHM) of $0.04\,{\rm pc}$, and finally adding noise. For each voxel, the noise is obtained by sampling from a Gaussian distribution with a mean of zero and a standard deviation of $0.1\,{\rm K}$. 

\section{Results}\label{SEC:RES}%

The {\sc Arepo} simulations presented here produce very similar morphologies to those presented in \citet{Cla17}, which used smoothed particle hydrodynamics. Fig. \ref{fig::column}a shows the column density from one frame of simulation {\sc Sim}02 (this frame is used throughout the paper to illustrate the procedures used). One can see numerous elongated sub-filaments. 

Fig. \ref{fig::column}b shows the corresponding C$^{18}$O integrated intensity map. The morphology of the filament is unchanged, but the sub-structure is much less sharp than in Fig. \ref{fig::column}a, and due to freeze-out some of the column-density peaks are missing. The lack of obvious signs of fibres in the integrated intensity map is similar to the results of \citet{Hac13}.

Figure \ref{fig::column}c shows the first moment map, i.e. the intensity-weighted mean radial velocity along each line of sight. Only velocity channels with greater than $5\sigma$ detections (i.e. $>\!0.5\,{\rm K}$) are included in the calculation.  The filament shows a complex velocity field with large alternating radial velocities, spanning a range of $\sim\,2.0\,{\rm km}\,{\rm s}^{-1}$. This corresponds to the range of velocities in the accretion flow, i.e. $\sim 1.0\,{\rm km}\,{\rm s}^{-1}$ towards the observer and $\sim 1.0\,{\rm km}\,{\rm s}^{-1}$ away from the observer, and is very similar to the range seen in L1495 by \citet{TafHac15}.

Figure \ref{fig::column}d shows the second moment map, i.e. the intensity weighted velocity dispersion along each line of sight. As with the first moment map, only velocity channels with greater than $5\sigma$ detections are included in the calculation. The second moment varies greatly over the filament, from regions with sub- or trans-sonic widths, to regions with highly supersonic widths. Small absolute radial velocities and large velocity dispersions tend to be concentrated near the spine of the filament. Conversely, large absolute radial velocities and small velocity dispersions tend to be found towards the edges of the filament. This is because the edges trace the shocks where the inflowing gas accretes onto the filament, while lines of sight near the spine are seeing through turbulent gas inside the filament.

Plots similar to Fig. \ref{fig::column} for frames from the other nine simulations are presented in Appendix B.   

\begin{figure*}
\centering
\includegraphics[width = 0.95\linewidth]{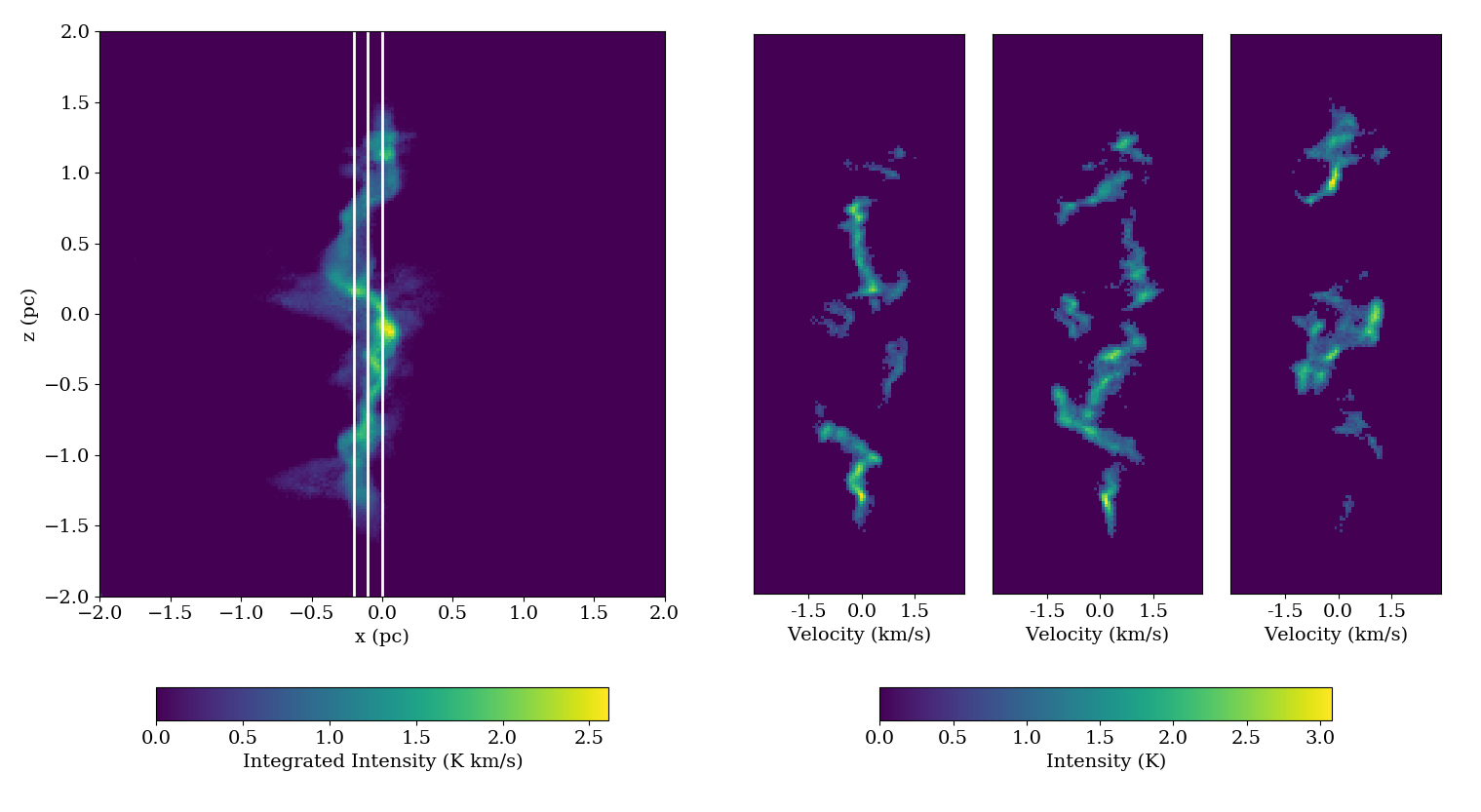}
\caption{On the left, the integrated intensity map of the C$^{18}$O emission, with white vertical lines showing the location of the three position-velocity plots displayed on the right.}
\label{fig::longslice}
\end{figure*}

\begin{figure*}
\centering
\includegraphics[width = 0.95\linewidth]{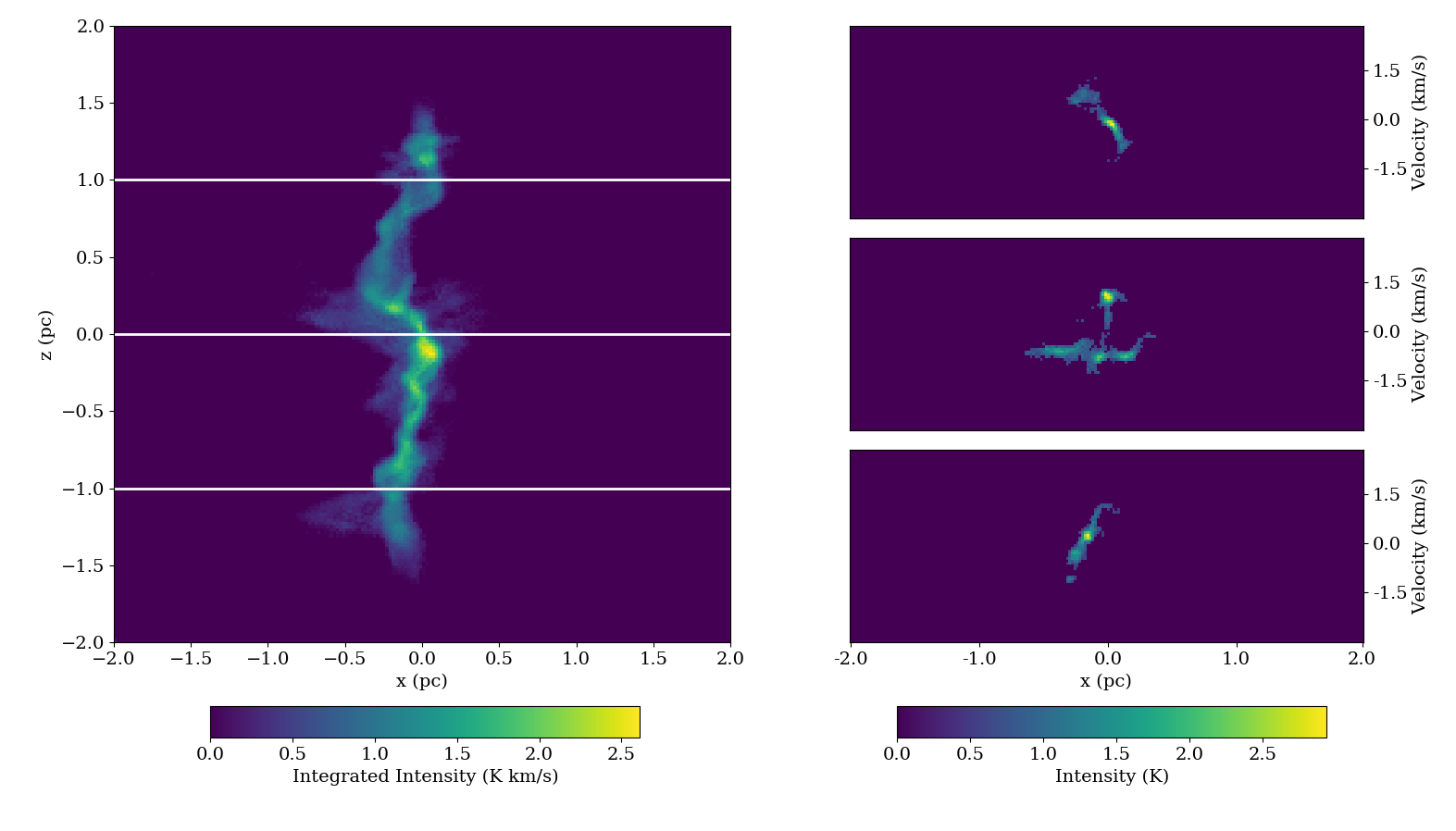}
\caption{On the left, the integrated intensity map of the C$^{18}$O emission, with white horizontal lines showing the location of the three position-velocity plots displayed on the right.}
\label{fig::horslice}
\end{figure*} 

\section{Discussion}\label{SEC:DIS}%

\begin{figure*}
\centering
\includegraphics[width = 0.95\linewidth]{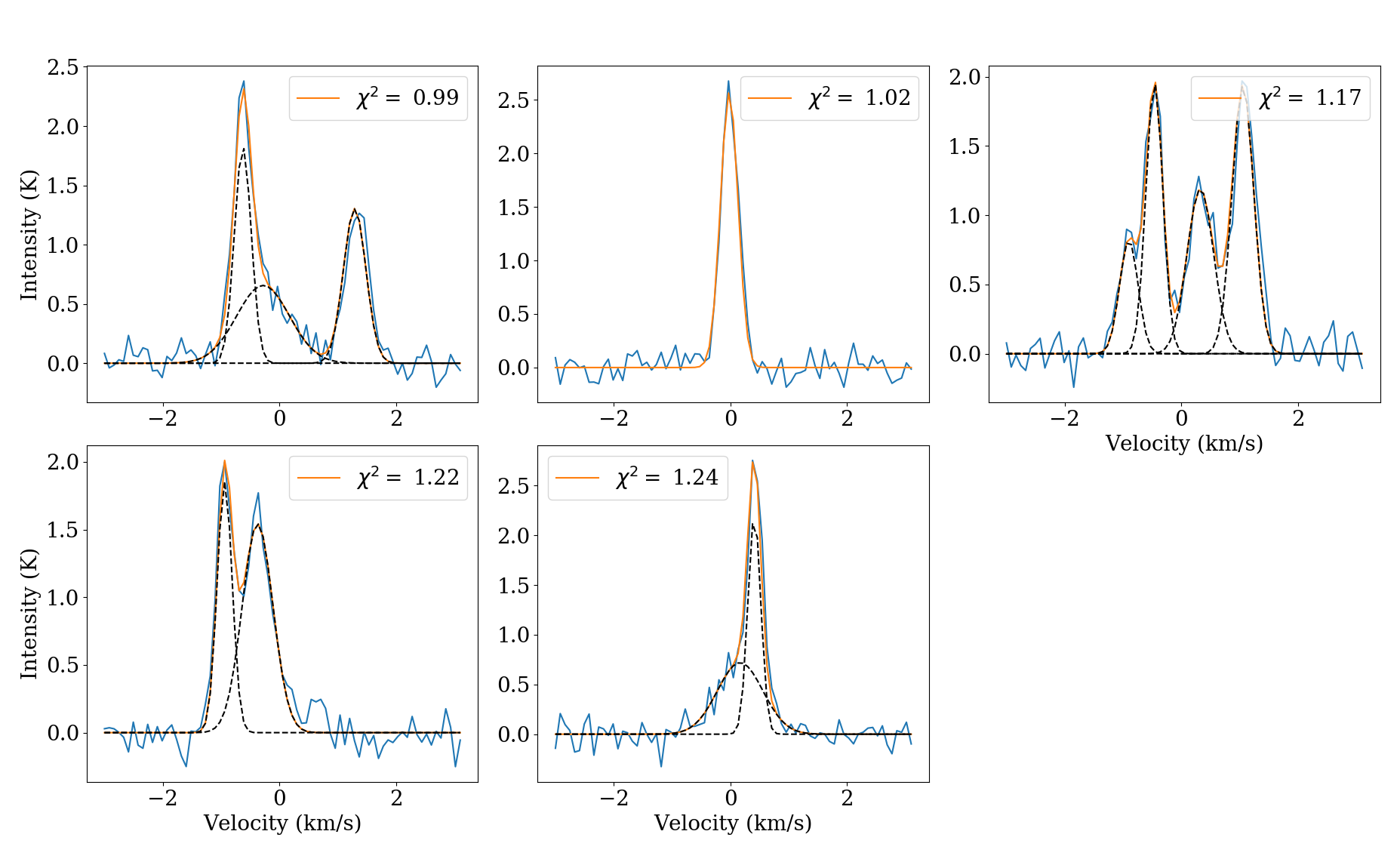}
\caption{The same five spectra displayed in Fig. \ref{fig::multvel} are plotted in blue, overlaid with the fits in yellow. The reduced $\chi^2$ value is shown on each plot. Individual velocity components are displayed with black dashed lines.}
\label{fig::fitting}
\end{figure*} 

To investigate whether the large second moment values in the interior of the filament are due to one single broad velocity component or multiple narrow velocity components, we show in Fig. \ref{fig::multvel} the spectra on five different representative lines of sight. Several spectra have multiple velocity components, and most components have sub- or trans-sonic widths. This prevalence of multiple narrow velocity components suggests the existence of fibres.

To distinguish between small localised regions with multiple velocity components (e.g. accreting cores) and regions which contain multiple elongated substructures at different velocities, we use position-velocity (PV) diagrams. Figure \ref{fig::longslice} displays longitudinal strips of the monochromatic intensity, $I_{_{\!v}}^{\rm obs}$, in PV space. What appears to be a single filament on the integrated intensity map is seen to be highly structured in velocity, and composed of fibres. This can also be seen in Fig. \ref{fig::horslice}, which displays horizontal strips of $I_{_{\!v}}^{\rm obs}$ in PV space. There are often high velocity gradients across a filament, and also multiple structures aligned along the line of sight but separated by $\sim\!1\,{\rm km}\,{\rm s}^{-1}$ in radial velocity. 

Due to the complex kinematics and number of multiple velocity components, it is apparent that the first and second moment maps can be misleading, and we need to fit the individual velocity components in each spectrum. 

\subsection{Fitting multiple velocity components}\label{SSEC::FITTING}%

We have develop a new, fully automated routine (Behind the Spectrum, BTS) for fitting multiple velocity components in optically thin lines. The routine does not assume the number of components in the spectrum \textit{a priori}, but uses the first, second and third derivatives to determine their number and positions. A least-squared fitting routine is then used to determine the best fit with that number of components, checking for over-fitting and over-lapping velocity centroids. A detailed explanation of BTS is given in Appendix A, along with tests.\footnote{BTS can be downloaded from https://github.com/SeamusClarke/BTS} Fig. \ref{fig::fitting} shows the fits for the five spectra presented in Fig. \ref{fig::multvel}.

\begin{figure*}
\centering
\includegraphics[width = 0.95\linewidth]{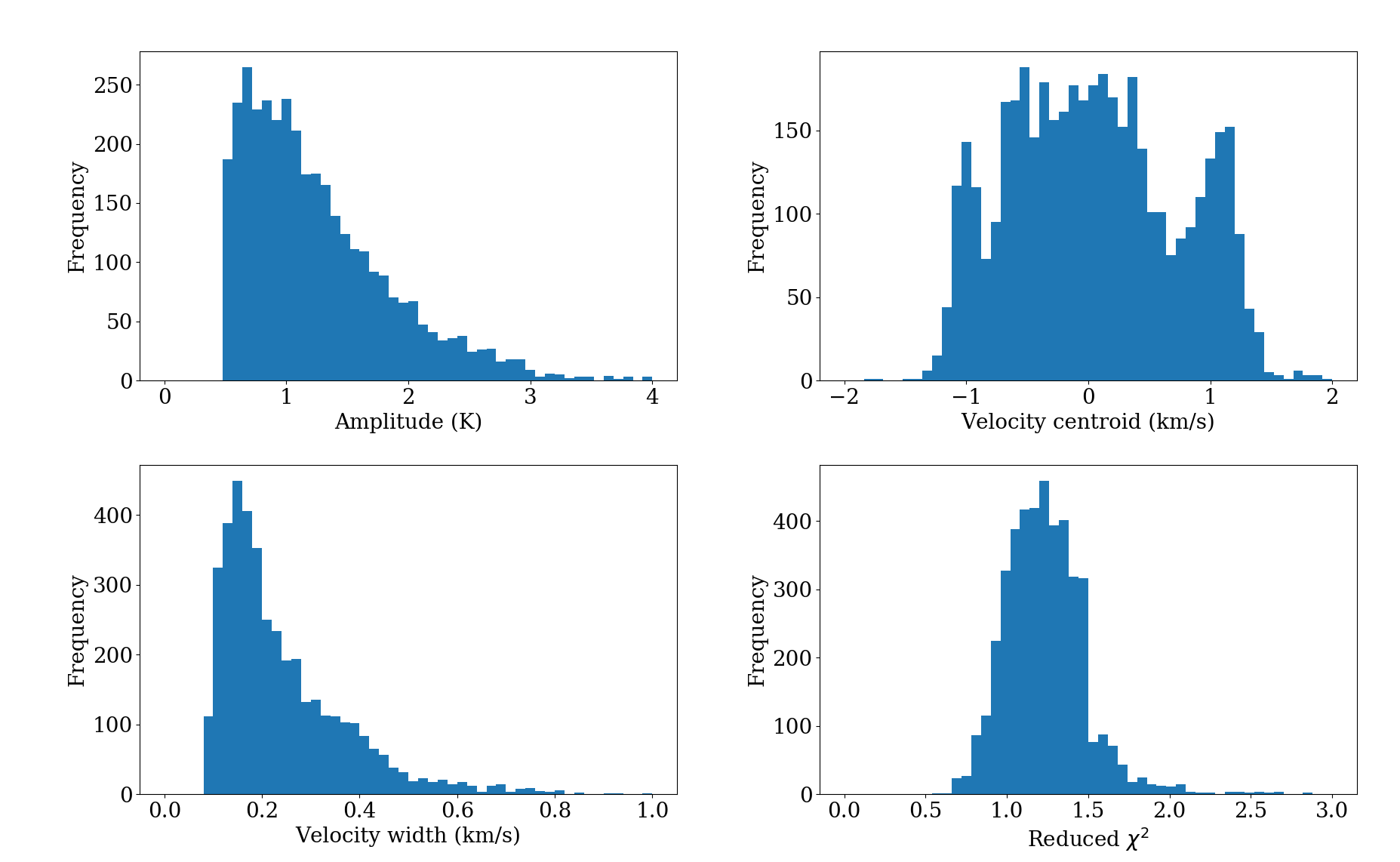}
\caption{Histograms showing the distributions of amplitude, velocity centroid, velocity width, and reduced $\chi^2$.}
\label{fig::hist}
\end{figure*}

For each component, $c$, BTS returns the amplitude (i.e. central intensity $I_{_c}^{\rm o}$), velocity centroid ($v_{_c}$) and dispersion ($\sigma_{_c}$). Fig. \ref{fig::hist} shows histograms of these parameters, and the reduced $\chi^{2}_{_c}$ values. The amplitudes peak just below $1\,{\rm  K}$. The distribution of velocity centroids is roughly symmetric, but non-Gaussian. The velocity dispersions are predominately sub- or trans-sonic, peaking at around $0.2\,{\rm km}\,{\rm s}^{-1}$. The reduced $\chi^{2}$ values peak just above $\sim 1$, which is the value for a good fit. Values less than 1 may be due to over-fitting, but they are rare. There are some fits with $\chi^{2}\!>\!2$, but these too are rare ($1.4\%$ of all fits), and are excluded from further analysis. 

The velocity centroid distribution can be approximated by a wide, approximately Gaussian distribution, centred on $0\, {\rm km}\,{\rm s}^{-1}$, which represents the turbulent gas inside the filament, plus two narrower outlying peaks at $\pm 1\,{\rm km}\,{\rm s}^{-1}$, which represent the inflowing gas. Combining the inflow velocity, $v_{\rm in}\!\simeq\!1\,{\rm km}\,{\rm s}^{-1}$, with the mean radius of the filament, $R_{\rm filament}\!\simeq\!0.2\,{\rm pc}$ and the critical density for C$^{18}$O$(J=1-0)$ excitation, $n_{\rm crit}\!\simeq\!1400\,{\rm H}_2\,{\rm cm}^{-3}$, we can estimate the mass inflow rate onto unit length of the filament, ${\dot M}\!\sim\!2 \pi R_{\rm filament}n_{\rm crit}\,{\bar m}\,v_{\rm in}\!\sim\!90 \,{\rm M}_{_\odot}\,{\rm pc}^{-1}\,{\rm Myr}^{-1}$; ${\bar m}\sim 5\times 10^{-24}\,{\rm g}$ is the mean mass associated with each H$_2$ molecule. The initial accretion rate onto the filament is $\sim\!70\,{\rm M}_{_\odot}\, {\rm pc}^{-1}\,{\rm Myr}^{-1}$, but due to gravitational acceleration this increases to $\sim\!100\,{\rm M}_{_\odot}\, {\rm pc}^{-1}\,{\rm Myr}^{-1}$ by the time at which the synthetic observations are produced. Thus, C$^{18}$O($J\!=\!1\!-\!0$) observations may afford a way of estimating accretion rates.          

The distribution of velocity dispersions is strongly peaked at the sound speed, $c_{s}\!\sim\! 0.2\,{\rm km}\,{\rm s}^{-1}$, and 89$\%$ of components have widths below the transonic limit at  $2c_{s}\!\sim\!0.4\,{\rm km}\,{\rm s}^{-1}$. This is in agreement with recent observational studies, which show that turbulence in filaments is typically sub- or trans-sonic \citep{Arz13,Hac13,Fer14,Kai16}, and with SPH simulations of forming filaments by \citep{Cla17}, which show that the low levels of turbulence in filaments can be maintained by the lumpy accretion flow from the surrounding turbulent medium.  Since on many lines of sight the spectrum has several distinct components, each with sub- or trans-sonic dispersion, this is macro-turbulence, i.e. bulk structures with sub- or trans-sonic internal velocity dispersion, moving at trans- or super-sonic velocities with respect to each other. Other simulations of turbulence driven by mass accretion show similar results, i.e.  the driven turbulence is not isotropic but highly structured \citep{Hei11}.

\subsection{Identifying fibres in PPV space}\label{SSEC::FIBRES}%

\begin{figure*}
\centering
\includegraphics[width = 0.95\linewidth]{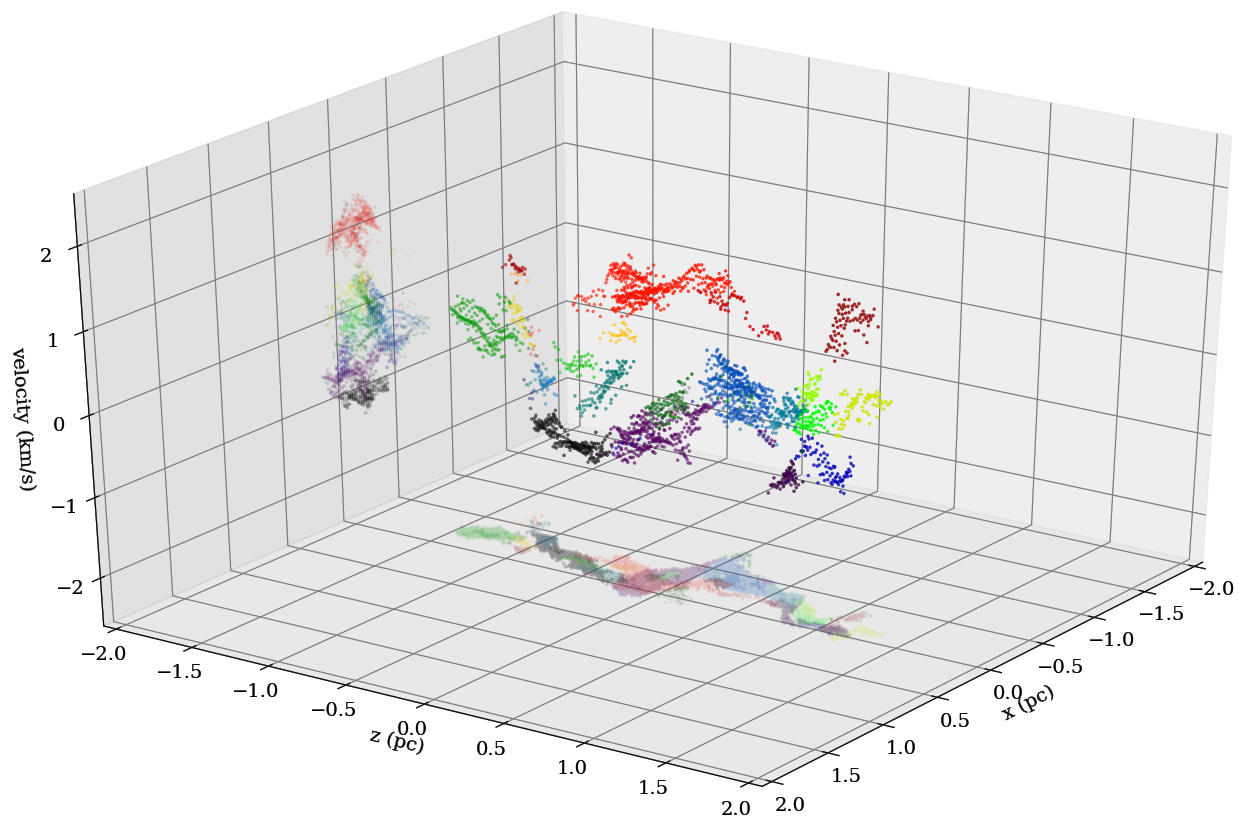}
\caption{A 3D plot showing the velocity centroids of the velocity components in PPV space, colour coded to show the individual groups identified using the friends-of-friends algorithm. The structures are generally filamentary and resemble the fibres in \citet{Hac13}}
\label{fig::fibres}
\end{figure*} 

\citet{Hac13} use a friends-of-friends (FoF) algorithm to identify fibres, and we follow their procedure as closely as possible. For this purpose, the PPV data comprise points representing the centres of pixels, $(x_{_p},y_{_p})$ and the centroids of velocity components identified on the associated lines of sight, $v_{_{p,c}}\,(1\!\leq\!c\!\leq\!{\cal C}_{_p})$, where ${\cal C}_{p}$ is the number of components along sight-line $p$. Those points, $(x_{_p},y_{_p},v_{_{p,c}})$, that have signal-to-noise ratio {\sc snr}$\,>6$ {\it and} at least 4 `good neighbours' are classified as `Grade 1', and the rest as `Grade 2'; a good neighbour is an adjacent pixel (one of 8) which has a velocity component, $(p',c')$, with (i) sufficiently close velocity centroid, $v_{_{p',c'}}$, that the gradient between them,
\begin{eqnarray}
\nabla v&=&(v_{_{p,c}}-v_{_{p',c'}})/\sqrt{(x_{_p}-x_{_{p'}})^2+(y_{_p}-y_{_{p'}})^2}\;,
\end{eqnarray}
satisfies $|\nabla v|\!<\!3\,{\rm km}\,{\rm s}^{-1}\,{\rm pc}^{-1}$, and (ii) {\sc snr}$\,>6$. Next, we run a FoF  search on the Grade 1 points, starting with the brightest one, and using a separation threshold of $0.04\,{\rm pc}$ (2 pixels) and a velocity-gradient threshold of $3\,{\rm km}\,{\rm s}^{-1}\,{\rm pc}^{-1}$. Once this search is complete, friendship groups with fewer than 8 Grade 1 points are discarded as being insignificant. Finally, starting from the friendship groups with more than 7 Grade 1 points, we extend the FoF search to the Grade 2 points, using the same separation and velocity-gradient thresholds as before; this dilates the existing fibres. Table \ref{tab::numfibres} gives the number of fibres identifed in each simulation.

Figure \ref{fig::fibres} shows the fibres from {\sc Sim 2}, demonstrating that they are elongated and similar in morphology to those identified by \citet{Hac13}. When we repeat this analysis without the CO freeze-out post-processing step, the fibres identified are almost identical, suggesting that the presence of fibres is not sensitive to the tracer used. Indeed, fibres have also been observed using N$_2$H$^+$, a tracer which is not affected by freeze-out \citep{Hac17}.

\subsection{Mapping fibres from PPV into PPP space}\label{SEC::PPVPPP}%

\begin{figure*}
\centering
\includegraphics[width = 0.90\linewidth]{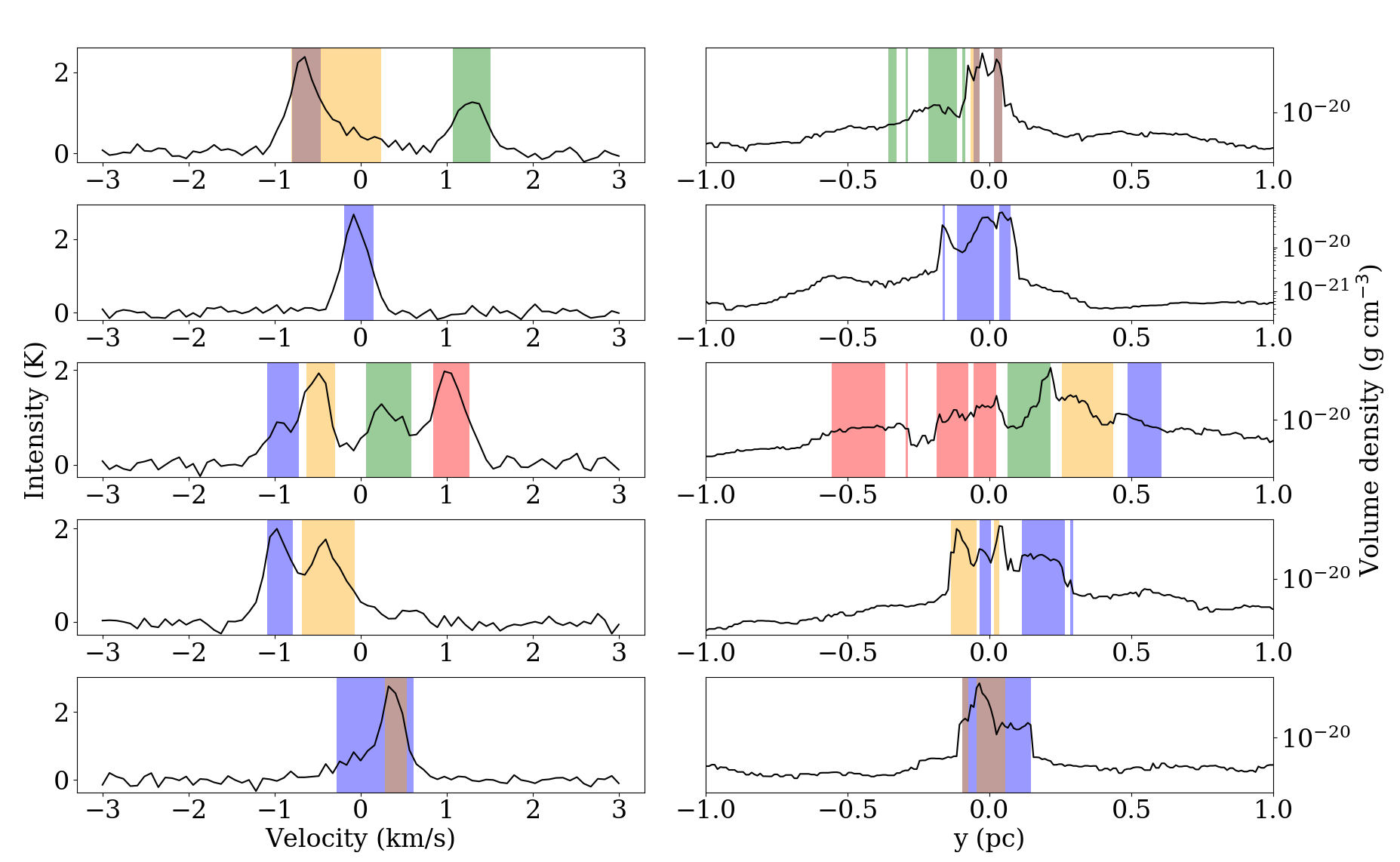}
\caption{The lefthand frames display the same five spectra presented in Fig. \ref{fig::multvel}, and the righthand frames display the density profiles along the corresponding lines-of-sight. The coloured bars identify the ranges contributing to the different velocity components ($v_{_c}\pm 1.15\sigma_{_c}$).}
\label{fig::singlemapping}
\end{figure*} 

\begin{figure*}
\centering
\includegraphics[width = 0.90\linewidth]{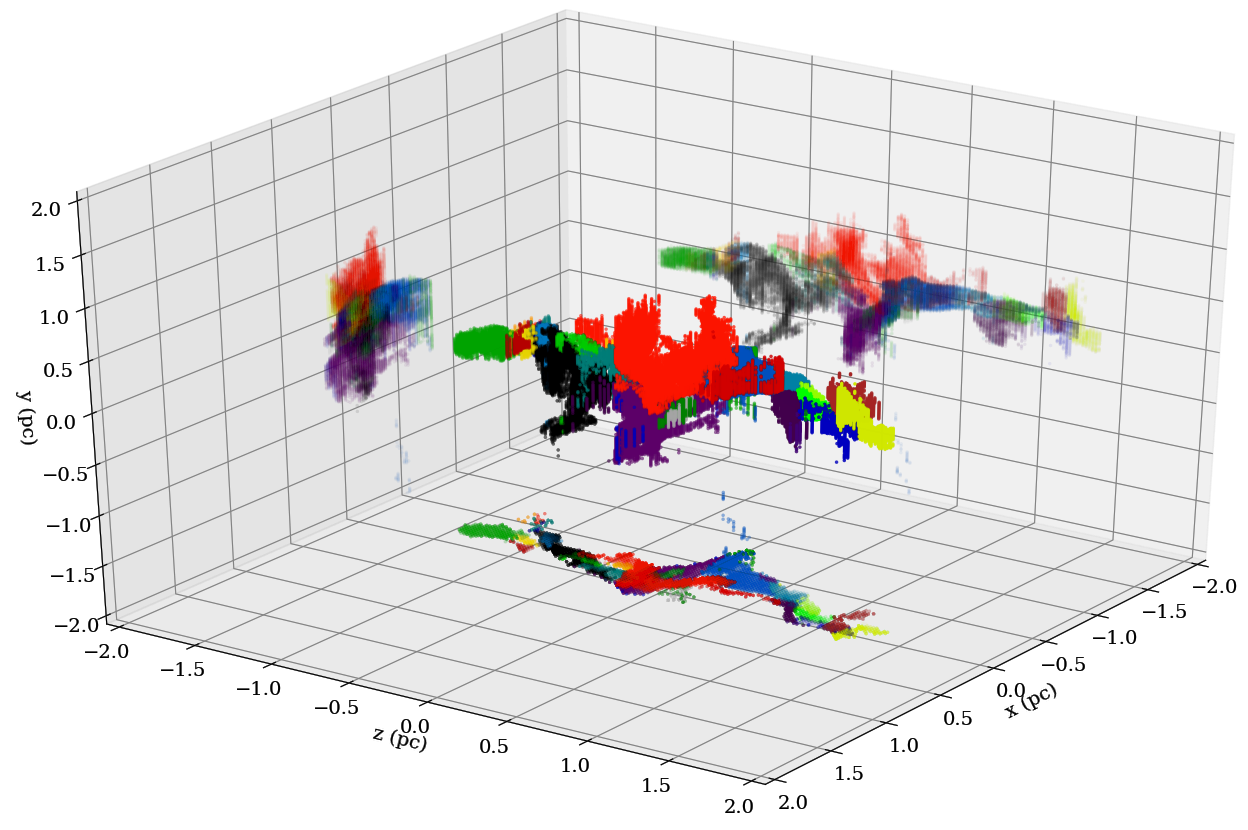}
\caption{A 3D plot showing the fibres identified in Fig. \ref{fig::fibres} mapped into PPP space, using the same colour code.}
\label{fig::PPPmapped}
\end{figure*}

It is sometimes assumed that coherent structures in PPV space correspond to coherent structures in PPP space. We test here how reliable this assumption is for simulations of filament formation. Similar tests have been performed on cloud scale simulations -- albeit without chemistry and radiative transfer -- showing that projection effects are important and can greatly complicate the mapping between PPV and PPP \citep{Moe15,Zam17}. Here we test the assumption on a single simple structure, an isolated filament. 

To isolate the gas in fibres, we apply two criteria. First, for each velocity component $c$ (defined by $I_{_c},v_{_c},\sigma_{_c}$) we consider only the gas along that line of sight having velocity in the interval $v_{_c}\pm 1.175\sigma_{_c}$, i.e. within the FWHM. Second, since we are using the C$^{18}$O($J\!=\!1\!-\!0$) line, we only consider gas which has a number density greater than the critical value for this transition, $n_{\rm crit}\!\sim\!1400\,{\rm cm}^{-3}$, in order to limit line of sight contamination by diffuse gas in the accretion flow. From the simulations we know that the median density of cells with at least 10$\%$ of their carbon in CO is $\sim\!1300\,{\rm cm}^{-3}$, so using $n_{\rm crit}$ as a density threshold is reasonable. 

Figure \ref{fig::singlemapping} illustrates the results of applying this procedure to the five spectra in Fig. \ref{fig::multvel}. In the lefthand panels, the coloured bands mark the FWHMs of the different velocity components, i.e. from $v_{_c}\!-\!1.175\sigma_{_c}$ to $v_{_c}\!+\!1.175\sigma_{_c}$. In the righthand panels, the density profiles along the corresponding lines of sight are presented, and the coloured bands mark the regions contributing to the different velocity components. Line of sight confusion is evident in several cases. In the bottom spectrum the overlapping narrow and wide components come from the same PPP feature, with the narrow component tracing quiescent material near the density peak, and the wide component tracing more extended material accreting onto this peak. In the middle spectrum, the four components originate from a region $\sim$1 pc wide; the main density feature along this line of sight (roughly between 0.1 and 0.4 pc) contains two velocity components (green and yellow bands), due to the convergent flow forming it, whilst the red and blue components are associated with density enhancements in the accretion flow, with the red component being associated with two distinct density features. The fourth spectrum shows severe line-of-sight confusion, with alternating velocity components (yellow and blue bands) due to acoustic oscillations.  
  
The fibres identified in PPV space (see Section \ref{SSEC::FIBRES}) are found by collating the centroids of the fitted velocity components, and so can be mapped into PPP space by summing all the voxels contributing to the FWHMs of those velocity components. Fig. \ref{fig::PPPmapped} shows the distribution in PPP space of all the fibres in Fig.  \ref{fig::fibres}. In PPP space, most fibres are compounded by continuous features, but they are also often fragmented or indistinct -- due to confusion, blending at boundaries, and overlaps.

In order to quantify this assertion, we estimate the proportions of fibres that are `complete', `contaminated' or `separate'. To do this we use a FoF algorithm (as in Section \ref{SSEC::FIBRES}) to identify groups of voxels in PPP space. To be friends, two voxels must be no more than $0.04\,{\rm pc}$ (2 voxels) apart. A fibre is `complete' if all the associated voxels belong to the same group. A fibre is `separate' if the associated voxels belong to more than one group, {\it and} two of these groups overlap on more than 7 lines of sight (i.e. 7 pixels). A fibre is `contaminated'  if the associated voxels belong to more than one group, but none of these groups overlap on more than 7 lines of sight. Table \ref{tab::numfibres} gives the number of fibres that fall into these categories; around 50\% of fibres are complete in PPP space, 30\% are contaminated, and 20\% are separated. 

We conclude that great caution must be exercised when discussing the properties of features identified in PPV space. In the simulations, only half of such features are free from contamination along the line of sight and attributable to a single feature in PPP space; one cannot know which ones without knowledge of the third spatial dimension. Moreover, many of those which are attributable to a single feature in PPP space would not be identified as coherent features in PPP space, as evidenced by the considerable overlap at the boundaries of features in Fig. \ref{fig::PPPmapped}. This is consistent with the results of \citet{Zam17}, who, using cloud-scale simulations, show that what is defined as a fibre is dependent on the viewing angle, and that fibres are often formed by density enhancements which are separated by over a parsec along the line-of-sight. 

While there appears to be rather poor correspondence between  fibres identified in PPV space, and sub-filaments identified in PPP space, the detection of fibres within a larger filament does indicate that there is a significant level of internal macro-turbulence. As the internal macro-turbulence is likely to be driven by lumpy accretion from the surrounding medium, fibres may be a good indicator of ongoing accretion. Conversely, a lack of fibres within a filament may reflect a low level of ongoing accretion.

\subsection{Identifying sub-filaments in PPP space}%

We use the \textsc{DisPerSE} \citep[Discrete PERsistent Structures Extractor,][]{Sou11} algorithm to locate filaments in PPP space. \textsc{DisPerSE} identifies critical points where the density gradient goes to zero, and integral lines connecting neighbouring critical points. These integral lines define the spine of a filaments; the ratio between the densities at either end of an integral line defines the persistence ratio, giving a measure of how robust that element of the spine is. 

To run \textsc{DisPerSE} we use the logarithm of the number density, in order to reduce the dynamic range. Spines are retained if the persistence ratio is greater than 0.3 (i.e. $\Delta\!\log_{_{10}}\!(n)\leq 0.3$), the density is greater than $10^3\,{\rm cm}^{-3}$, and the spine connects at least 10 points. Before analysis, spines are smoothed using \textsc{DisPerSE}'s inbuilt function \textsc{skelconv} and the option \textsc{smooth} with a smoothing length of 5 points. 

Once the spine of a sub-filament has been found, we determine the gas that is associated with it, by producing a radial density profile at every spine point and collating all points within the FWHM of this profile. The profile at a spine point is obtained by first defining the plane perpendicular to the spine at the point, and then using points that lie on or near this plane to produce an azimuthally averaged Gaussian radial density profile, from which the FWHM can be obtained.

\begin{figure*}
\centering
\includegraphics[width = 0.95\linewidth]{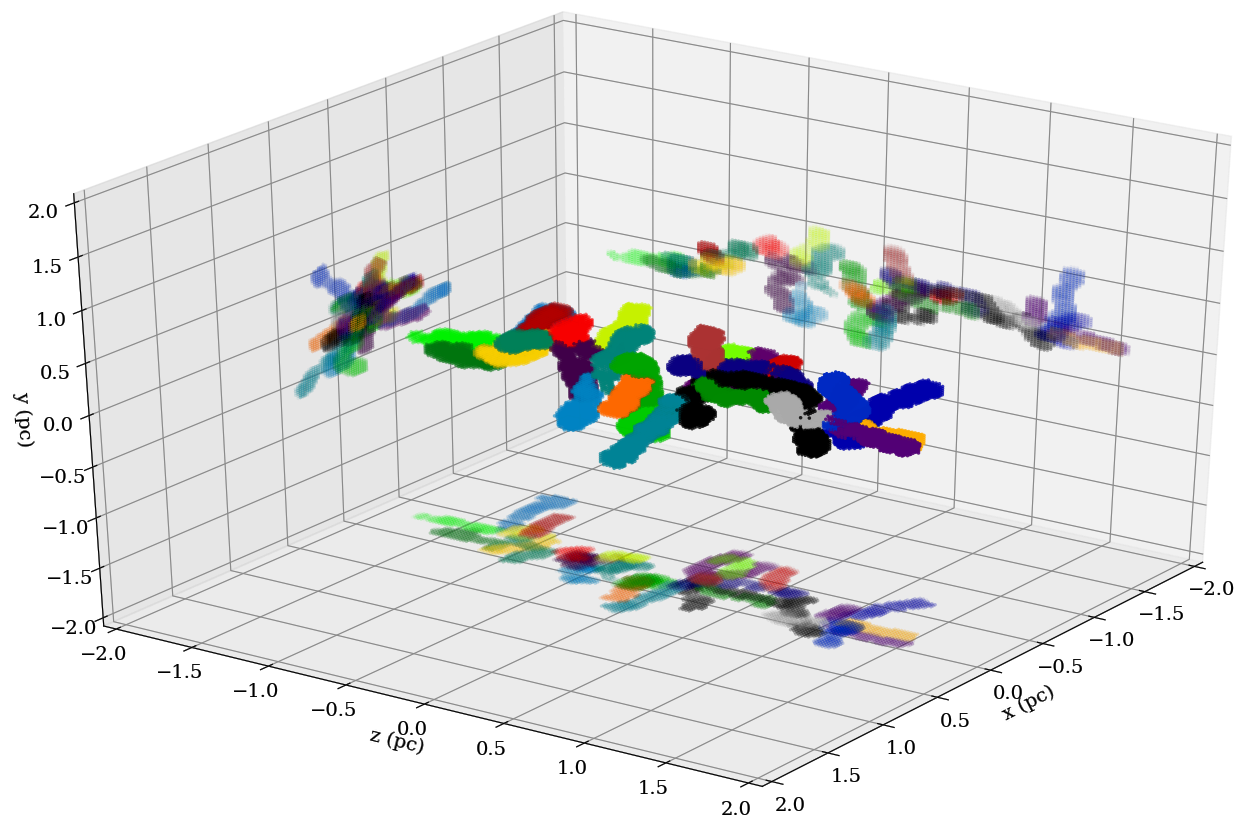}
\includegraphics[width = 0.95\linewidth]{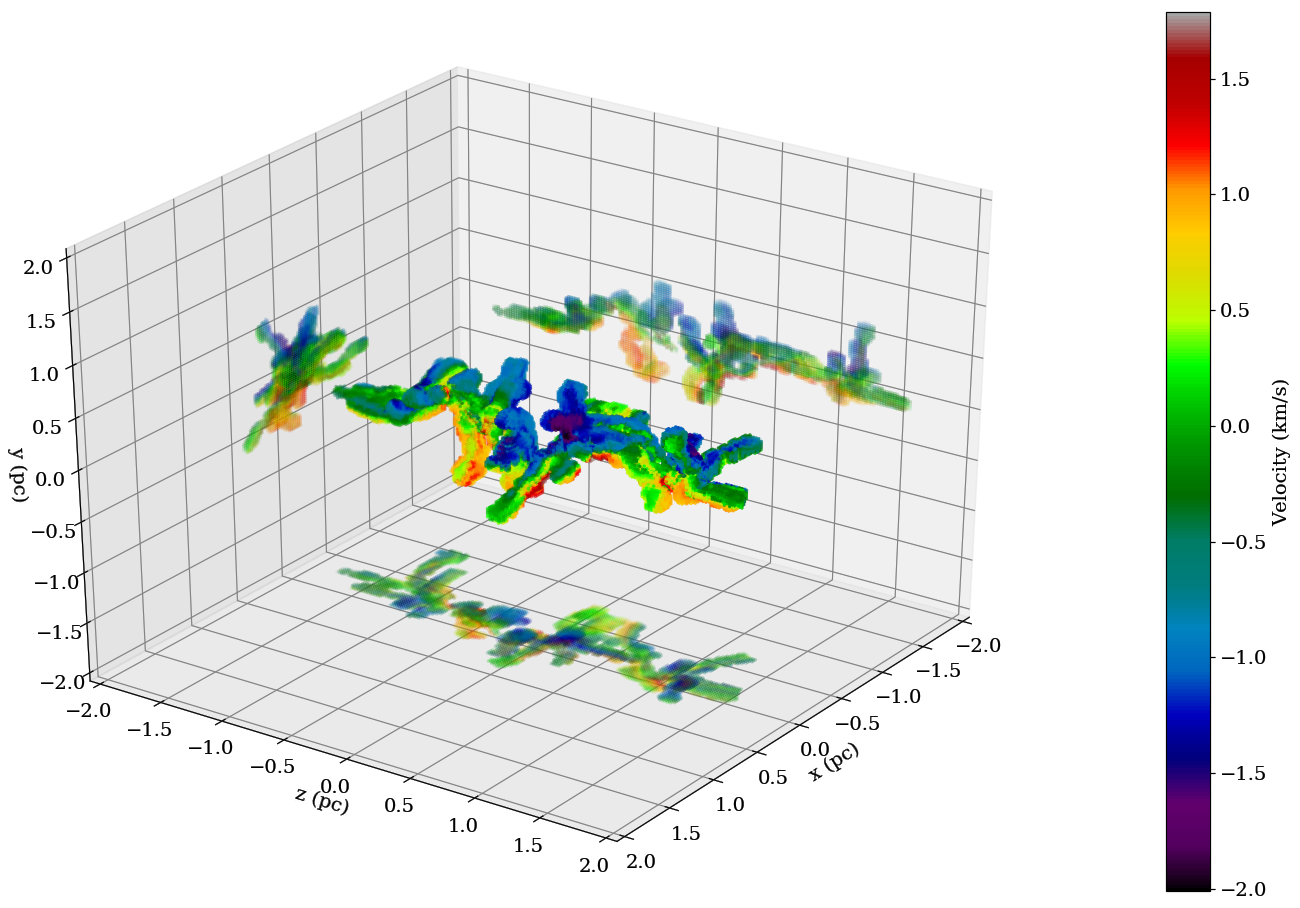}
\caption{A 3D plot showing the sub-filaments found using \textsc{DisPerSE}. In the top panel, the different sub-filaments are colour coded. In the bottom panel the colour shows the line-of-sight velocity from the simulation at that position.}
\label{fig::subfils}
\end{figure*}

Fig. \ref{fig::subfils} shows the gas associated with the 28 highly tangled sub-filaments identified by \textsc{DisPerSE} in {\sc Sim}02. The bottom panel of Fig. \ref{fig::subfils} shows the mean line-of-sight velocity at each pixel aligned with a sub-filament point, demonstrating that some sub-filaments possess a large velocity range, $\la\!3\,{\rm km}\,{\rm s}^{-1}$. This is one reason why features identified in PPP space do not always correspond to well-defined features in PPV space. 

\begin{table}
\centering
\begin{tabular}{@{}*5l@{}}
\hline\hline
& \multicolumn{4}{c}{Number of:}\\ \hline
{\sc Sim ID} & fibres & complete & contaminated & separate \\ \\
01  & 18 & 10 (55.5\%) & 4  (22.2\%) & 4 (22.2\%)\\
02  & 25 & 14 (56.0\%) & 7  (28.0\%) & 4 (16.0\%)\\
03  & 21 & 7  (33.3\%) & 6  (26.8\%) & 8 (38.1\%)\\
04  & 17 & 7  (41.2\%) & 5  (29.4\%) & 5 (29.4\%)\\
05  & 26 & 15 (57.7\%) & 8  (30.8\%) & 3 (11.5\%) \\
06  & 26 & 10 (38.5\%) & 10 (38.5\%) & 6 (23.1\%)\\
07  & 15 & 6  (40.0\%) & 5  (33.3\%) & 4 (26.7\%)\\
08  & 23 & 13 (56.5\%) & 6  (26.1\%) & 4 (17.4\%)\\
09  & 24 & 12 (50.0\%) & 7  (29.2\%) & 5 (20.8\%)\\
10 & 24 & 13 (54.2\%) & 5  (20.8\%) & 6 (25.0\%)\\ \hline
Total & 219 & 107 (48.9\%) & 63 (28.8\%) & 49 (22.3\%)\\ \hline
\end{tabular}
\centering
\caption{A table showing the number of fibres identified in each of the 10 simulations, and the number of fibres which when mapped into PPP space are `complete' continuous features, or `contaminated' by gas along the line-of-sight, or a combination of `separate' PPP features (see Section \ref{SEC::PPVPPP} for discussion).}
\label{tab::numfibres}
\end{table}

\subsection{Mapping sub-filaments into PPV space}%

\begin{figure}
\centering
\includegraphics[width = 0.90\linewidth]{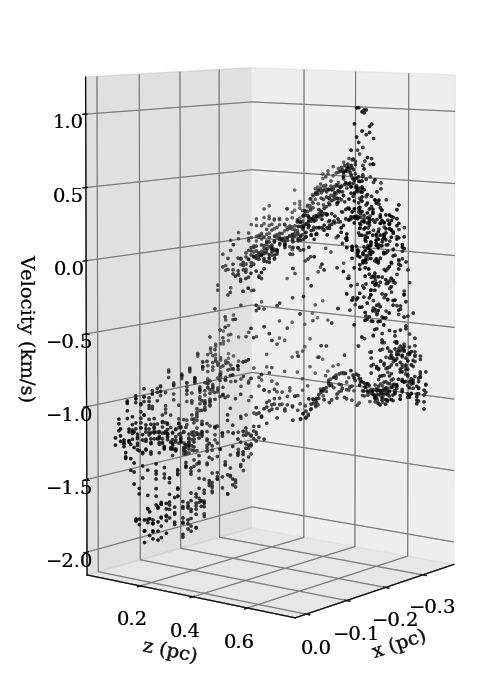}
\caption{A 3D plot showing the voxels from a single sub-filament mapped into PPV space.}
\label{fig::firstPPV}
\end{figure} 

\begin{figure*}
\centering
\includegraphics[width = 0.95\linewidth]{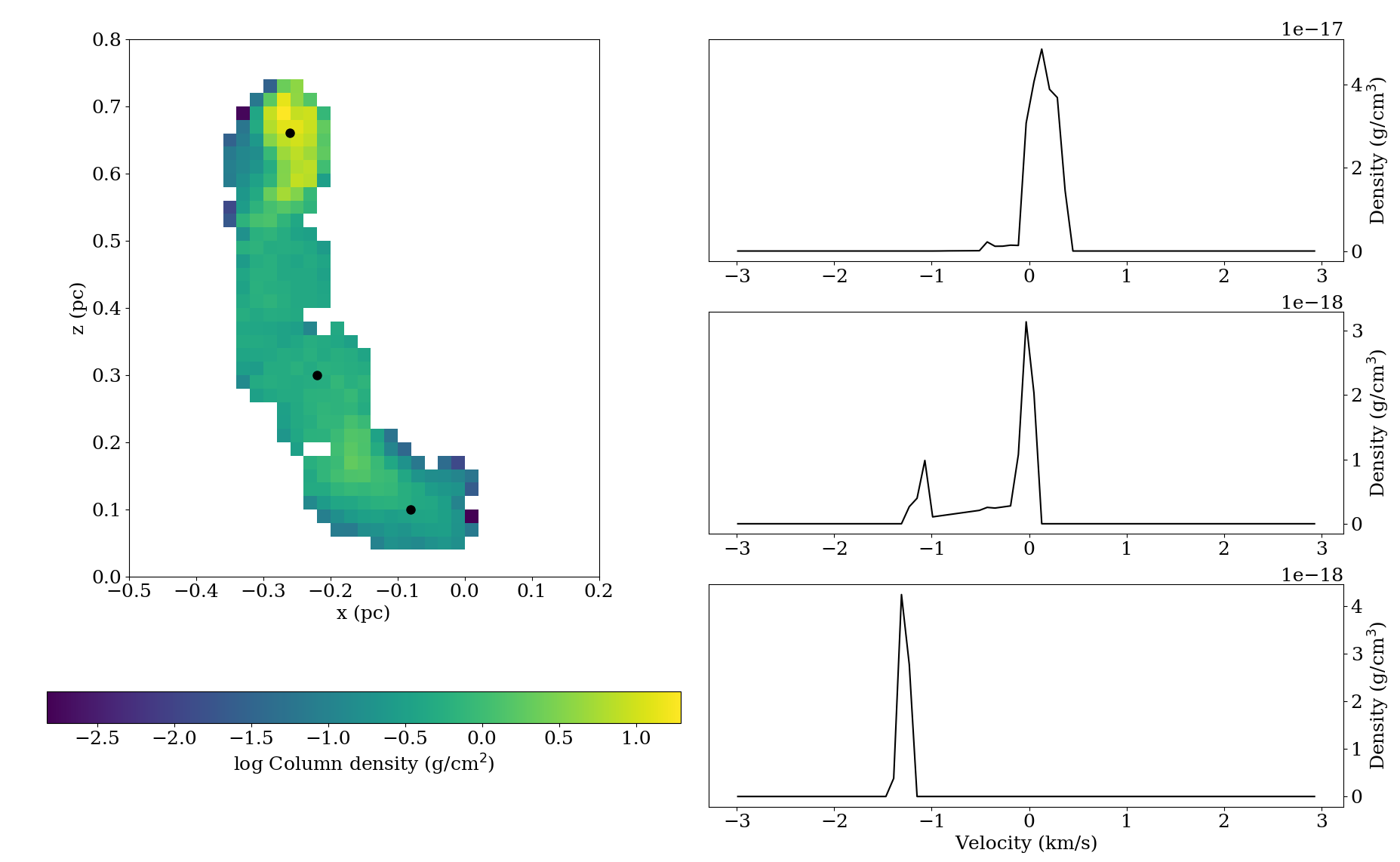}
\caption{The lefthand frame displays the column density of the sub-filament from Fig. \ref{fig::firstPPV}, and the black dots show the positions of the three `spectra' displayed on the right, showing multiple velocity components, separated by $\sim\!1\,{\rm km}\,{\rm s}^{-1}$.}
\label{fig::mapped2PPV}
\end{figure*} 

Each voxel from the simulation data-cube has position, $(x,y,z)$ and velocity, $(v_{_x},v_{_y},v_{_z})$, and therefore defines a point in, for example, the $(x,y,v_{_z})$ PPV space. Fig. \ref{fig::firstPPV} shows the points for a single sub-filament, demonstrating that it exists over a wide range of velocities, from $\sim\! -2.0$ to $\sim\!+1.5\,{\rm km}\,{\rm s}^{-1}$. In some places it is split into 2 distinct branches, one between $\sim\!0$ and $\sim\!1\,{\rm km}\,{\rm s}^{-1}$, and the other between $\sim\!-1$ and $\sim\!-2\,{\rm km}\,{\rm s}^{-1}$. Sub-filaments frequently exhibit large velocity gradients, abrupt breaks, and multiple strands in PPV space, like this one.

Using these points, we produce an approximate PPV data-cube for this sub-filament. We define velocity bins $0.08\,{\rm km}\,{\rm s}^{-1}$ wide (matching the velocity resolution of the synthetic spectra), and add the volume density associated with each point to the corresponding bin. Any points that have density below the critical density ($1400\,{\rm cm}^{-3}$; $\sim\!10\%$ of points), or above the freeze-out density ($10^5\,{\rm cm}^{-3}$; $\sim\!1\%$ of points) are discarded. This procedure avoids doing radiation transport, and is therefore only meaningful because the C$^{18}$O line is thermally excited and optically thin. Fig. \ref{fig::mapped2PPV} shows the column density of this sub-filament, and spectra at the three positions marked with black dots, showing multiple velocity components, one at $\sim\!-1\,{\rm km}\,{\rm s}^{-1}$ and the other at $\sim\! 0\,{ \rm km}\,{\rm s}^{-1}$. High velocity ranges and steep velocity gradients within individual sub-filaments are the main reason why they can not be identified reliably in PPV space. 

\begin{figure}
\centering
\includegraphics[width = 0.90\linewidth]{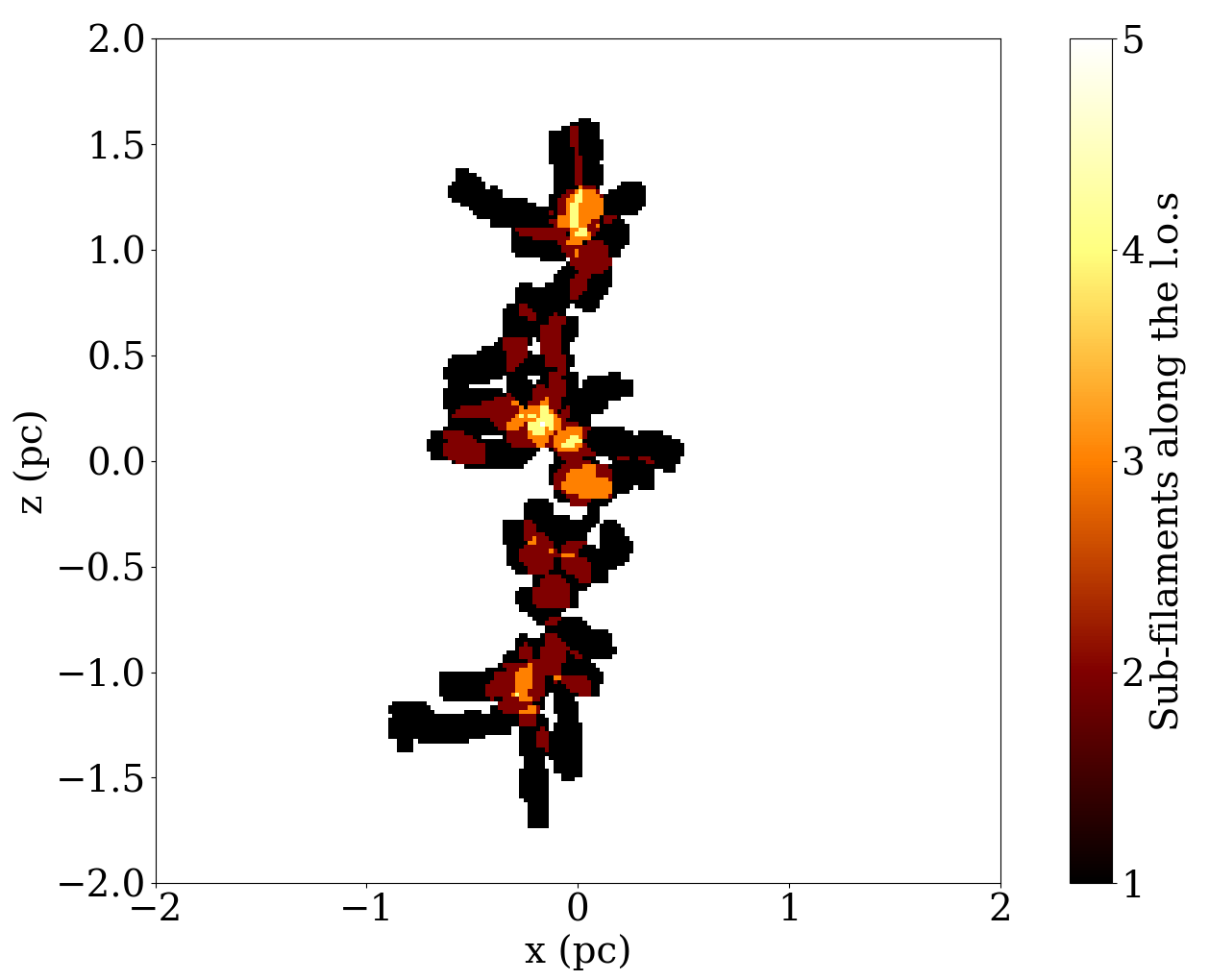}
\caption{A map of the number of different sub-filaments along each line-of-sight. Over 30$\%$ of lines-of-sight intercept more than one sub-filament. }
\label{fig::losCon}
\end{figure}

Moreover, there are often multiple sub-filaments along the same line-of-sight and occupying the same velocity range. Figure \ref{fig::losCon} shows the number of sub-filaments along each line-of-sight in {\sc Sim}02. $30\%$ of lines of sight intercept  more than one sub-filament, and in the immediate vicinity of  a dense core, there can be as many as 5 unique sub-filaments along a single line-of-sight. The possibility of multiple sub-filaments along a line-of-sight and the fact that individual sub-filaments are often not velocity coherent and distinct, means that it is extremely difficult to recover them from PPV cubes. 

\subsection{The statistics of fibres and sub-filaments}%

We use $J$-moments \citep{Jaf18} to classify the morphologies of fibres and sub-filaments from their 2D projection on the sky. To determine $J$-moments, we first construct dendrograms to identify structures. Dendrograms identify hierarchically nested structures and can be visualised as a tree-like structure. The largest structure is termed the trunk which is subsequently split into smaller structures called branches. These branches continue to be split until they reach local maxima which cannot be split again, termed leaves. The dendrograms are built in three-dimensional space (either PPP or PPV), using the {\sc astrodendro} Python package \footnote{http://www.dendrograms.org/}, and then these structures are projected into the same two-dimensional space (PP). This allows for a direct comparison between structures in different three-dimensional spaces.   

To construct dendrograms, we need to set three parameters: the minimum intensity that a pixel must have to be considered when building the dendrogram, \verb|min\_value|; the minimum intensity excess that a leaf or branch must have relative to its parent branch or trunk to be retained, \verb|min\_delta|; and the minimum number of pixels that a leaf or branch must have to be retained, \verb|min\_npix|. For sub-filaments in volume-density cubes we set \verb|min\_value|$\,=\! 10^{3}\,{\rm cm}^{-3}$, \verb|min\_delta|$\,=\!10^{3}\,{\rm cm}^{-3}$, and \verb|min\_npix|$\,=\!65$. For fibres in the velocity cubes we set \verb|min\_value|$\,=\!0.5\,{\rm K}$, \verb|min\_delta|$\,=\!0.3\,{\rm K}$, and \verb|min\_npix|$\,=\!60$. These choices are dictated by the noise levels in the data-cubes.

For the purpose of explaining $J$-moments, we assume that the intensity of a pixel measures surface-density, i.e. mass per unit area. For each element of the dendrogram (leaf, branch or trunk), we determine the area, $A$, the mass, $M$, a notional moment of inertia, ${\cal I}_{_0}\!=\!AM/4\pi$, and the principal moments of inertia, ${\cal I}_{_1}$ and ${\cal I}_{_2}\;\;(\geq\! {\cal I}_{_1})$. From these we construct the $J$-moments, $J_{_i}\!=\!({\cal I}_{_0}\!-\!{\cal I}_{_i})/({\cal I}_{_0}\!+\!{\cal I}_{_i})$. Elements with $0\!<\!J_{_2}\!\leq\!J_{_1}\!<1$ represent centrally concentrated structures, like cores; elements with $-1\!<\!J_{_2}\!\leq\!J_{_1}\!<0$ represent centrally rarefied structures, like shells; and elements with $J_{_1}\!>0,\;J_{_2}\!<\!0$ represent elongated structures, like filaments.

\begin{figure*}
\centering
\includegraphics[width = 0.45\linewidth]{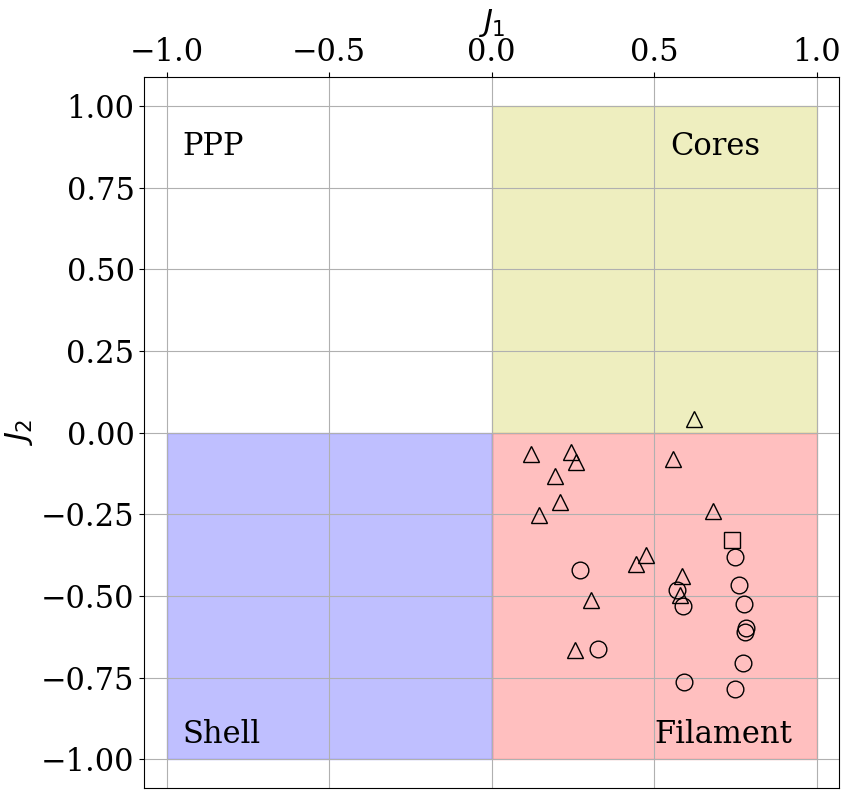}
\includegraphics[width = 0.45\linewidth]{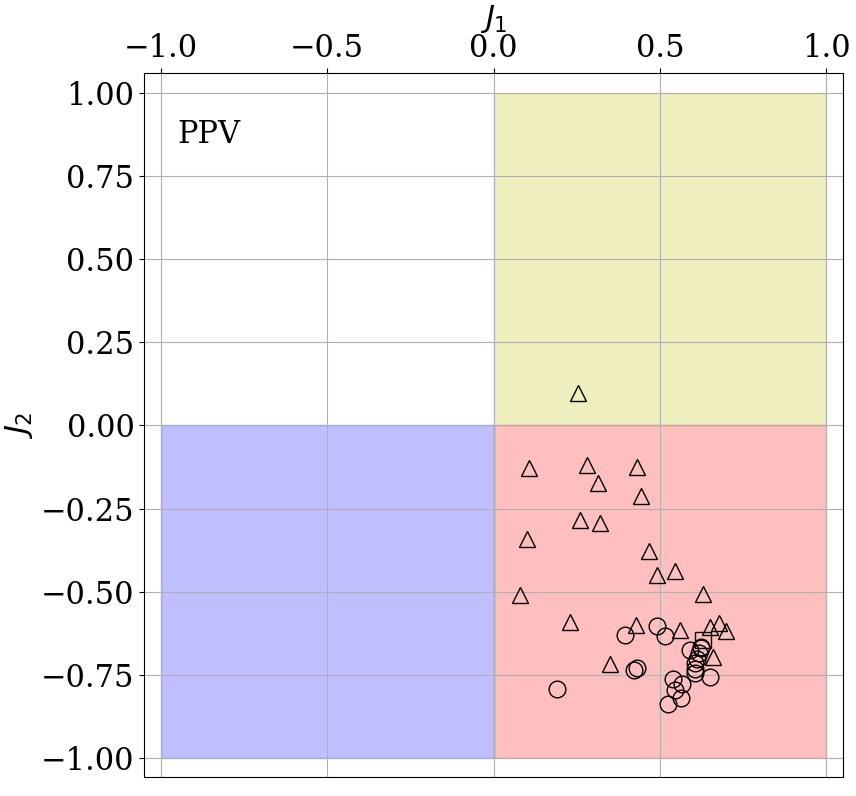}
\caption{$J$ moments of the sub-filaments (lefthand plot) and fibres (righthand plot) from {\sc Sim}02, confirming that at all levels they are elongated. The square represents the trunk of the dendrogram, circles the branches, and triangles the leaves.}
\label{fig::singleJ}
\end{figure*}

\begin{figure*}
\centering
\includegraphics[width = 0.45\linewidth]{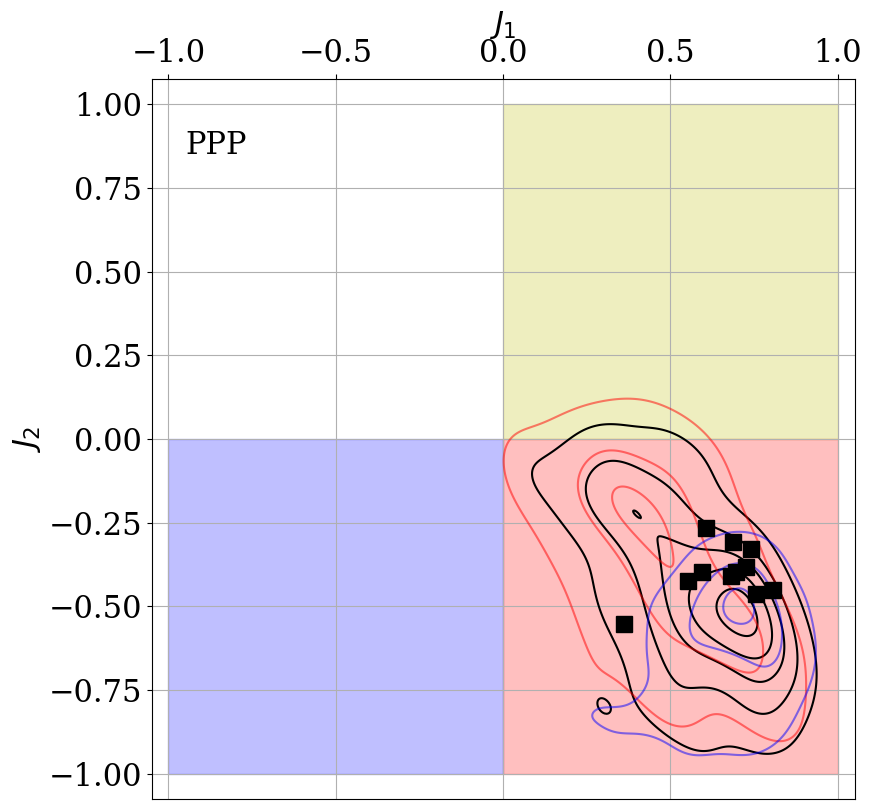}
\includegraphics[width = 0.45\linewidth]{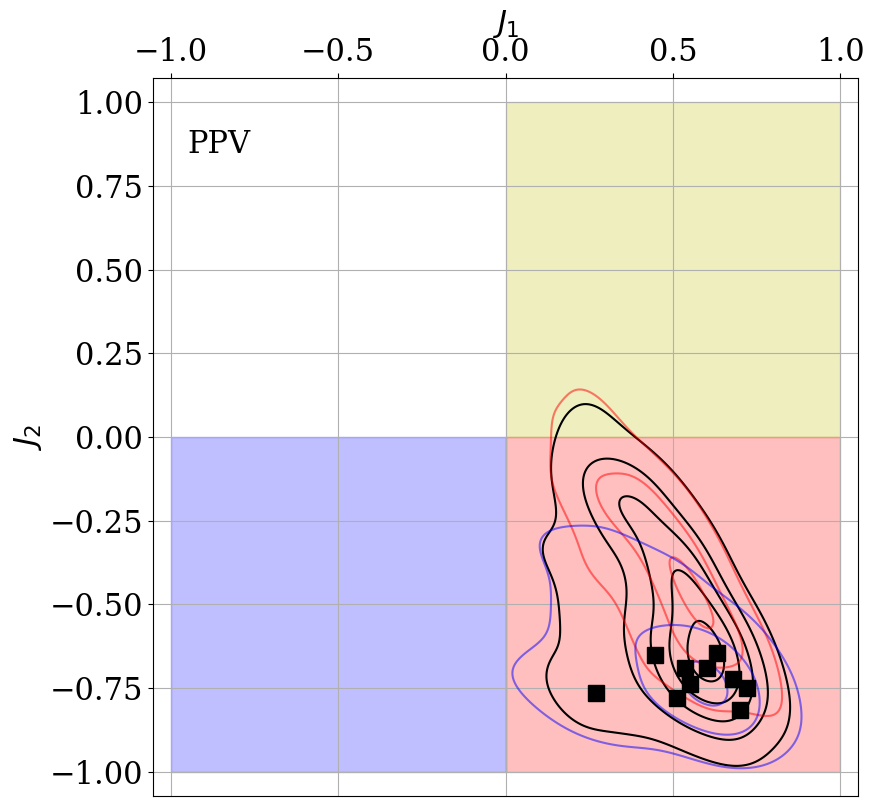}
\caption{The distribution of $J$-moments for the sub-filaments (lefthand plot) and fibres (righthand plot) from all 10 simulations, constructed using a KDE. Filled black squares represent the trunks of the dendrograms. Black contours delineate the distribution of all structures; red and blue contours delineate the distributions of leaves and branches respectively.}
\label{fig::allJ}
\end{figure*}

Fig. \ref{fig::singleJ} shows the $J$-moments of the sub-filaments (lefthand plot) and fibres (righthand plot) from {\sc Sim}02. Almost all structures lie in the lower-right (pink) quadrant of the $J$-plot indicating, objectively, that at all levels they are elongated. Fig. \ref{fig::allJ} shows a Kernel Density Estimate (KDE) plot of $J$ values for the structures identified in all 10 simulations.\footnote{A KDE converts a set of discrete points, here given by $(J_{_1},J_{_2})$, into a continuous distribution, by convolving them with a kernel, in this case a Gaussian whose width has been computed using the method described by \citet{Sil86}.} The sub-filaments contain 288 structures: 162 leaves and 115 branches. The fibres contain 296 structures: 157 leaves and 129 branches. 

Some statistical trends can be inferred from Fig. \ref{fig::allJ}. Several of these trends are attributable to the fact that sub-filaments are defined using a larger range of volume-densities, $\Delta\!\log_{_{10}}\!(n)\!\la\!4$, whereas fibres are defined using a much smaller range, $\Delta\!\log_{_{10}}\!(n)\!\la\!2$; this is because the C$^{18}$O emission used to identify fibres is concentrated between the critical density, $n_{\rm crit}\!\simeq\!1400\,{\rm cm}^{-3}$, and the freeze-out density, $n_{\rm freeze-out}\!\simeq\!10^5\,{\rm cm}^{-3}$. First, fibres are -- at all levels of the dendrogram -- systematically narrower than sub-filaments, as evidenced by the fact that their elements (trunk, branches, leaves) are more concentrated towards the bottom righthand corner of the pink quadrant of the $J$-plot than those of sub-filaments. Second, there are a few fibre branches that are curved, and hence populate the bottom lefthand corner of the pink quadrant of the $J$-plot, with small $J_{_1}$ and large negative $J_{_2}$; this has to do with the braiding observed in fibres. Third, there are a few sub-filament leaves that are not very elongated, and hence populate the top lefthand corner of the pink quadrant of the $J$-plot; this has to do with the fact that the sub-filament analysis is better able to pick up small not very elongated but very dense condensations along the spine of a filament (proto-cores), which are not accounted for in the  fibre analysis because it is assumed that their CO has frozen out.

\subsection{High density tracers: N$_2$H$^+$ and NH$_3$}%

fibres have also been detected in NGC1333 \citep{Hac17} and the Orion Integral Shaped Filament \citep{Hac18} using N$_2$H$^+$, which is a high density tracer. However, the Orion fibres are spatially distinct and easily identified on the integrated intensity map, unlike the Taurus fibres which overlap on the sky, and are only distinct in PPV space; thus in our terminology, the Orion fibres would actually be classified as sub-filaments. N$_2$H$^+$ should be a better tracer of sub-filaments, since N$_2$H$^+$ emission is less influenced by shocks than C$^{18}$O, and the volume-filling factor of N$_2$H$^+$-bright gas is much lower than C$^{18}$O-bright gas, leading to less line-of-sight confusion.

Ammonia (NH$_3$) is also a  dense gas tracer, and is thought to trace similar gas to N$_2$H$^+$ \citep{John10,Hac17}. NH$_3$ has been used to detect elongated features in the TMC-1 cloud in Taurus, using the \textsc{NbClust} algorithm \citep{Feh16}, but it is unclear how these features relate to fibres as defined by \citet{Hac13}. NH$_3$ has also been used by \citet{Wil18} to study the more massive filaments ({\it aka} spokes) in the hub-and-spoke system SDC13. However, they find no evidence for fibres; each of the four filaments (spokes) shows only a single velocity component. 

It will be important to understand better how fragmenting filaments appear when observed with nitrogen bearing molecules. These molecules trace different gas from C$^{18}$O, and therefore mapping from PPV space to PPP space may be more straightforward. Synthetic N$_2$H$^+$ and NH$_3$ observations are outside the scope of this paper, but will be the focus of future work.     

\section{Conclusions}\label{SEC:CON}%
 
Filament formation and fragmentation is a complex process, characterised by tangled, interconnected sub-structures and complicated kinematics, on both large and small scales. While it is the real density structures in PPP space that ultimately reflect the process of fragmentation and star-formation, molecular-line observations only reveal features in PPV space. Unfortunately, the mapping from PPV space to PPP space is compromised by confusion along the line-of-sight, and the identification criteria for features in PPV space (i.e. velocity coherence) do not guarantee that such features are physically continuous and distinct in PPP space. 

Synthetic C$^{18}$O observations of simulated filaments show complex spectra, with multiple velocity components on many lines-of-sight, similar to real observations of filaments \citep[][Suri et al. in prep.]{Hac13,TafHac15,Dha18}. Most of these velocity components have widths less than $\sim\!0.4\,{\rm km}\,{\rm s}^{-1}$, the transonic limit at $\sim\!10\,{\rm K}$, in agreement with observational studies showing that turbulence within filaments in typically sub- or trans-sonic \citep{Arz13,Hac13,Fer14,Kai16}.

Fibres, defined by \citet{Hac13} as velocity coherent structures in PPV space, are numerous in synthetic C$^{18}$O observations of the simulations presented here, with on average 22 fibres in a filament $\sim\!3\,{\rm pc}$ in length. Moreover, these fibres are not strongly affected by whether CO freeze-out at high densities is included. 

The identification of velocity coherent fibres in filaments has led to the suggestion that fibres are discrete structures, and act as building blocks for filaments; hence that by studying individual fibres and groups of fibres one can infer the internal structure of a filament. However, mapping from PPV space to PPP space is often compromised by line-of-sight confusion; $\sim\!50\%$ of fibres have some form of contamination from a physically separate parcel of gas, or consist of two or more physically continuous but separate features that happen to reside at the same velocity. It is impossible for an observer to know which features in PPV space belong to the $\sim50\%$ that are continuous in PPP space. Furthermore, those features that are continuous in PPP space may only appear distinct in velocity space because of internal shocks. It is therefore unclear what the properties of individual fibres and groups of fibres can tell us about the underlying structure of a filament, beyond the presence of internal shocks. 

Fibres identified in PPV space do not correspond closely with sub-filaments identified in PPP space. As discussed in \citet{Cla17}, sub-filaments are a consequence of internal turbulence within the parent filament, driven by accretion. Distinct sub-filaments do not appear as distinct velocity coherent fibres, because they are imprinted with the large-scale convergent flow onto the parent filament; they cover a large velocity range and can be made up of several distinct features in PPV space. Combined with the fact that there are many lines-of-sight which intercept more than one sub-filament, it is clear that identifying sub-filaments observationally is challenging. 

Although fibres identified in PPV space are not closely related to sub-filaments identified in PPP space, the 2D projected morphologies of fibres and sub-filaments are broadly similar. Using $J$-plots (Jaffa et al. 2018), we show that fibres are on average somewhat narrower than sub-filaments (because the C$^{18}$O tracer selects gas in a relatively narrow density range), and occasionally more curved.

Synthetic observations of these simulations in N$_2$H$^+$ and NH$_3$ will be presented in a future paper. The lines from these nitrogen-bearing molecules have the advantage that they trace higher-density gas, which should occupy a smaller volume, and therefore be less influenced by either shocks (hence more velocity-coherent), or line-of-sight confusion. In principle this should make mapping from PPV space into PPP space more straightforward, but it seems likely that the non-correspondence between fibres identified in PPV space and sub-filaments identified in PPP space will persist.

\section{Acknowledgments}\label{SEC:ACK}%
SDC and SW acknowledges support from the ERC starting grant No. 679852 `RADFEEDBACK'. APW, ADC and PCC gratefully acknowledge the support of a consolidated grant (ST/N000706/1) from the UK Science and Technology Facilities Council. STS acknowledges funding by the Deutsche Forschungsgemeinschaft (DFG) via the Sonderforschungs\-bereich SFB 956 Conditions and Impact of Star Formation (subproject A4) and the Bonn-Cologne Graduate School. RLS gratefully acknowledges the support of a UK Science and Technology Facilities Council postgraduate studentship. SEJ gratefully acknowledge the support of postgraduate scholarships from the School of Physics $\&$ Astronomy at Cardiff University and the UK Science and Technology Facilities Council. SW further thanks the DFG for funding through the Collaborative Research Center (SFB956) on the `Conditions and impact of star formation'. PCC acknowledges support from the European Community's Horizon 2020 Programme H2020-COMPET-2015, through the StarFormMapper Project (number 687528). SDC would like to thank Volker Ossenkopf for useful discussions about radiation transfer. The authors would also like to thank the anonymous referee for their helpful comments on the paper. This work was performed using the facilities of the Advanced Research Computing at Cardiff Division, Cardiff University. 

\bibliographystyle{mn2e}
\bibliography{SynObs_arxiv} 

\appendix

\section{Behind The Spectrum (BTS), an automated multiple velocity component fitting code}\label{APP:FITTING}%

We describe Behind The Spectrum (BTS), a new automated routine for fitting line profiles, which uses the first, second and third derivatives of the intensity to estimate objectively the number and positions of the components. A least-squares fitting routine is then used to determine the best fit with that number of components, checking for over-fitting and over-lapping velocity centroids. The code is freely available for download at https://github.com/SeamusClarke/BTS.

\subsection{Code methodology}\label{APP::METHOD}%

\begin{figure}
\label{fig::3Gauss}
\centering
\includegraphics[width = 0.93\linewidth]{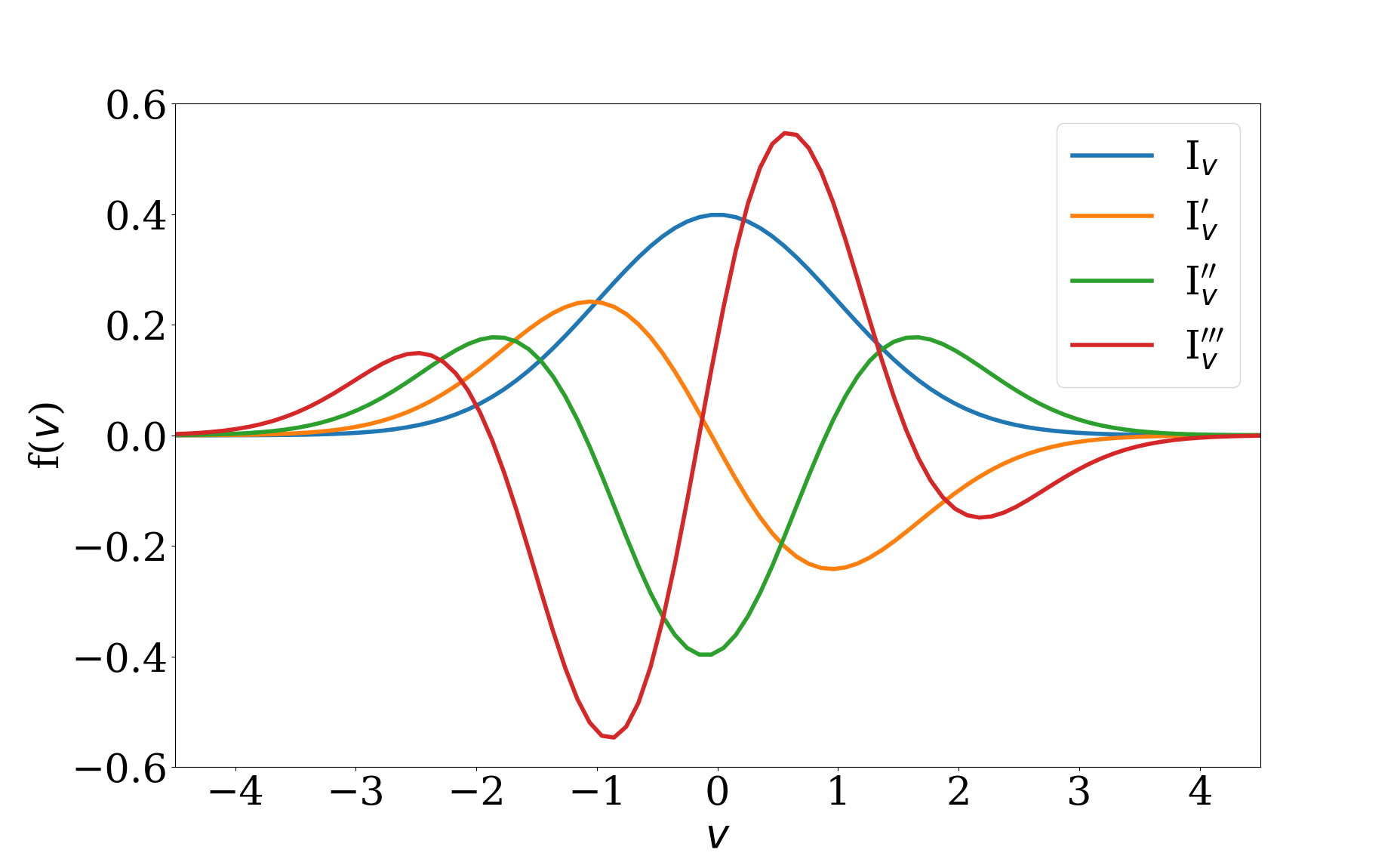}
\includegraphics[width = 0.93\linewidth]{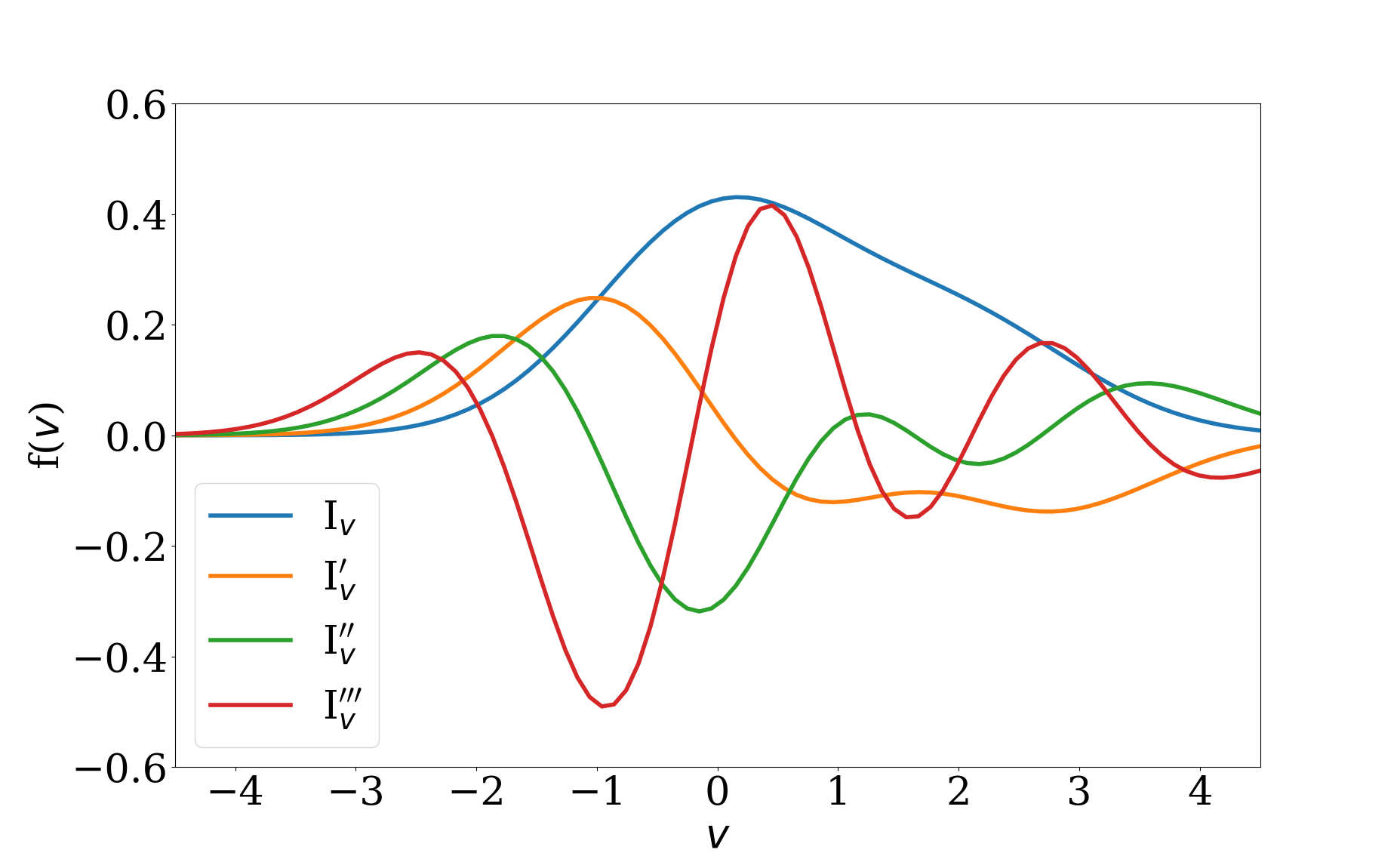}
\includegraphics[width = 0.93\linewidth]{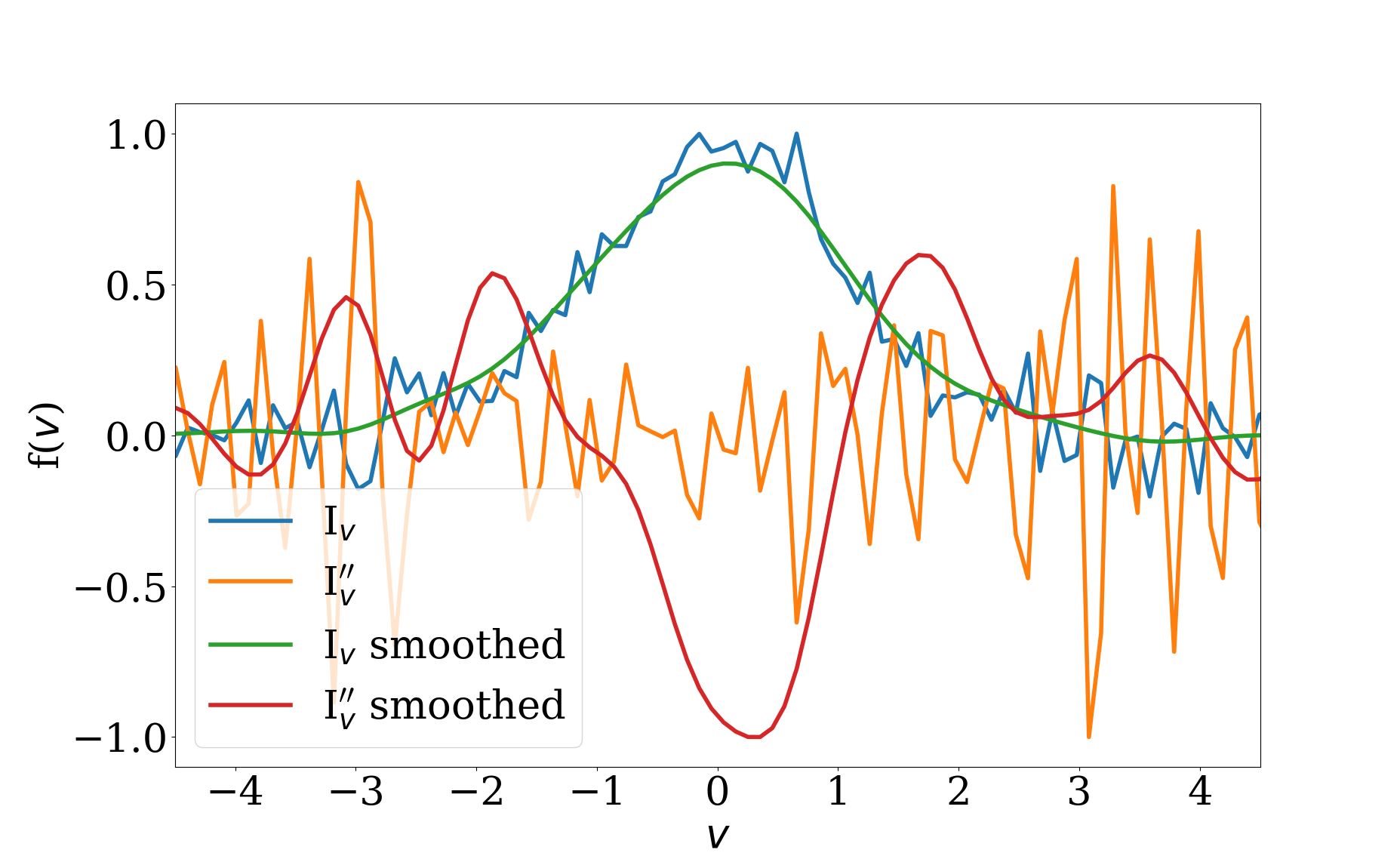}
\caption{{\sc Top.} A Gaussian line profile centred on $v\!=\!0$, along with the first three derivatives. {\sc Middle.} Two overlapping Gaussian line profiles, centred on $v\!=\!0$ and $v\!=\!2$, along with the first three derivatives. {\sc Bottom.} A noisy Gaussian centred on $v\!=\!0$ (blue) along with its second derivative (orange); and the same Gaussian after smoothing (green) and its second derivative (red).}
\end{figure}

The top panel of Fig. \ref{fig::3Gauss} shows a perfect Gaussian line profile, $I_{_v}$, centred on $v\!=\!0$, and its first three derivatives. At the maximum, the first derivative, $I_{_v}'$, is 0 and decreasing, the second derivative, $I_{_v}''$, has a minimum, and therefore the third derivative, $I_{_v}'''$, is 0 and increasing. The local minimum in $I_{_v}''$ is used as the primary indicator for the centroid of a line (hereafter a `velocity component'); $I_{_v}'$ and $I_{_v}'''$ are used as secondary checks. This is because $I_{_v}''$ appears to be better able to locate additional line components than $I_{_v}'$ or $I_{_v}'''$. This is demonstrated in the middle panel of Fig. \ref{fig::3Gauss} where a second Gaussian profile has been added, centred on $v\!=\!2$, with the same width as the first, and half the amplitude. $I_{_v}'$ is not zero at this location, but $I_{_v}''$ shows a local minimum at $x\!\approx\! 2.2$ and $I_{_v}'''$ is close to 0.  

\begin{figure}
\label{fig::singletest}
\centering
\includegraphics[width = 0.93\linewidth]{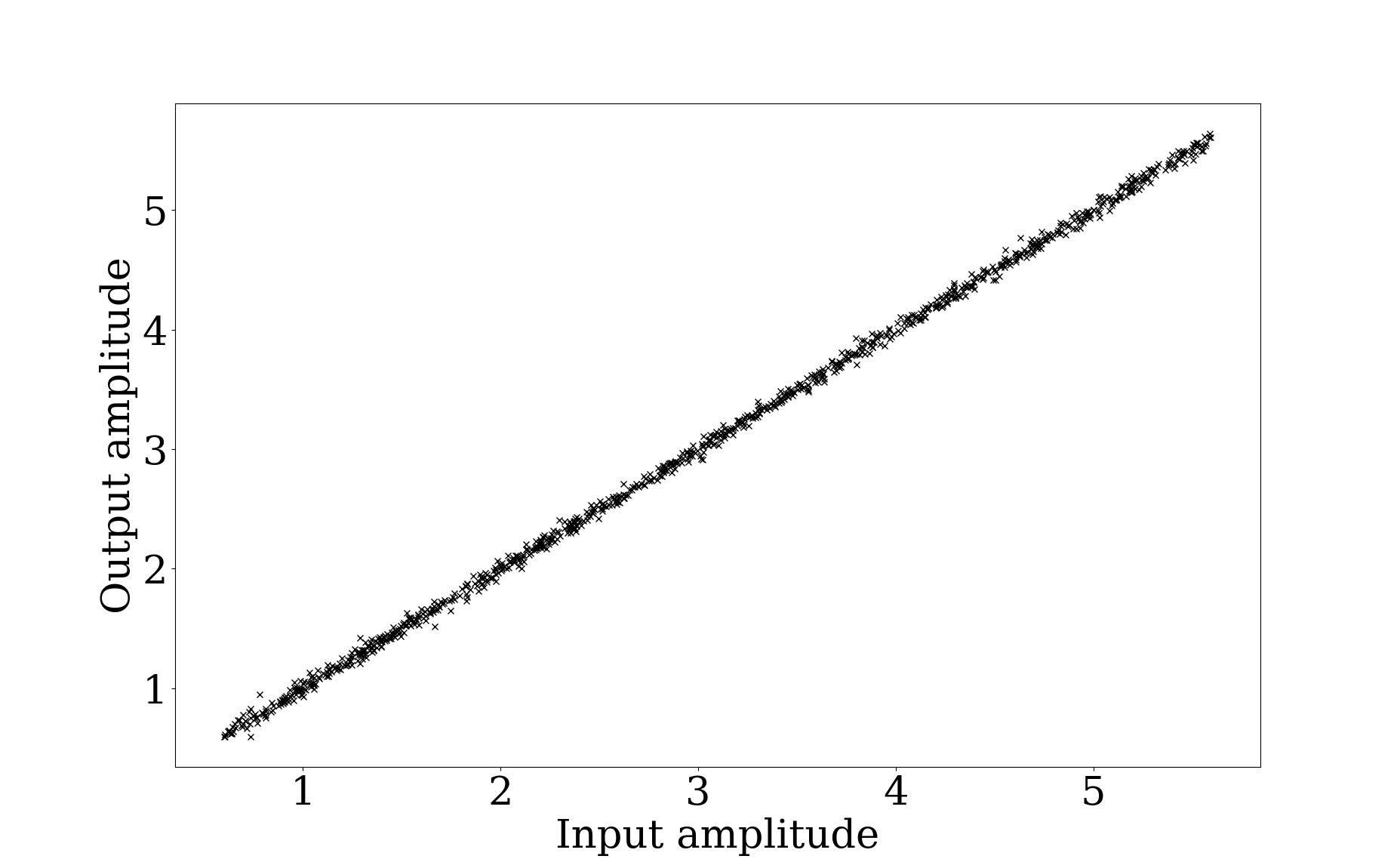}
\includegraphics[width = 0.93\linewidth]{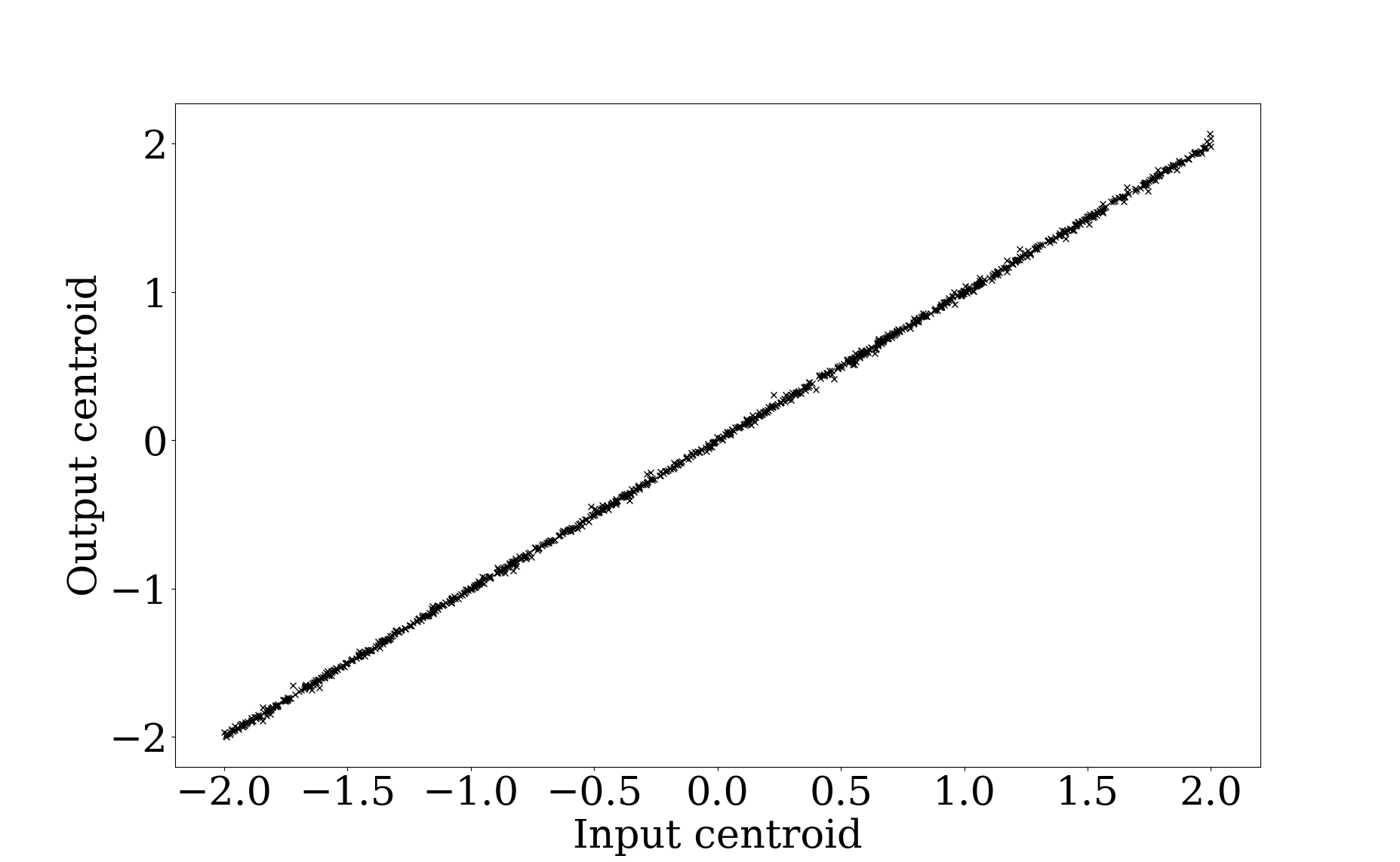}
\includegraphics[width = 0.93\linewidth]{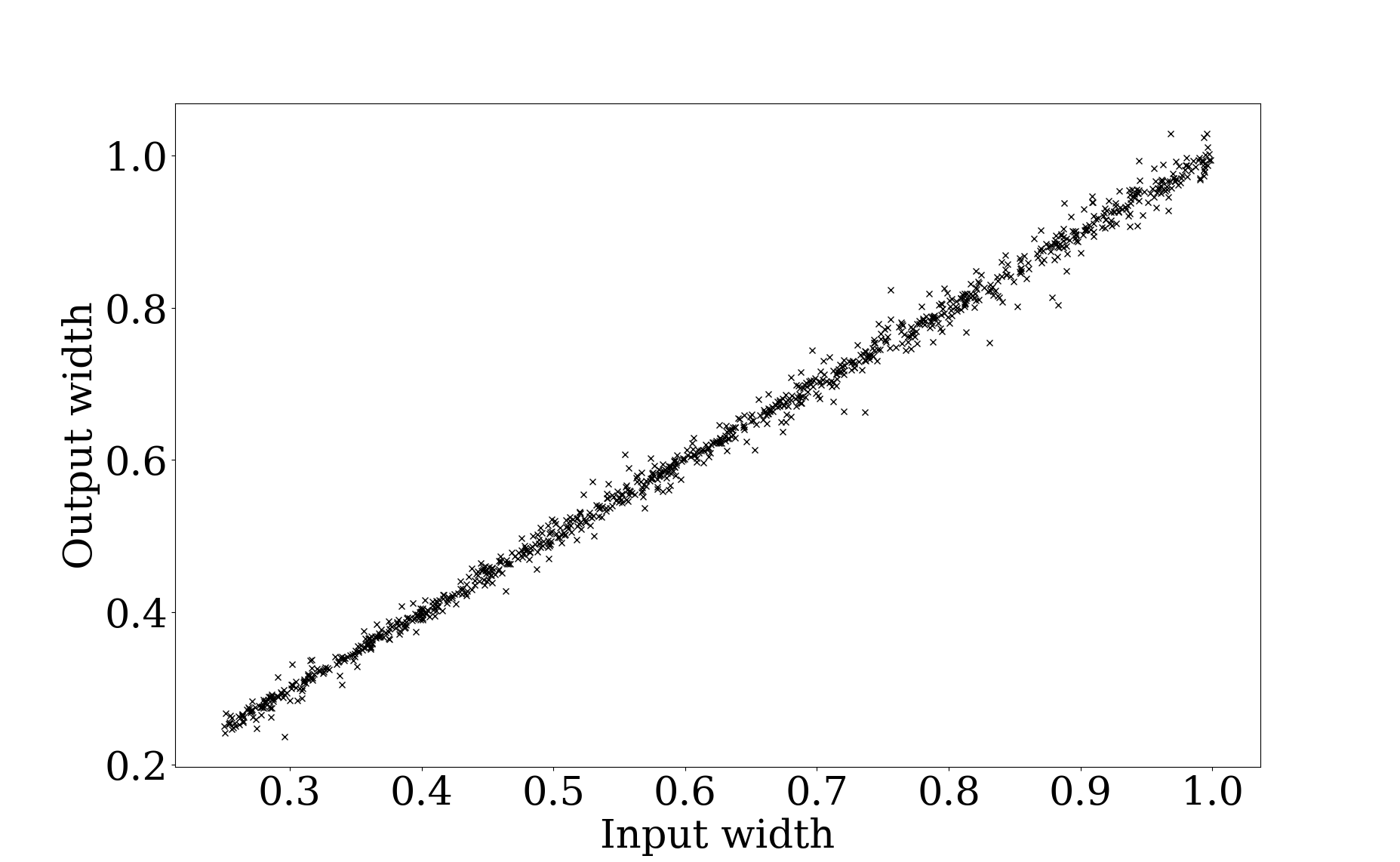}
\caption{Plot of the input parameters against the output estimates from the BTS fitting routine applied to a single noisy Gaussian profile. The top, middle and bottom panels show, respectively, the amplitude, velocity centroid and velocity dispersion. The median percentage error on all three parameters is $\leq\!1\%$.}
\end{figure}

Observed spectra are noisy, and this seriously distorts higher derivatives of the intensity. The bottom panel of Fig. \ref{fig::3Gauss} shows the same Gaussian profile as the top panel, but with noise added to each velocity channel (of width $\Delta v_{\rm channel}\!=\!0.08\,{\rm km}\,{\rm s}^{-1}$). The noise is generated by sampling from a Gaussian distribution with a mean of 0 and a standard deviation of $I_{\rm noise}\!=\!0.04\,{\rm K}$, leading to a peak signal-to-noise ratio of $\sim\!10$; $I_{_v}''$ is now dominated by noise. To combat this, BTS smooths noisy spectra by convolving them with a Gaussian kernel having a standard deviation of $\alpha_{_{\rm BTS}}\,\Delta v_{\rm channel}$, before determining the derivatives. The smoothed spectrum in the bottom panel of Fig. \ref{fig::3Gauss} is obtained in this way, with the default $\alpha_{_{\rm BTS}}\!=\!3$. $I_{_v}''$ still has a few minima, but these are not identified as extra velocity components if $I_{_v}$ at these positions is below a signal-to-noise threshold of $I_{\rm min}\!=\!\beta_{_{\rm BTS}}I_{\rm noise}$ where the default setting is $\beta_{_{\rm BTS}}\!=\!5$. We note that the BTS default values, $\alpha_{_{\rm BTS}}\!=\!3$ and $\beta_{_{\rm BTS}}\!=\!5$, can be overwritten by the user.

In this way we obtain an estimate of the number of velocity components, ${\cal C}$, and a first estimate of their velocity centroids, ${\tilde v}_{_c}\;(1\!\leq\!c\!\leq\!{\cal C})$. The intensity at velocity $v_{_c}$ is used as a first estimate of the amplitude of that component ${\tilde I}^{\rm o}_{_c}\!=\!I\!\left({\tilde v}_{_c}\right)$. A first estimate of the velocity dispersion of the component is given by ${\tilde\sigma}_{_c}\!=\![v_{_+}\!-\!v_{_-}]/\sqrt{8\ln{2}}\Delta\!{\cal N}$, where $v_{_+}$ ($v_{_-}$) is the first velocity above (below) ${\tilde v}_{_c}$ where the intensity falls below ${\tilde I}^{\rm o}_{_c}/2$ (thus $[v_{_+}\!-\!v_{_-}]$ is a sort of {\sc fwhm}) and $\Delta\!{\cal N}$ is the number of velocity centroids between $v_{_-}$ and $v_{_+}$.

These initial estimates (${\cal C},{\tilde v}_{_c},{\tilde I}^{\rm o}_{_c},{\tilde\sigma}_{_c}$; distinguished by tildes) are given to the least-squares fitting routine \textsc{curve\_fit} from the \textsc{Python} library \textsc{scipy}. \textsc{curve\_fit} is run with the Trust Region Reflective option, which allows the following bounds to be placed on the fitting parameters:
\begin{eqnarray}\nonumber
\hspace{1.5cm}\beta_{_{\rm BTS}}I_{_{\rm NOISE}}\;\leq\!&\!I^{\rm o}_{_c}\!&\!\leq\;2I_{_{\rm MAX}};\\\nonumber
v_{\rm min}\;\leq\!&\!v_{_c}\!&\!\leq\;v_{\rm max};\\\nonumber
2\Delta v_{\rm channel}\;\leq\!&\!\sigma_{_c}\!&\!\leq\;v_{\rm max} - v_{\rm min}.
\end{eqnarray}
\noindent Here $I_{_{\rm MAX}}$ is the maximum observed intensity, $ v_{_{\rm MIN}}$ and $ v_{_{\rm MAX}}$ are the minimum and maximum velocity in the spectrum, and $2\Delta v_{\rm channel}$ is the spectral resolution. \textsc{curve\_fit} returns the best fit parameters, ($v_{_c},I^{\rm o}_{_c},\sigma_{_c}$), and the estimated covariance matrix for these parameters, $\mathbf{C}_{_{cc'}}$. Standard deviation errors on the fit parameters are then given by $e_c = \sqrt{\mathbf{C}_{cc}}$. There are cases for which \textsc{curve\_fit} is unable to find a best fit, either because the initial estimates for the fit parameters are poor, or the $\chi^2$ landscape is complicated. In these cases the lack of convergence is noted and no fit is recorded. However, these cases are rare, $\ll\!1\%$. In the future, a Monte Carlo Markov Chain routine will be added to fit those spectra which cannot be fit using \textsc{curve\_fit}. 

Once a fit,  
\begin{eqnarray}
I^{\rm fit}\left(v_{_n}\right)&=&\sum\limits_{c=1}^{c={\cal C}}\left\{\!I^{\rm o}_{_c}\exp\!\left(\!\frac{-\,(v_{_n}-v_{_c})^2}{2\sigma_{_c}^2}\!\right)\!\right\}\!,
\end{eqnarray}
has been found, its reduced $\chi^2$ is calculated,
\begin{eqnarray}
\label{eq::chi_2}
\chi^2_{\rm reduced}\!\!&\!\!=\!\!&\!\!\frac{1}{({\cal N}-3{\cal C})} \sum\limits_{n=1}^{n={\cal N}}\left\{\!\frac{\left(I^{\rm obs}\left(v_{_n}\right)-I^{\rm fit}\left(v_{_n}\right)\right)^2}{\sigma_{_{\rm noise}}^2}\!\right\}\!.
\end{eqnarray} 
Here ${\cal N}$ is the number of data points being fit, $3{\cal C}$ the number of parameters for the fit, $I^{\rm obs}\left(v_{_n}\right)$ is the observed intensity at velocity $v_{_n}$, and $\sigma_{\rm noise}$ is the noise in the observed spectrum. If $\chi^2_{\rm reduced}\!>\!\gamma_{_{\rm BTS}}\!=\!1.5$, an extra velocity component, $c\!=\!{\cal C}+1$ is added. The initial estimate for the extra component's centroid, ${\tilde v}_{_{{\cal C}+1}}$, is the velocity of the channel for which the absolute residual is largest. The initial estimate for the extra component's amplitude is the intensity in this channel. The initial estimate for the extra component's velocity dispersion is the velocity resolution, $\Delta v_{\rm channel}$. If the new fit delivers $\chi^2_{\rm reduced}\!<\!\gamma_{_{\rm BTS}}$, the new fitting parameters are retained; if they do not then the old fitting parameters are reinstated. To avoid over-fitting, fits which have $\chi^2_{\rm reduced}\!<\!\gamma_{_{\rm BTS}}$, are re-fitted with the  component having the smallest amplitude removed; if the fit with fewer components still has $\chi^2_{\rm reduced}\!<\!\gamma_{_{\rm BTS}}$, then the reduced set of fitting parameters are retained; otherwise the old fit is reinstated. 

The code also checks for overlapping velocity components. Such components may appear in spectra for physical reasons (e.g. jets), so this check can be disabled. However, if overlapping velocity components are not desired, the code checks if any two component centroids lie within one velocity channel of each other, and, if they do, the weaker of the two is removed and the fit repeated.  

\subsection{Code testing}\label{APP::TESTS}%

To test the BTS code, we use noisy spectra with a known number of velocity components and known parameters for each component. The test spectra have a velocity range of $-3\,{\rm km}\,{\rm s}^{-1}$ to $+3\,{\rm km}\,{\rm s}^{-1}$, a velocity resolution of $\Delta v_{\rm channel}\!\simeq\!0.08\,{\rm km}\,{\rm s}^{-1}$, and a noise level of $0.1\,{\rm K}$ per velocity channel, like the synthetic spectra from the simulations. Unless stated otherwise, we use the default parameter settings, $\alpha_{_{\rm BTS}}\!=\!3$ (spectrum smoothed over 3 velocity channels),  $\beta_{_{\rm BTS}}\!=\!5$ (signal-to-noise threshold for a velocity component) and $\gamma_{_{\rm BTS}}\!=\!1.5$ (fitting acceptance threshold). 

The first test involves a single velocity component, with parameters randomly sampled from uniform distributions: amplitude, $0.6\,{\rm K}\!\leq\!I^{\rm o}_{_1}\!\leq\!5.6\,{\rm K}$; centroid, $-2.0\,{\rm km}\,{\rm s}^{-1}\!\leq\!v_{_1}\!\leq\!+2.0\,{\rm km}\,{\rm s}^{-1}$; and velocity dispersion, $0.25\,{\rm km}\,{\rm s}^{-1}\!\leq\!\sigma_{_1}\!\leq\!1.00\,{\rm km}\,{\rm s}^{-1}$. Fig. A2 demonstrates the close correspondence between the input parameters and those fitted by BTS, for 1000 realisations; the median errors on the amplitude, centroid and dispersion are, respectively, $0.84^{+ 0.72}_{-0.47} \%$, $0.66^{+ 1.03}_{-0.38}\%$ and $0.93^{+ 0.94}_{-0.51} \%$, where $+/-$ denotes the interquartile range. The median reduced $\chi^2$ is $0.99^{+ 0.10}_{-0.11}$. In these tests the code only ever fitted a single component; it never attempted to fit multiple components.

The second test addresses the ability of BTS to detect the correct number of velocity components. The number of components, ${\cal C}$, is randomly sampled from a uniform distribution, $1\leq{\cal C}\leq 4$. To avoid attempting to fit unresolved components, which would skew the results, we require that every pair of components be separated by their mean FWHM, i.e. $|v_{_c}\!-\!v_{_{c'}}|\geq 1.175(\sigma_{_c}+\sigma_{_{c'}})$. To accommodate multiple components, the velocity centroid range is increased to $-2.5\,{\rm km}\,{\rm s}^{-1}\!\leq\!v_{_1}\!\leq\!+2.5\,{\rm km}\,{\rm s}^{-1}$, and the velocity dispersion range is decreased to $0.25\,{\rm km}\,{\rm s}^{-1}\!\leq\!\sigma_{_1}\!\leq\!0.50\,{\rm km}\,{\rm s}^{-1}$. In 1000 realisations, BTS always identifies the correct number of components, and the median errors on the fitted parameters are essentially the same as in the first test, with no dependence on ${\cal C}$.

The third test addresses how sensitive BTS is to the user-defined parameters, $\alpha_{_{\rm BTS}}$,  $\beta_{_{\rm BTS}}$, $\gamma_{_{\rm BTS}}$, by repeating the second test with non-default values. (a) {\sc Smoothing.} If the smoothing length is decreased from $\alpha_{_{\rm BTS}}\Delta v_{\rm channel}\!=\!3\Delta v_{\rm channel}$ to $2\Delta v_{\rm channel}$, BTS is successful in 983 out of 1000 tests. The 17 mis-identified spectra are fitted with two many components, because $2\Delta v_{\rm channel}$ is too small a smoothing length to remove all the noise fluctuations. However, these spectra are easily identifiable as they have $\chi^2_{\rm reduced}\!<\!0.8$. If the smoothing length is increased from $\alpha_{_{\rm BTS}}\Delta v_{\rm channel}\!=\!3\Delta v_{\rm channel}$ to $5\Delta v_{\rm channel}$, BTS is successful in 995 out of 1000 tests; 5 spectra are mis-identified because they have components with dispersions of only 3 velocity channels, and are consequently over-smoothed and missed, but these spectra are easily identifiable as they have $\chi^2_{\rm reduced}\!>\!3.5$. (b) {\sc SNR threshold.} If the signal-to-noise threshold is reduced from  $\beta_{_{\rm BTS}}\!=\!5$ to $\beta_{_{\rm BTS}}\!=\!3$, BTS has 100\% success rate in 1000 tests. If it is reduced further to $\beta_{_{\rm BTS}}\!=\!2$, BTS is successful in 998 out of 1000 tests. We therefore recommend $\beta_{_{\rm BTS}}\!\geq\!3$. (c) {\sc Fit acceptance.} If the acceptance threshold is increased from $\chi^2_{\rm reduced}\!<\!\gamma_{_{\rm BTS}}\!=\!1.5$ to $\chi^2_{\rm reduced}\!<\!\gamma_{_{\rm BTS}}\!=\!2.0$, BTS is successful in 994 out of 1000 tests, and if it is increased further to 2.5, BTS is successful in 986 out of 1000 tests. This is due to poorer fits with fewer components being accepted as they lie below the $\chi^2$ limit. Conversely, if the limit is reduced to $\gamma_{_{\rm BTS}}\!=\!1.2$, the success rate is 99.4$\%$, because a few spectra have to be overfit to get $\chi^2_{\rm reduced}$ below this limit. Thus, all 3 user-defined parameters have a weak effect on the reliability of BTS, and with sensible choices the success rate is $\gtrsim\!99\%$. 

To determine the best choices in a given situation, the code includes a testing routine which allows the user to run the tests described here for sample spectra with parameters similar to their observations (velocity resolution, noise level, expected amplitudes, centroids and widths). We note that, since BTS fits lines with Gaussians it ought not be used on spectra dominated by non-Gaussian components, e.g. optically thick spectra or highly skewed spectra from outflows. 

\section{Results from all 10 simulations}\label{APP:PICS}%

\begin{figure*}
\centering
\includegraphics[width = 0.35\linewidth]{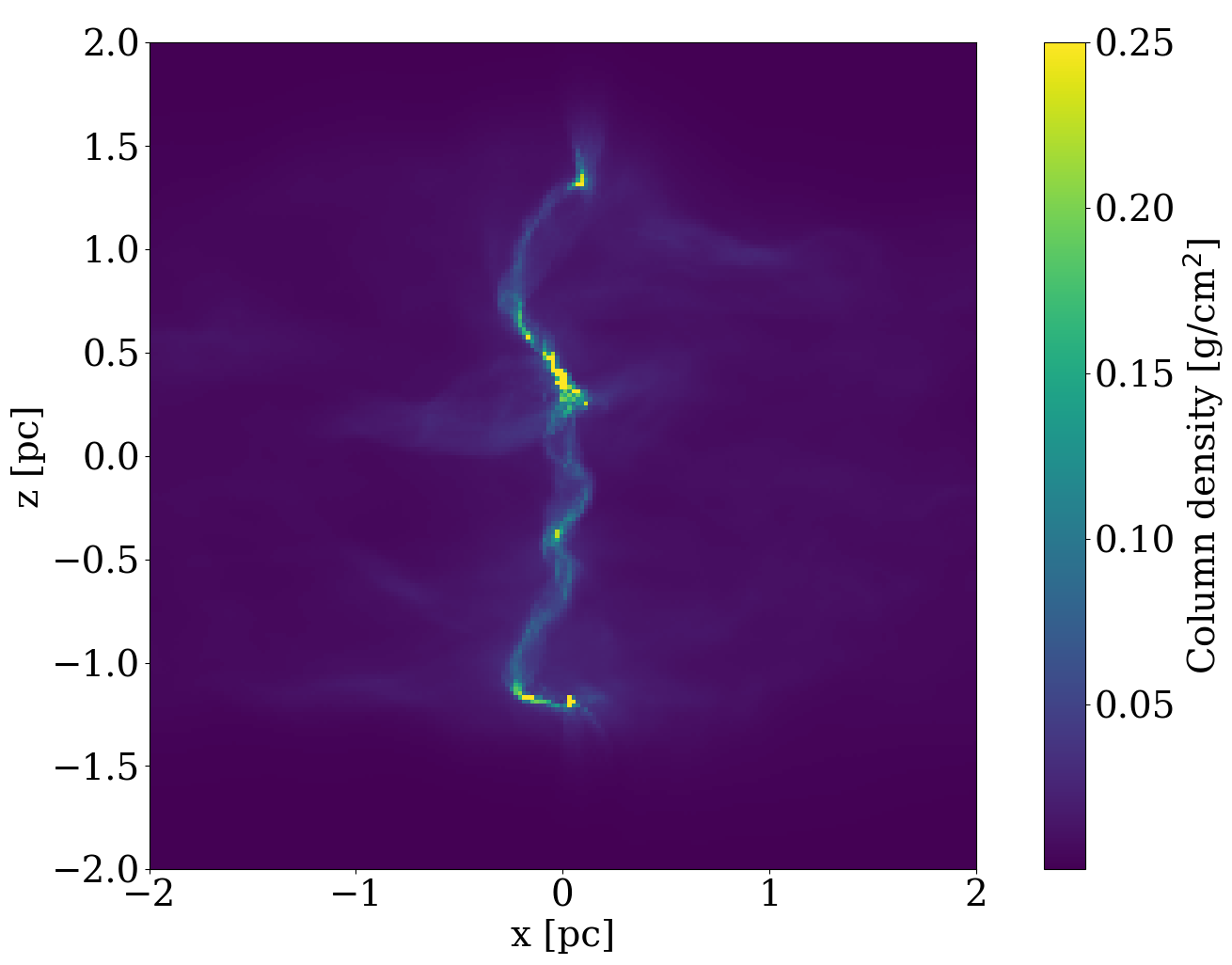}
\includegraphics[width = 0.35\linewidth]{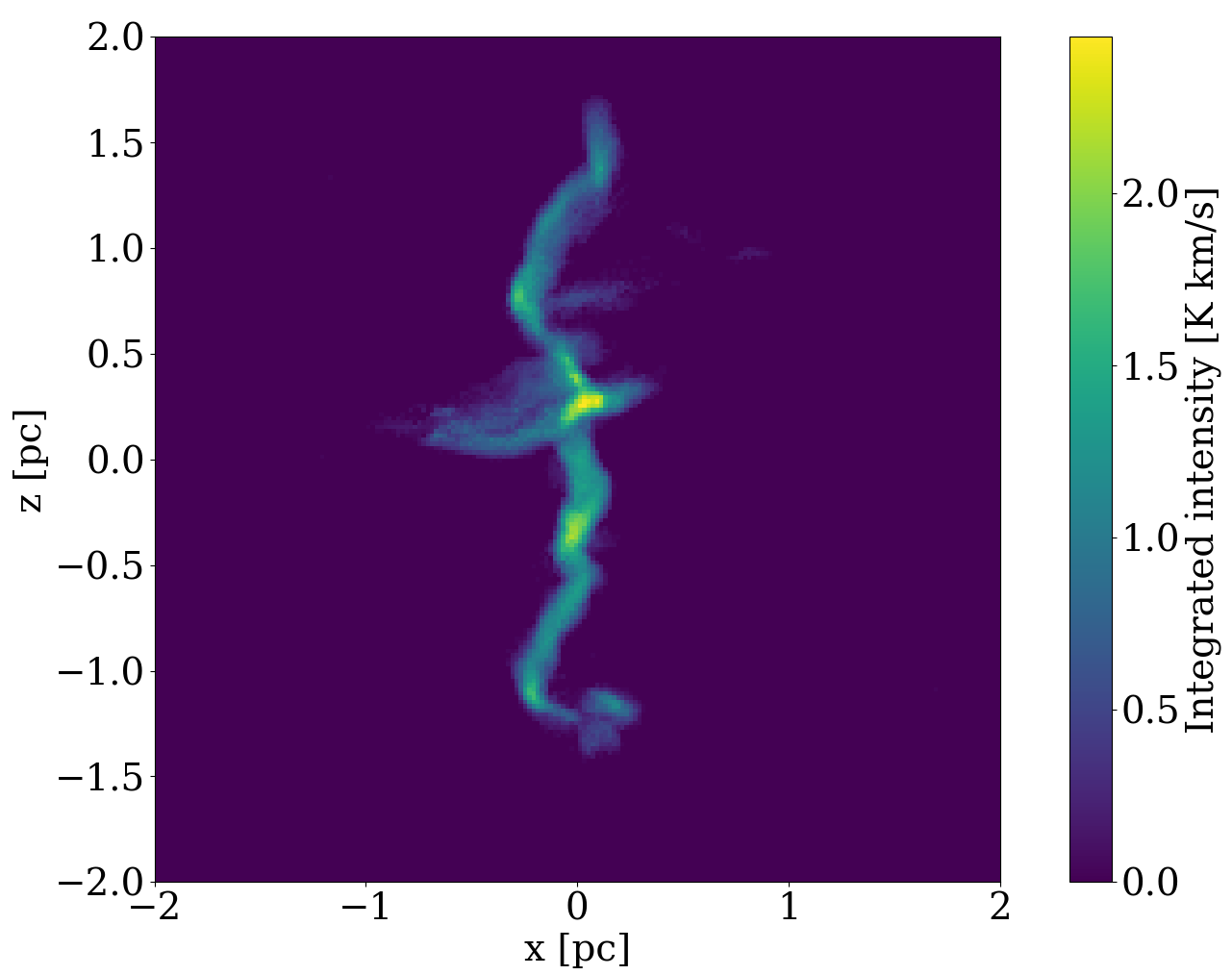}
\includegraphics[width = 0.35\linewidth]{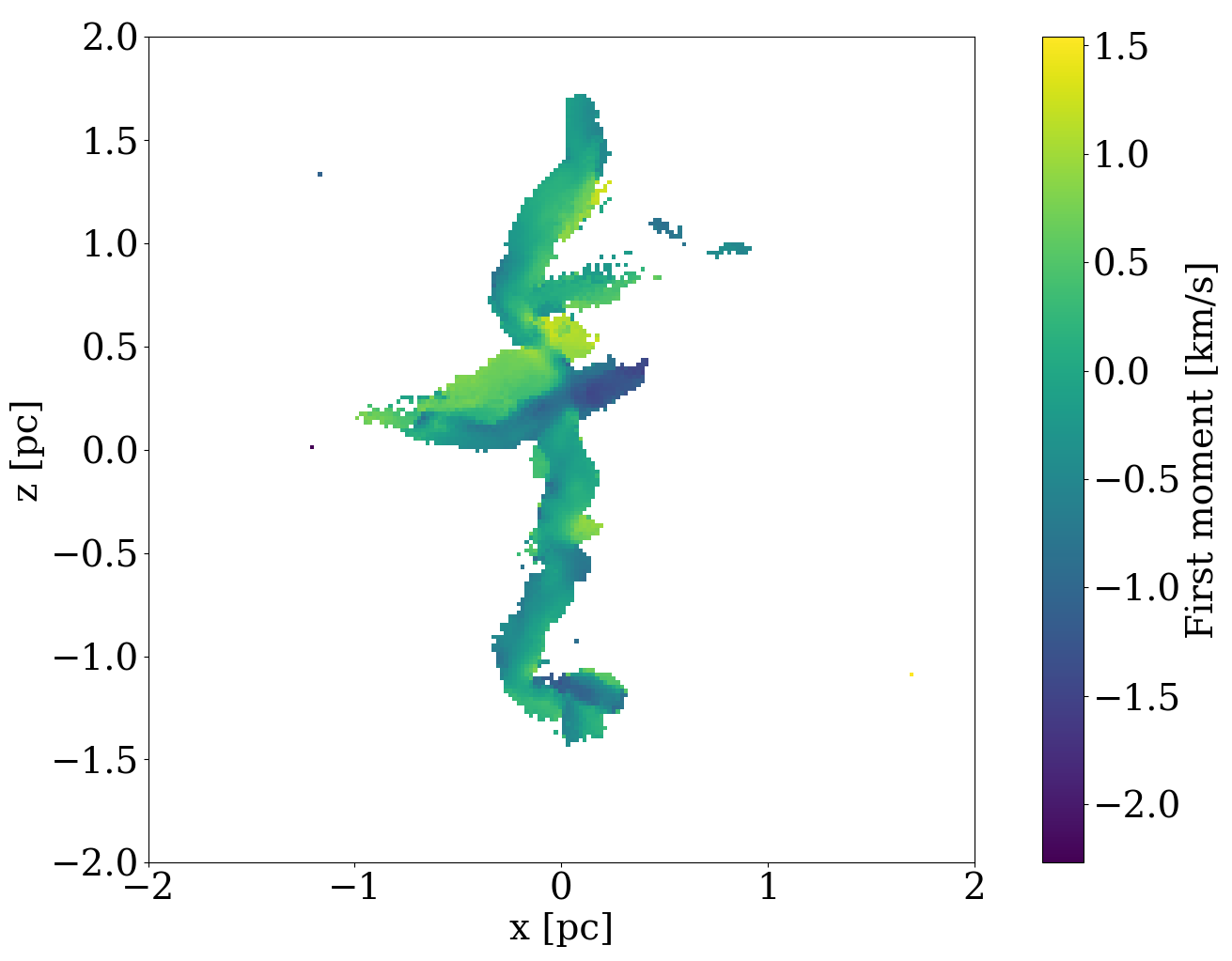}
\includegraphics[width = 0.35\linewidth]{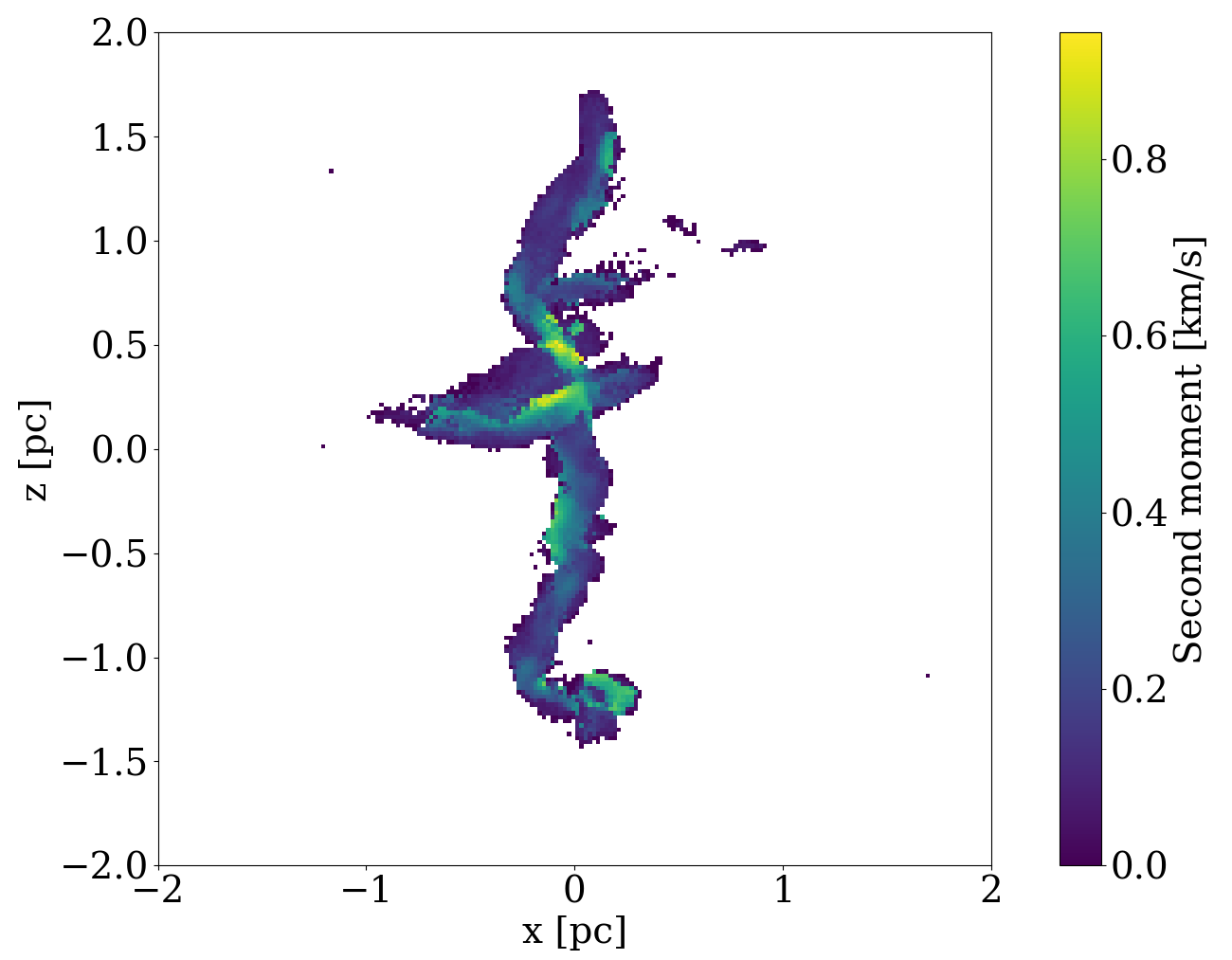}
\caption{Maps from simulation {\sc Sim}01, showing (a) the column density, (b) the integrated intensity of synthetic C$^{18}$O emission, (c) the intensity-weighted velocity centroid of synthetic C$^{18}$O emission, and (d) the intensity-weighted velocity dispersion of synthetic  C$^{18}$O emission. All maps have the same resolution, $0.02\,{\rm pc}$, and have not been convolved with a beam.}
\label{fig::seed1}
\end{figure*}

\begin{figure*}
\centering
\includegraphics[width = 0.35\linewidth]{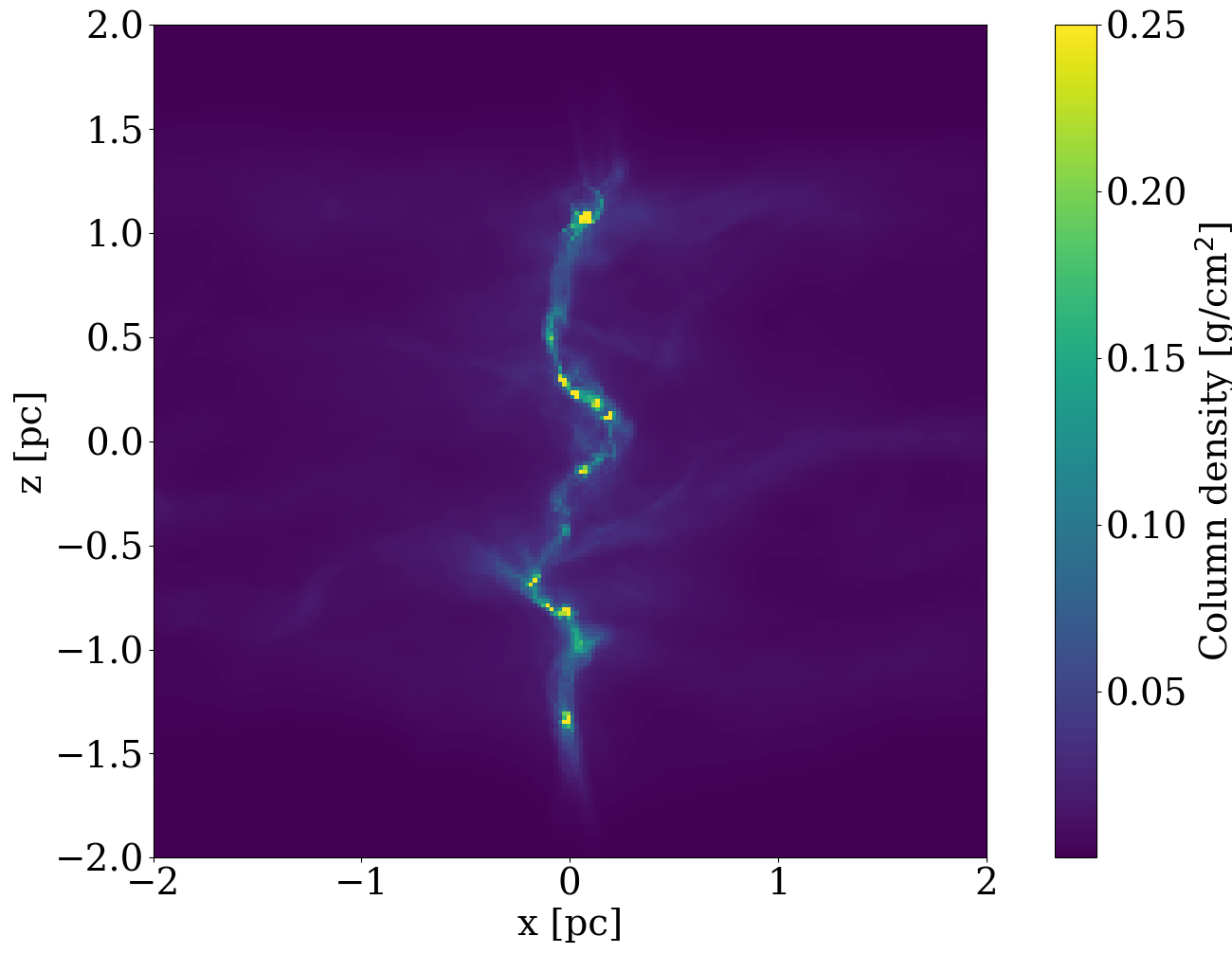}
\includegraphics[width = 0.35\linewidth]{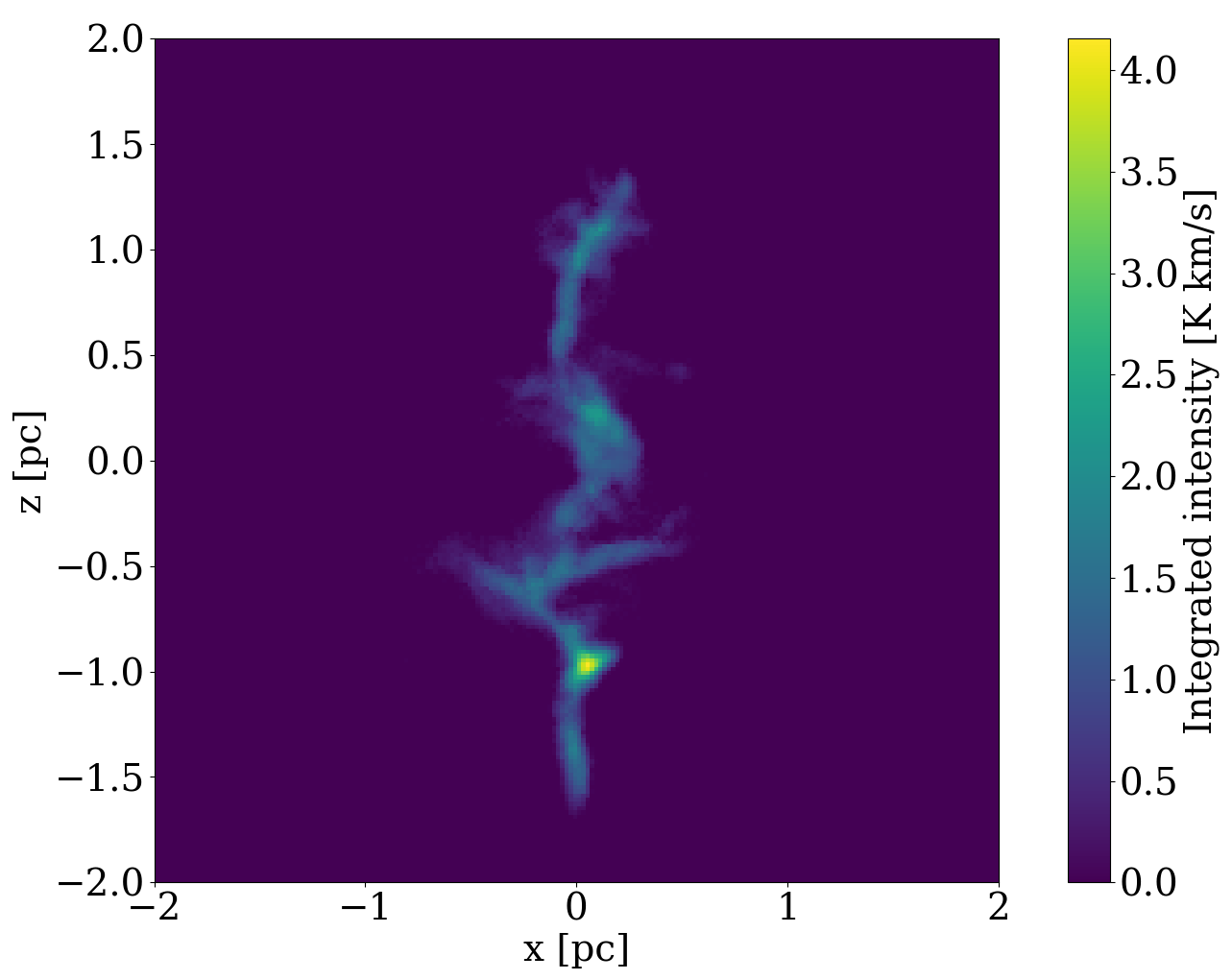}
\includegraphics[width = 0.35\linewidth]{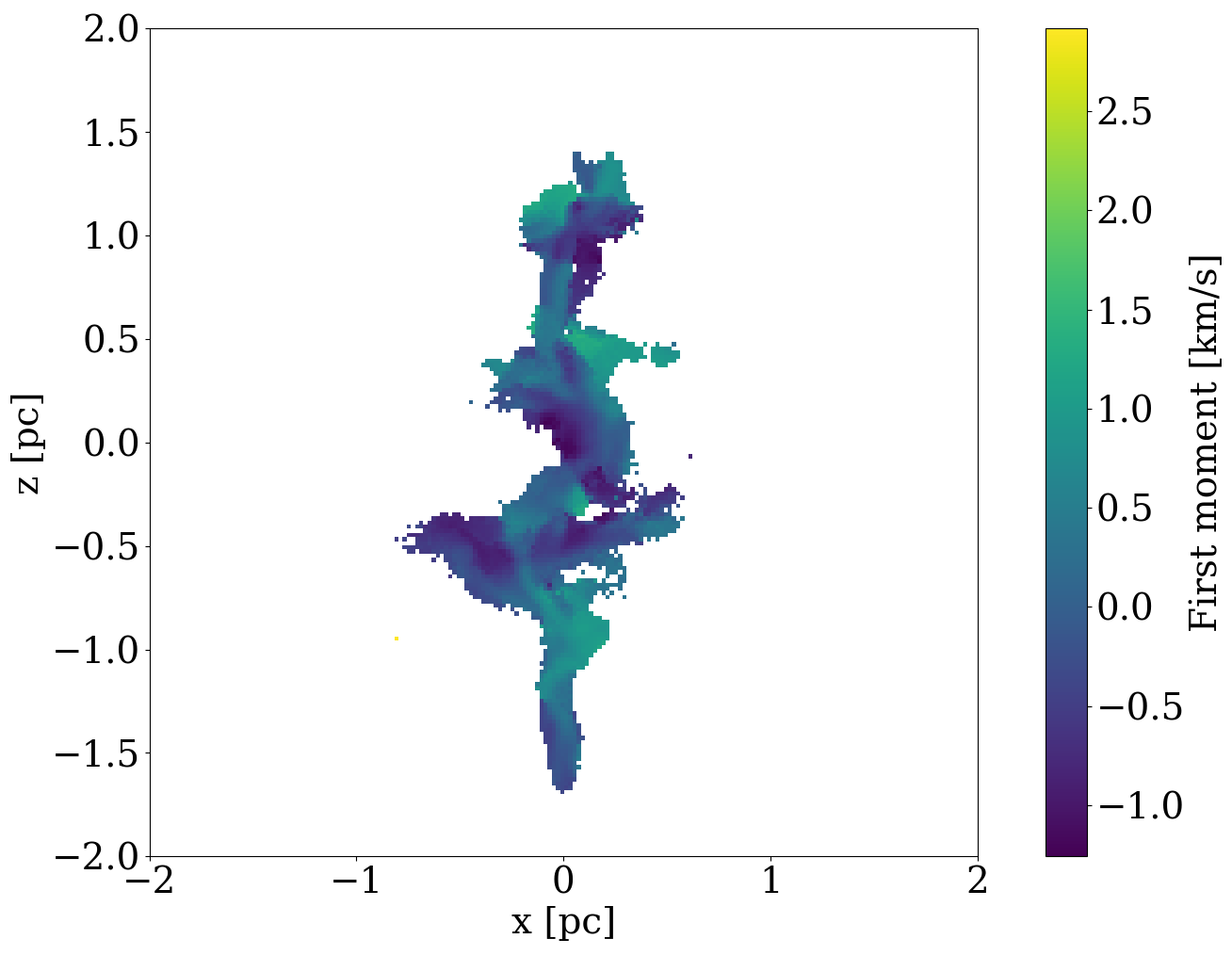}
\includegraphics[width = 0.35\linewidth]{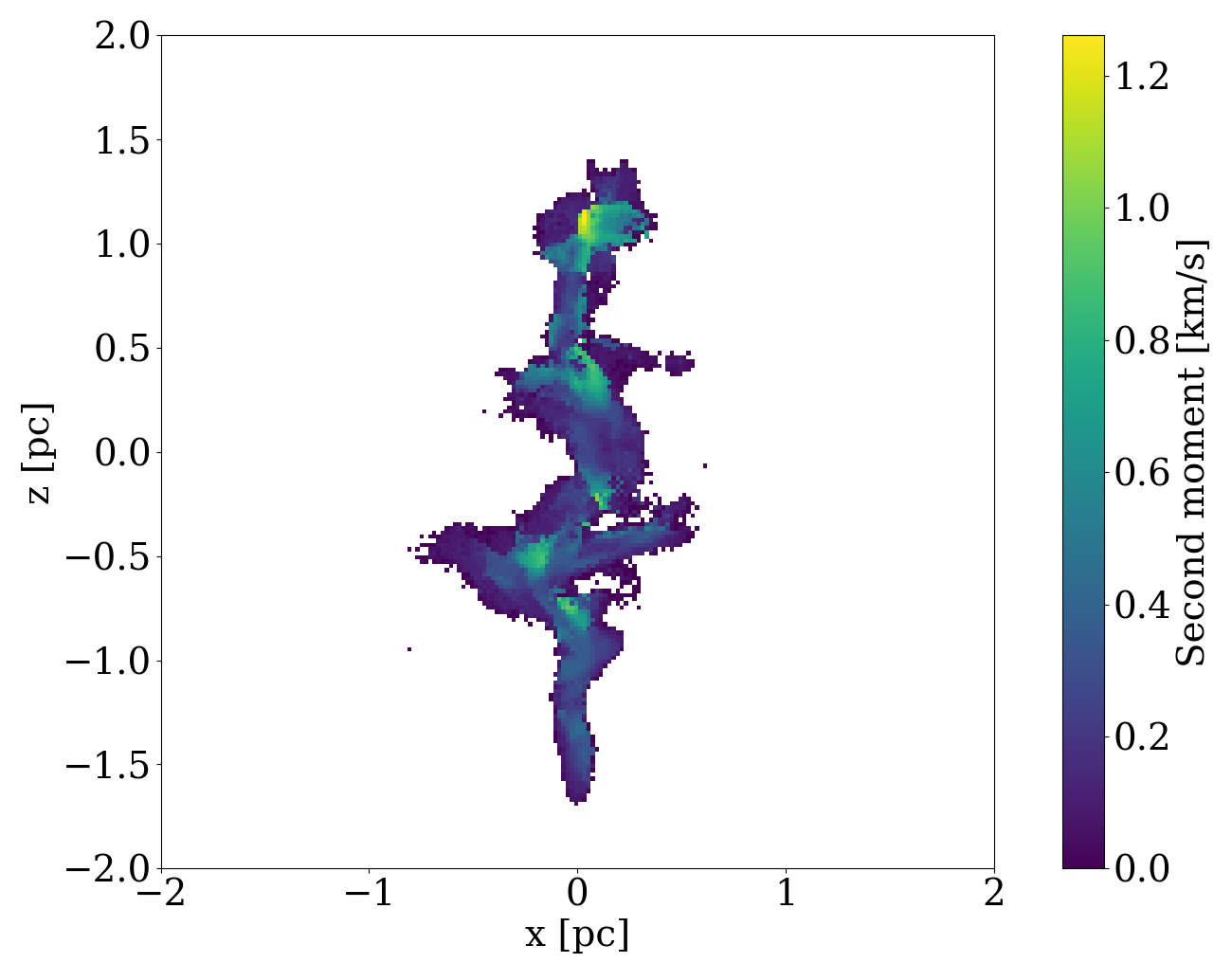}
\caption{As Fig. \ref{fig::seed1}, but for {\sc Sim}03.}
\label{fig::seed3}
\end{figure*}

\begin{figure*}
\centering
\includegraphics[width = 0.35\linewidth]{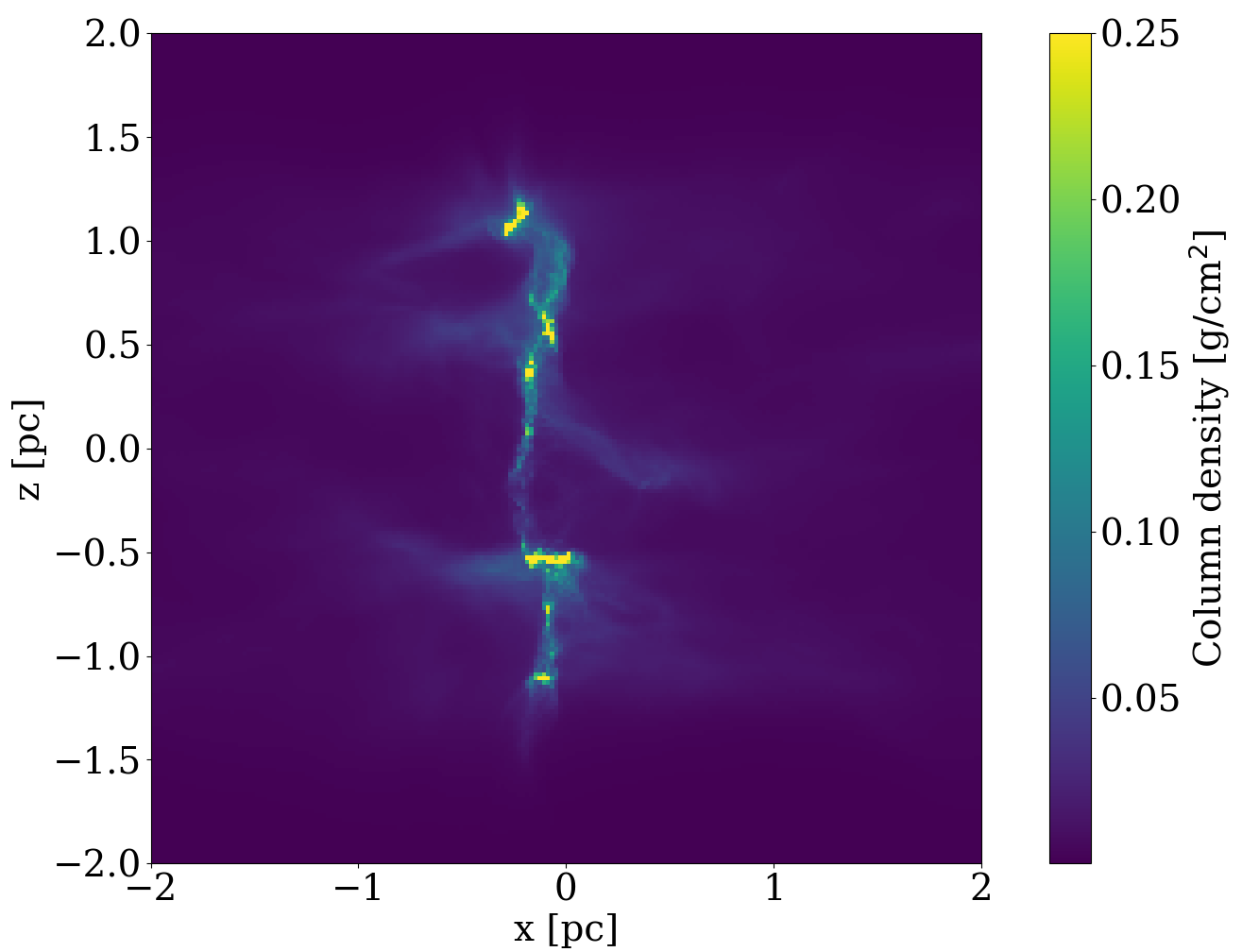}
\includegraphics[width = 0.35\linewidth]{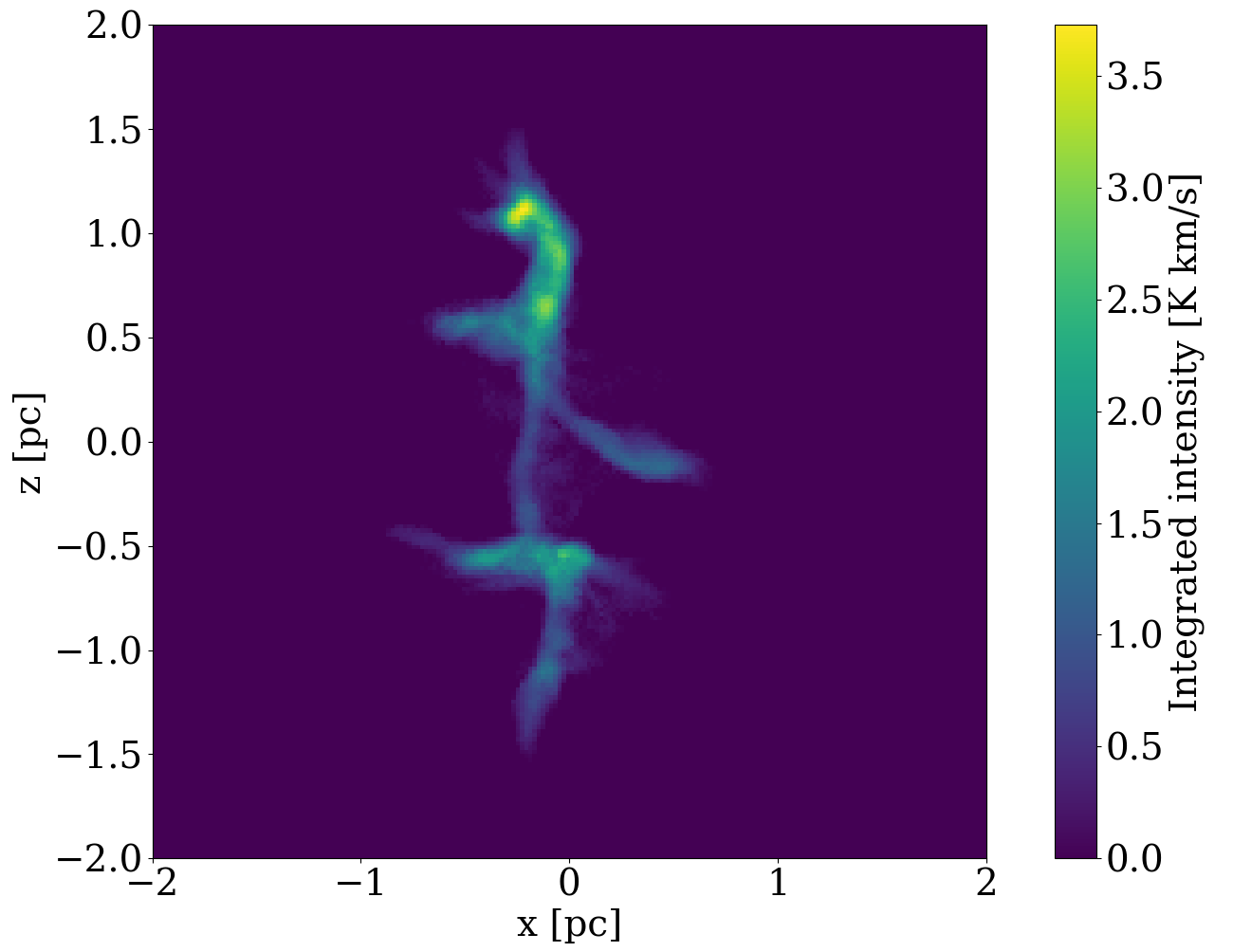}
\includegraphics[width = 0.35\linewidth]{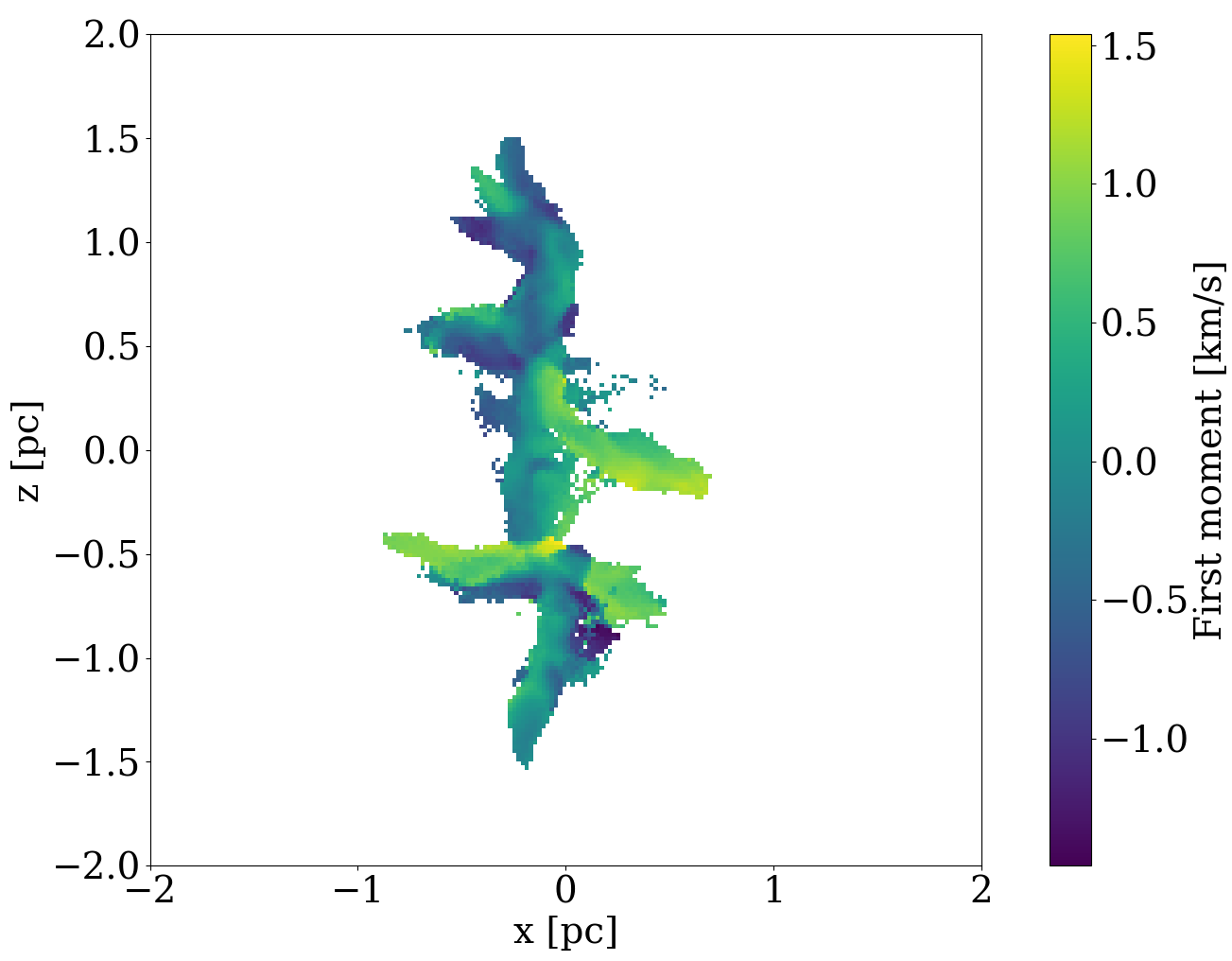}
\includegraphics[width = 0.35\linewidth]{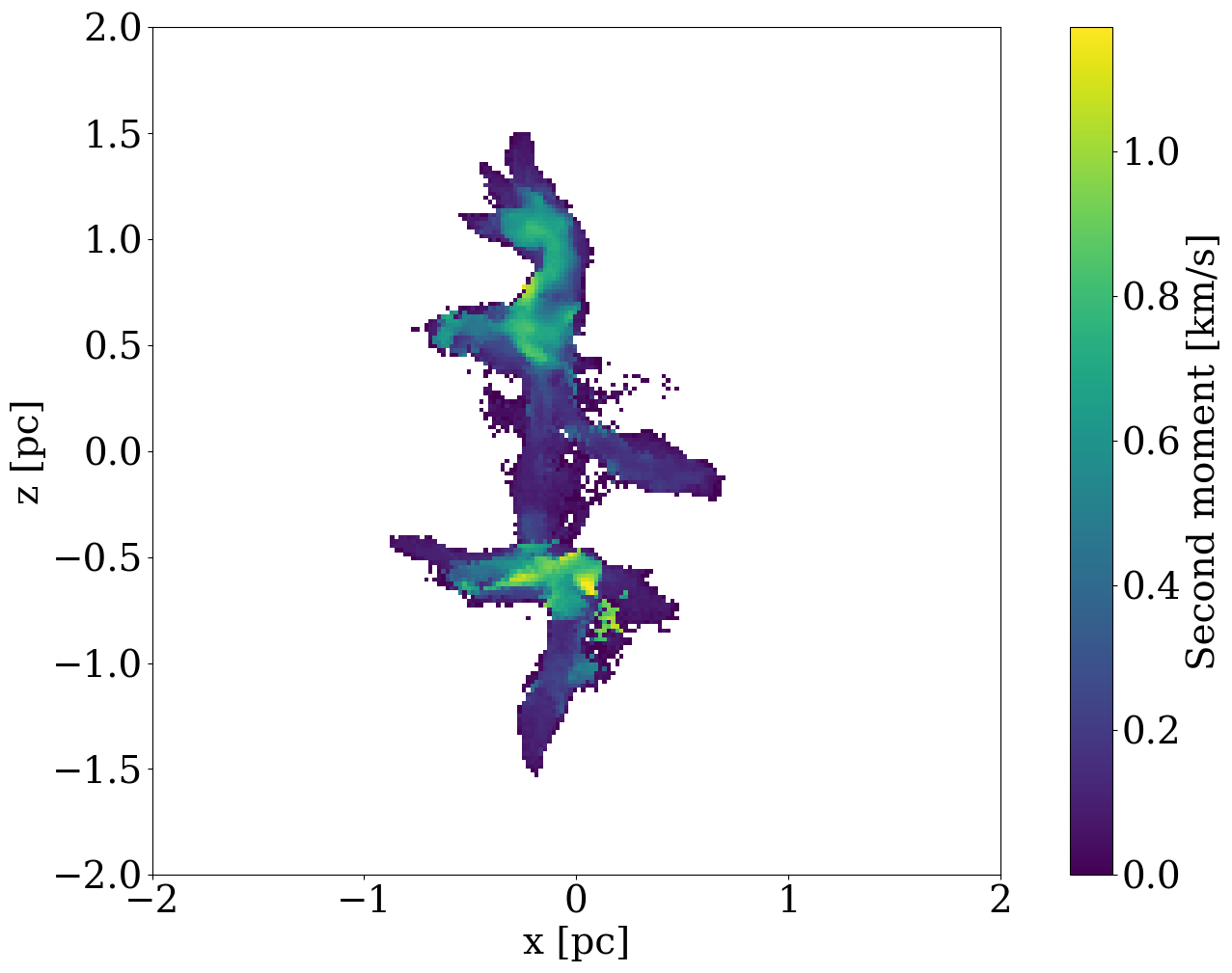}
\caption{As Fig. \ref{fig::seed1}, but for {\sc Sim}04.}
\label{fig::seed4}
\end{figure*}

\begin{figure*}
\centering
\includegraphics[width = 0.35\linewidth]{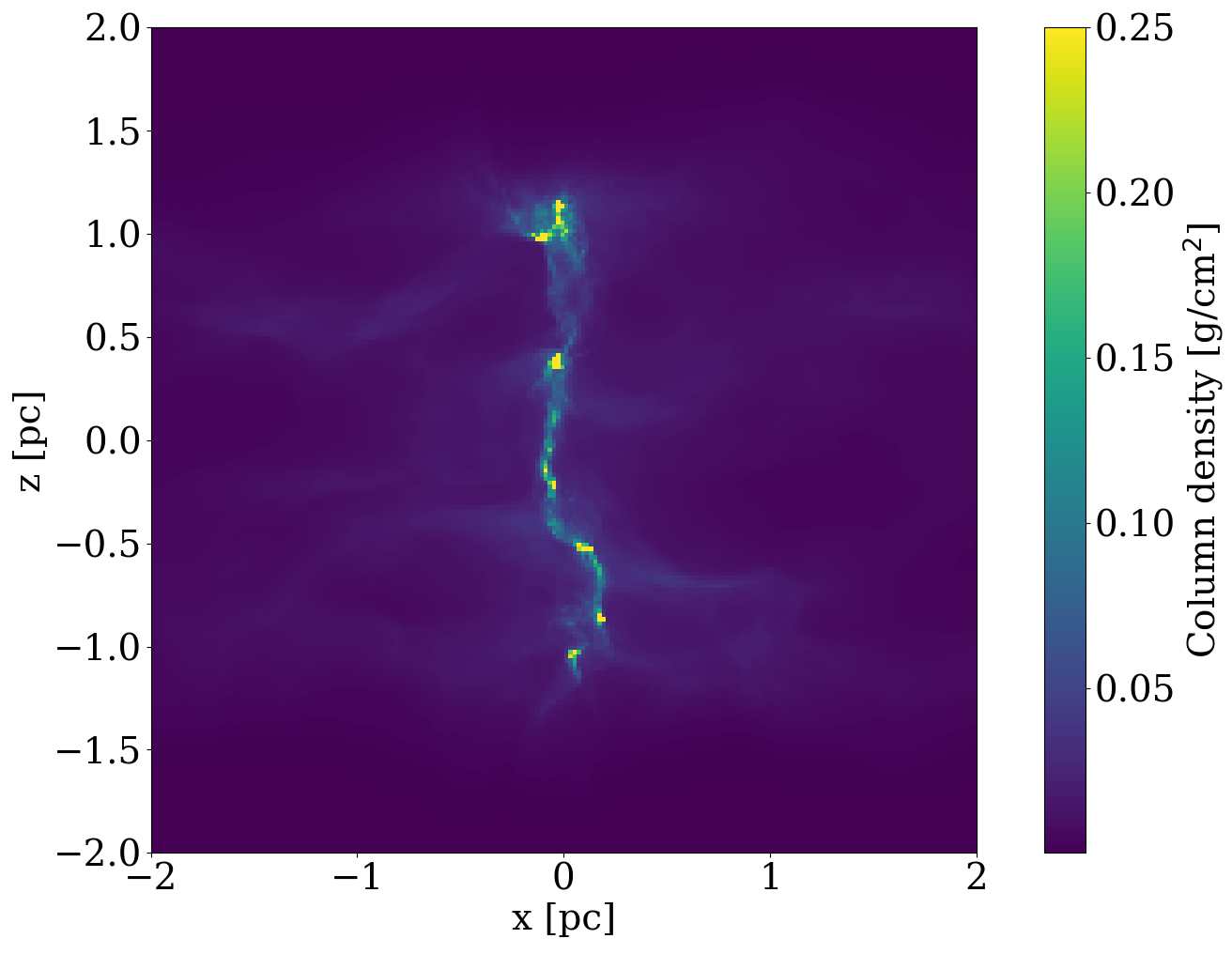}
\includegraphics[width = 0.35\linewidth]{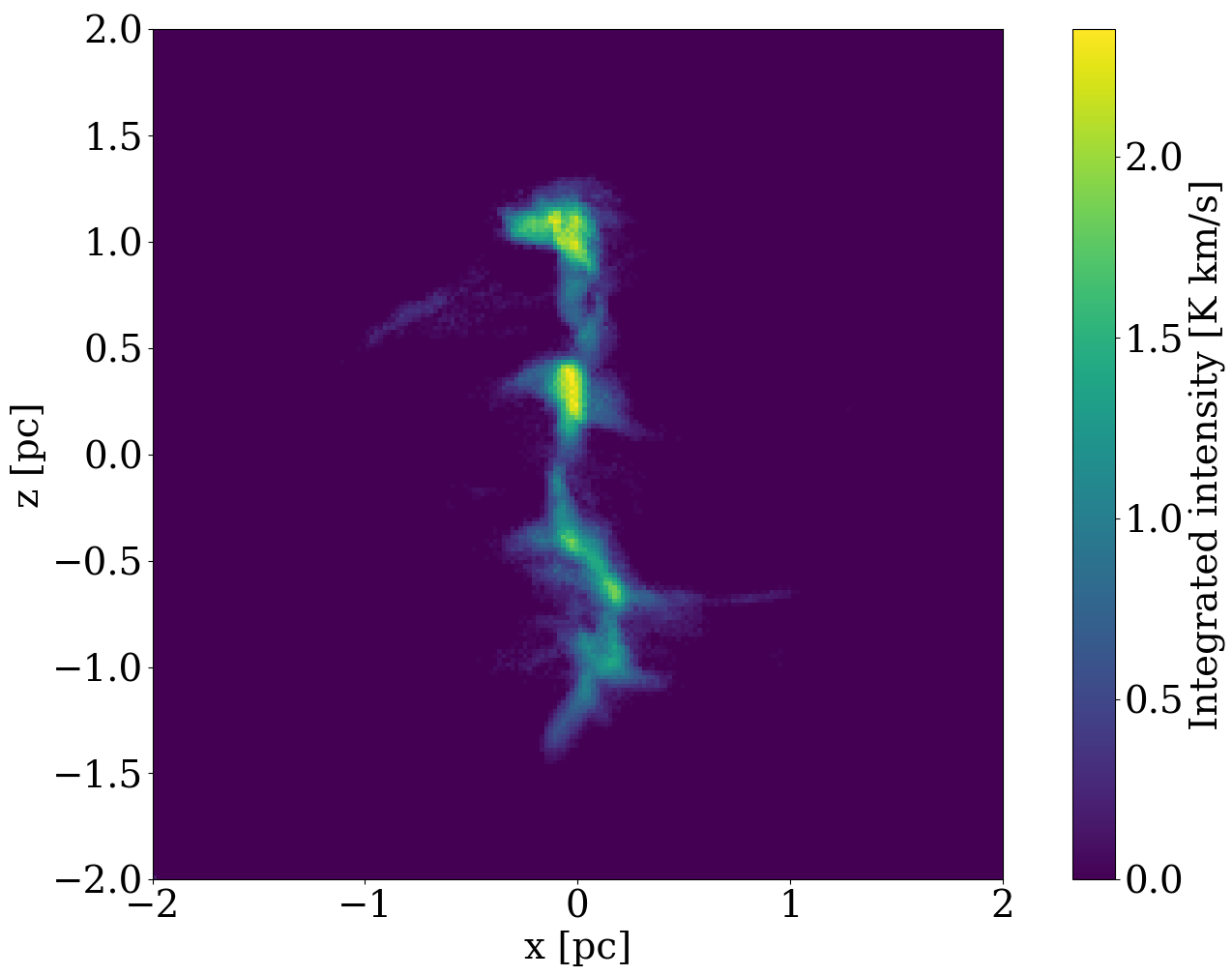}
\includegraphics[width = 0.35\linewidth]{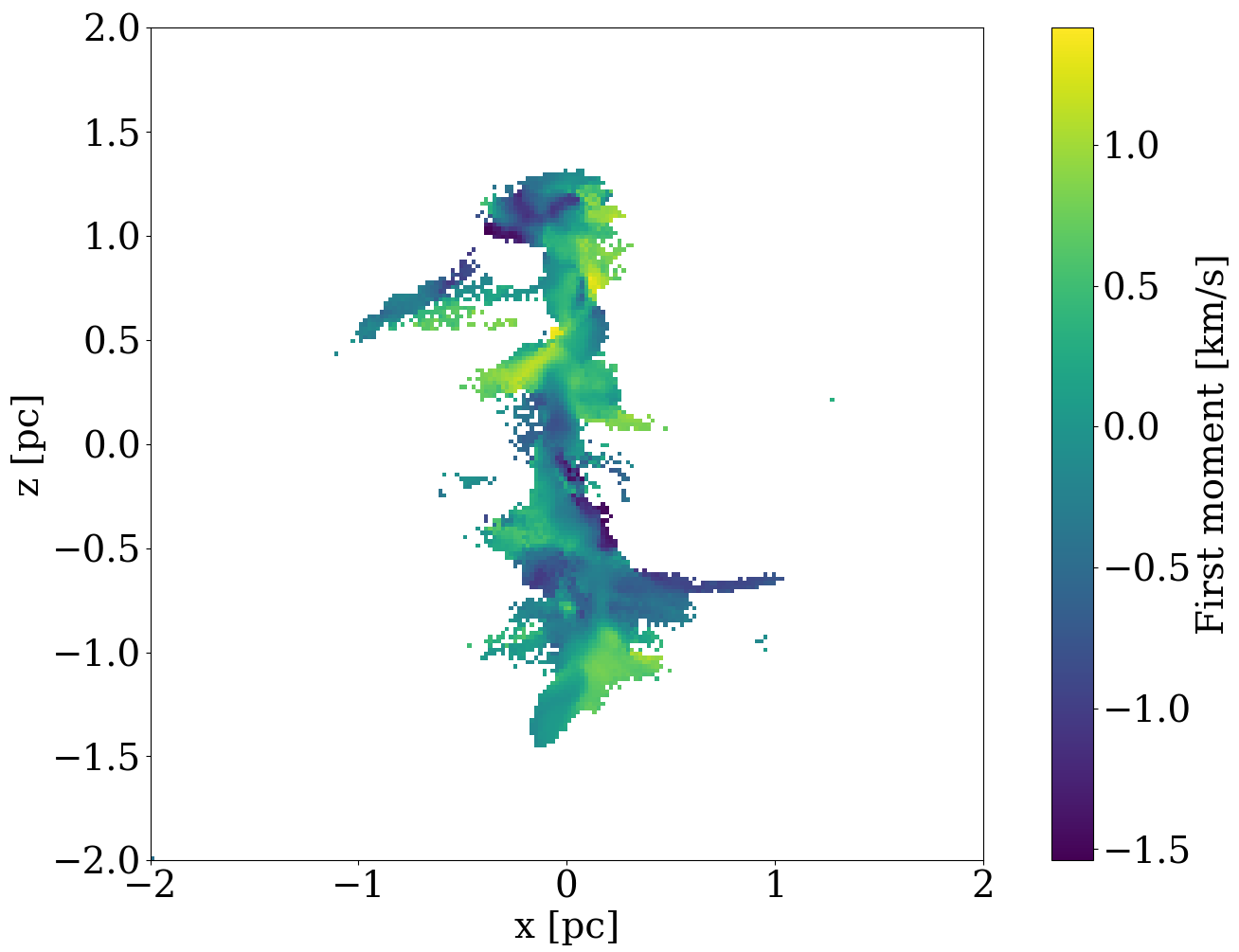}
\includegraphics[width = 0.35\linewidth]{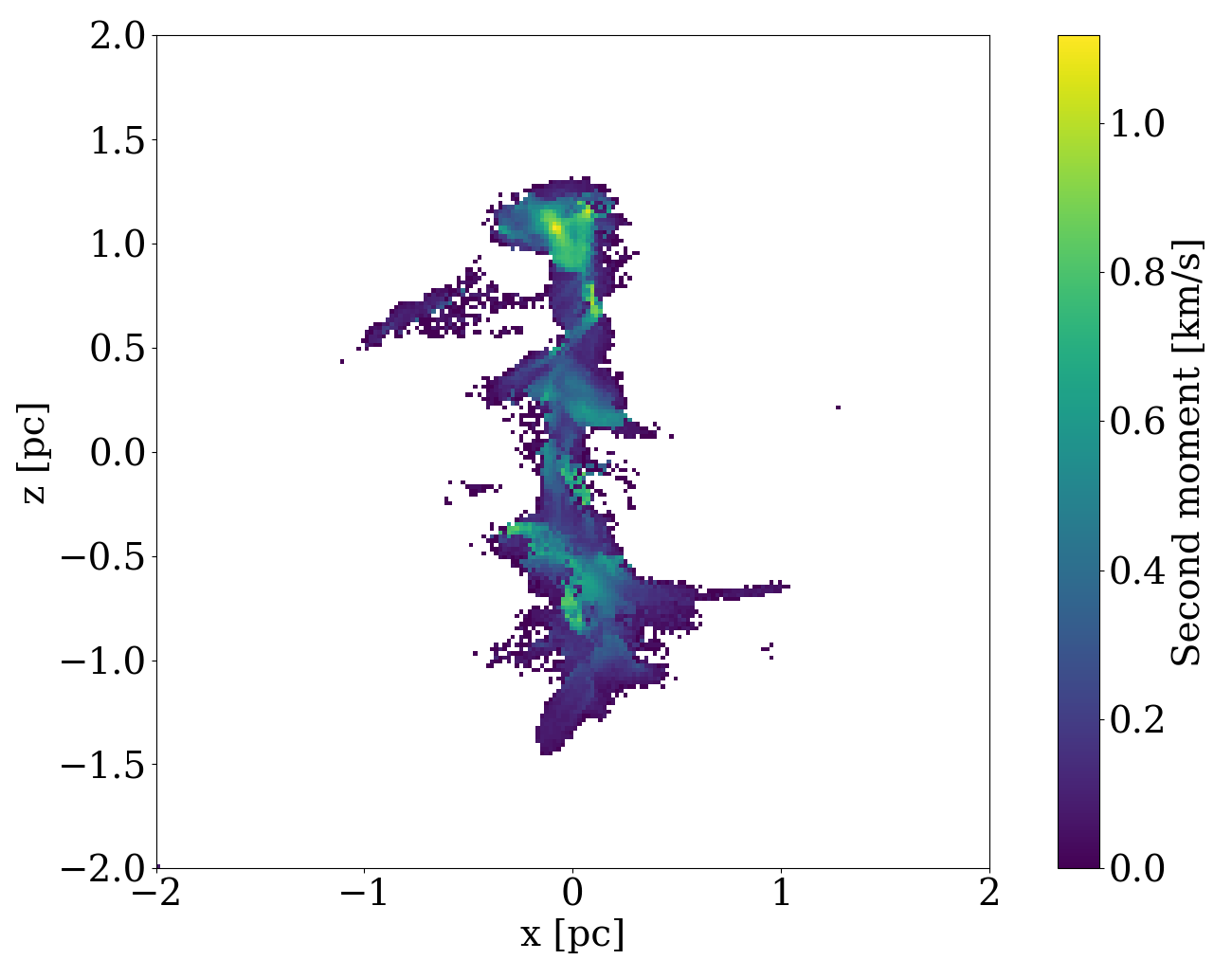}
\caption{As Fig. \ref{fig::seed1}, but for {\sc Sim}05.}
\label{fig::seed5}
\end{figure*}

\begin{figure*}
\centering
\includegraphics[width = 0.35\linewidth]{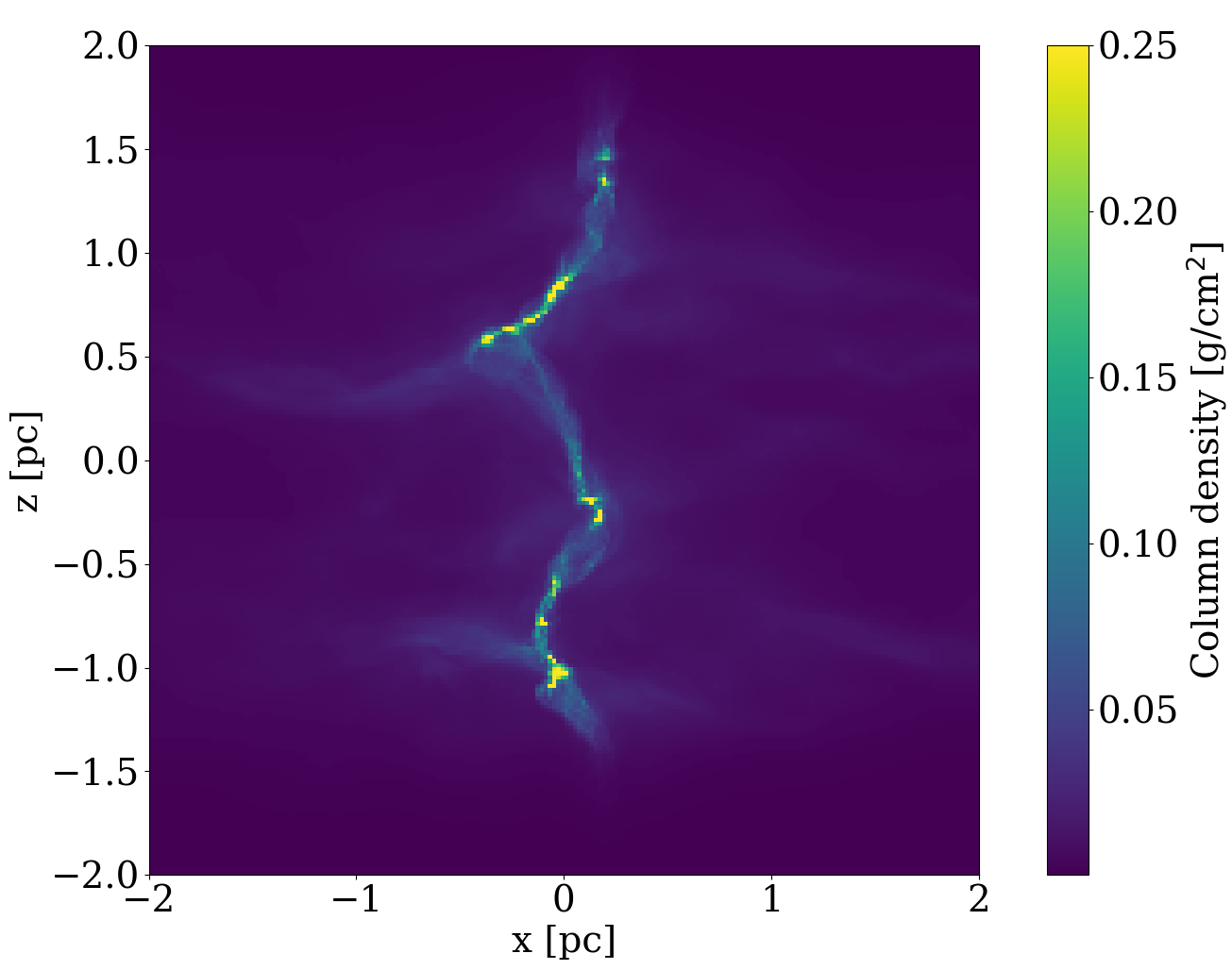}
\includegraphics[width = 0.35\linewidth]{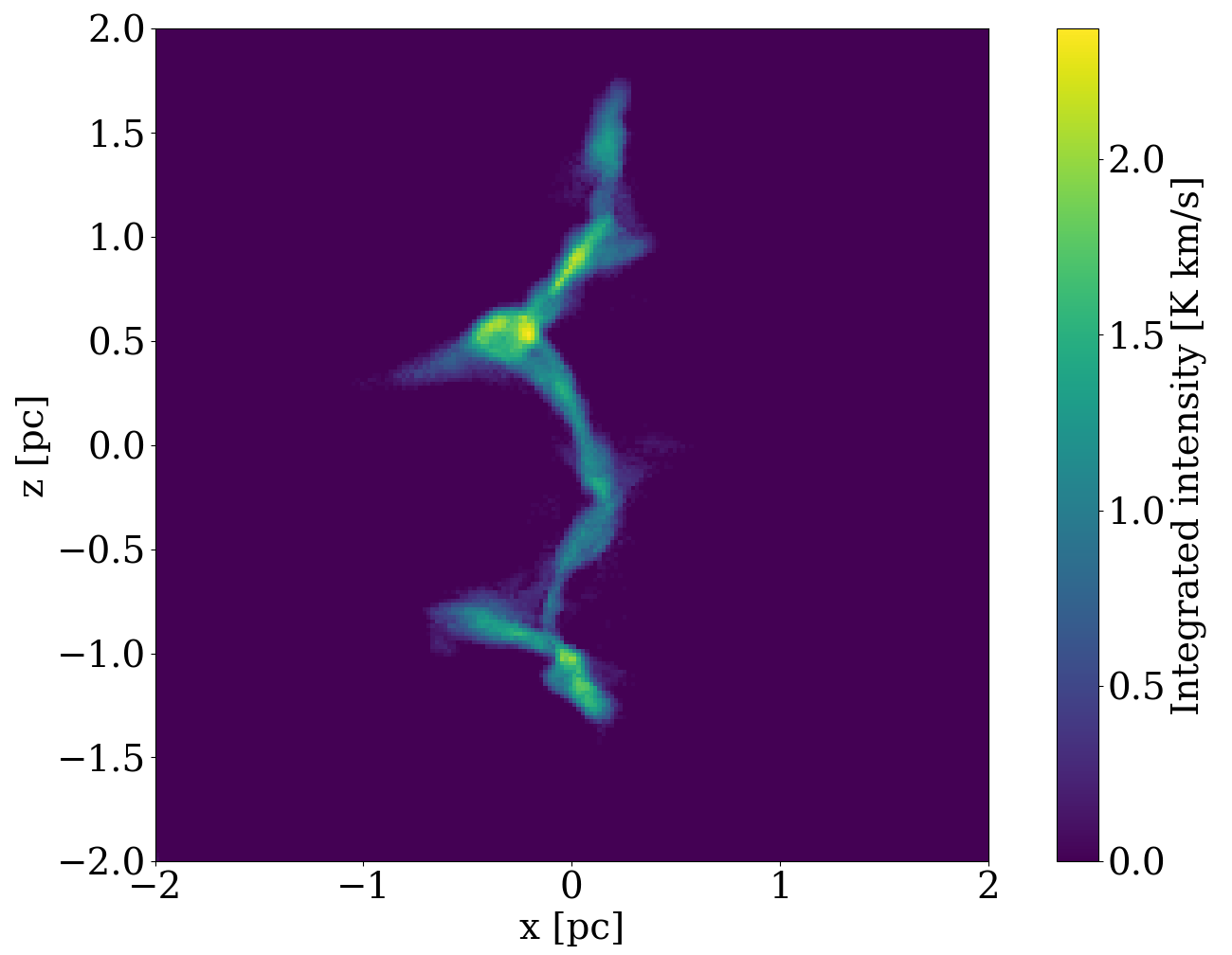}
\includegraphics[width = 0.35\linewidth]{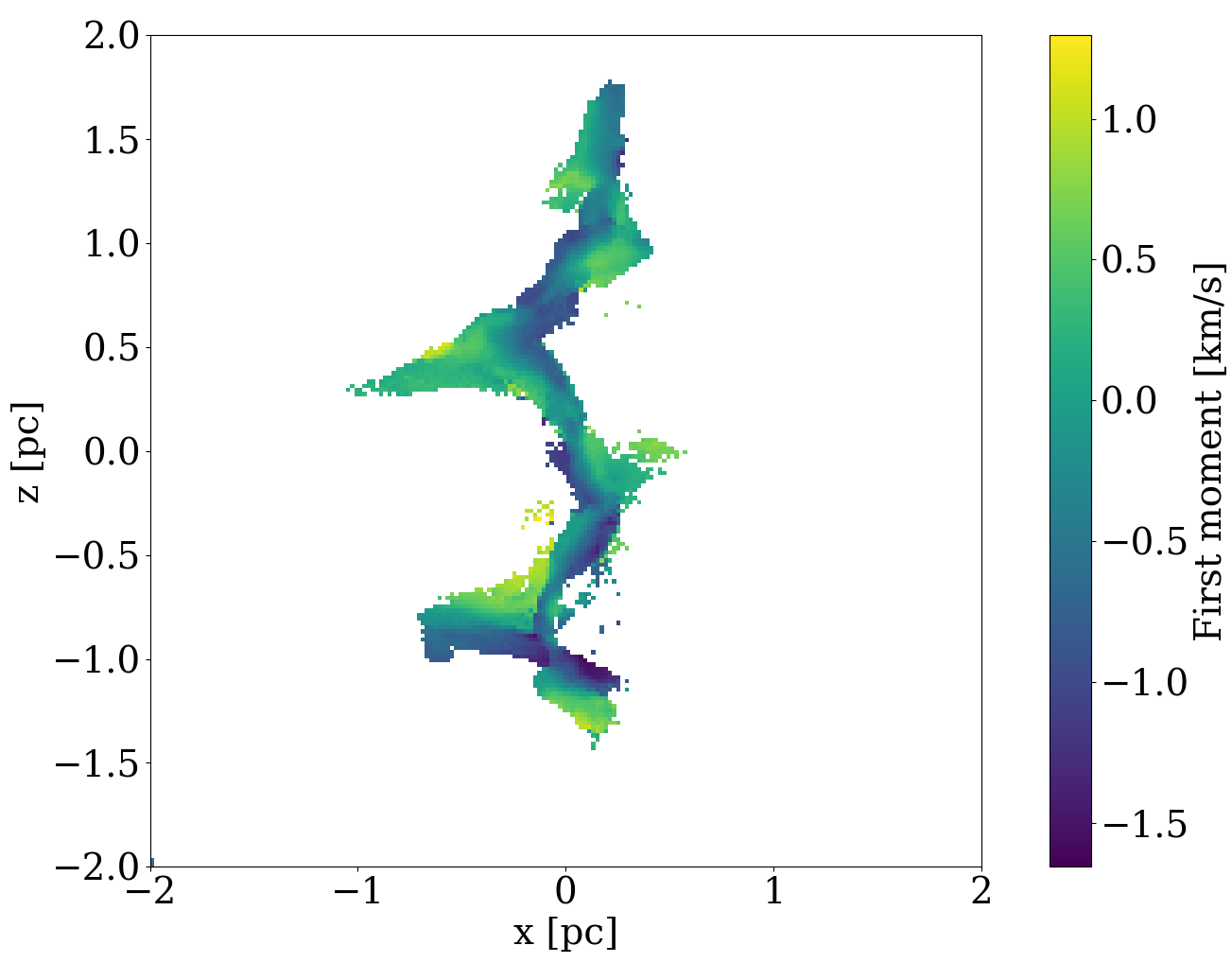}
\includegraphics[width = 0.35\linewidth]{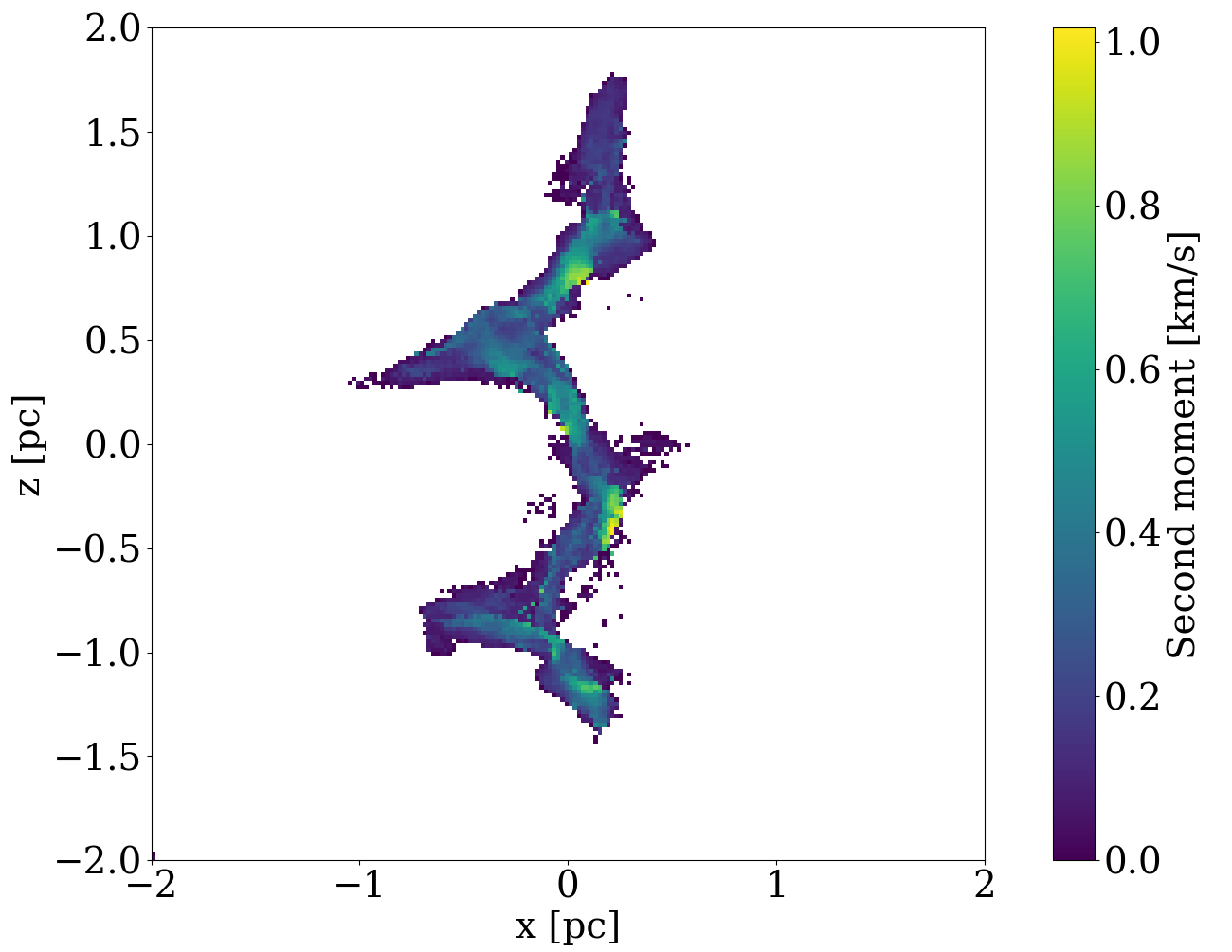}
\caption{As Fig. \ref{fig::seed1}, but for {\sc Sim}06.}
\label{fig::seed6}
\end{figure*}

\begin{figure*}
\centering
\includegraphics[width = 0.35\linewidth]{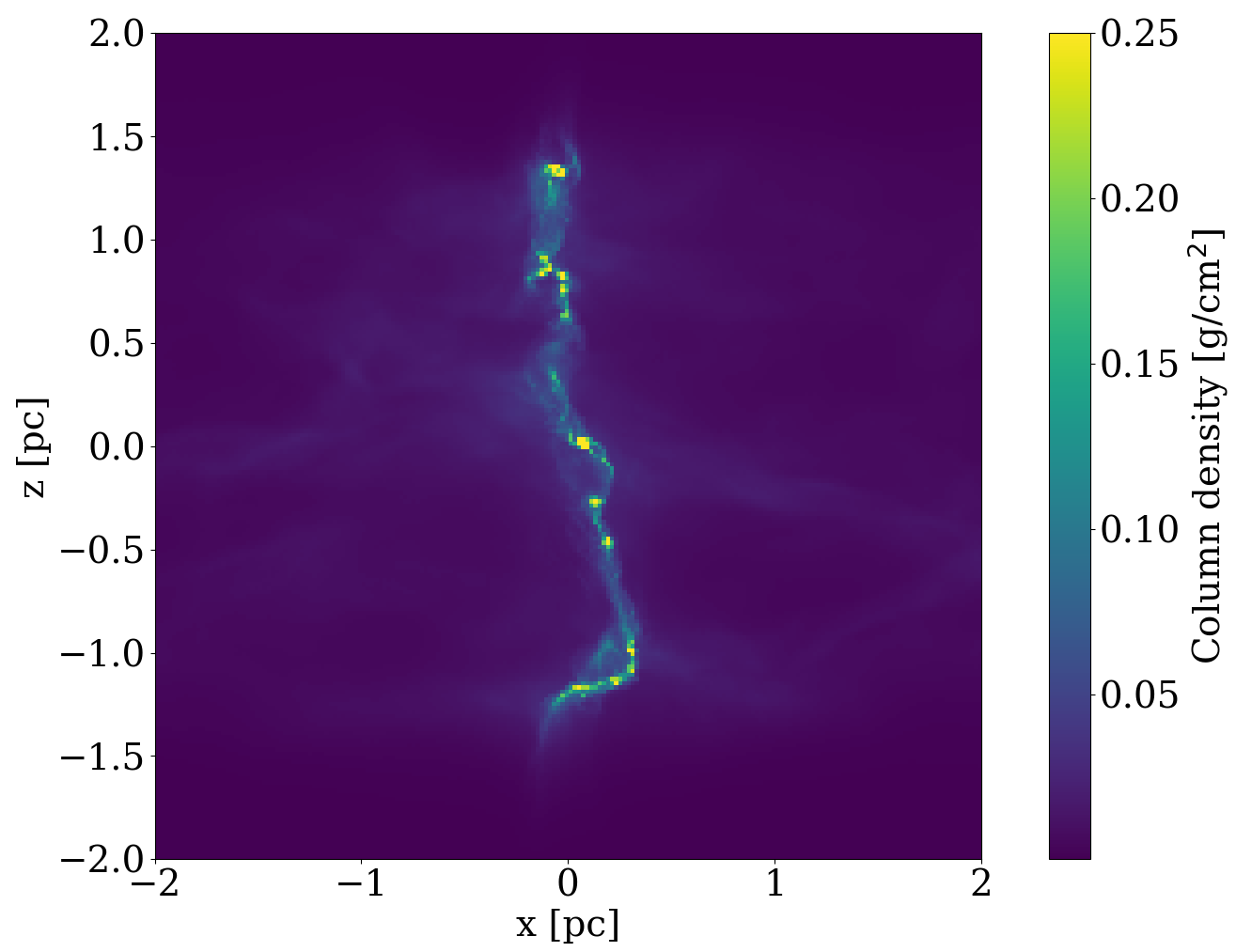}
\includegraphics[width = 0.35\linewidth]{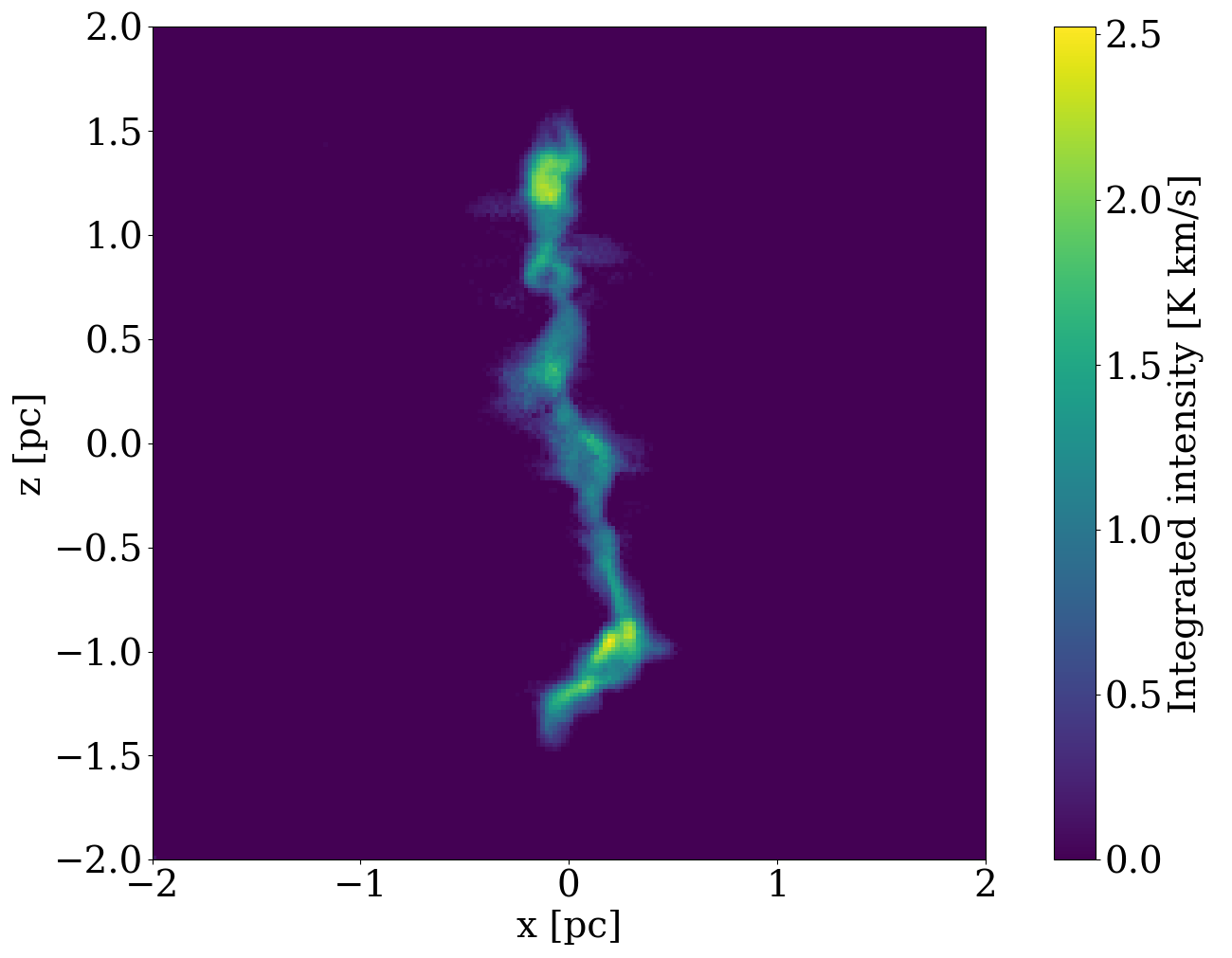}
\includegraphics[width = 0.35\linewidth]{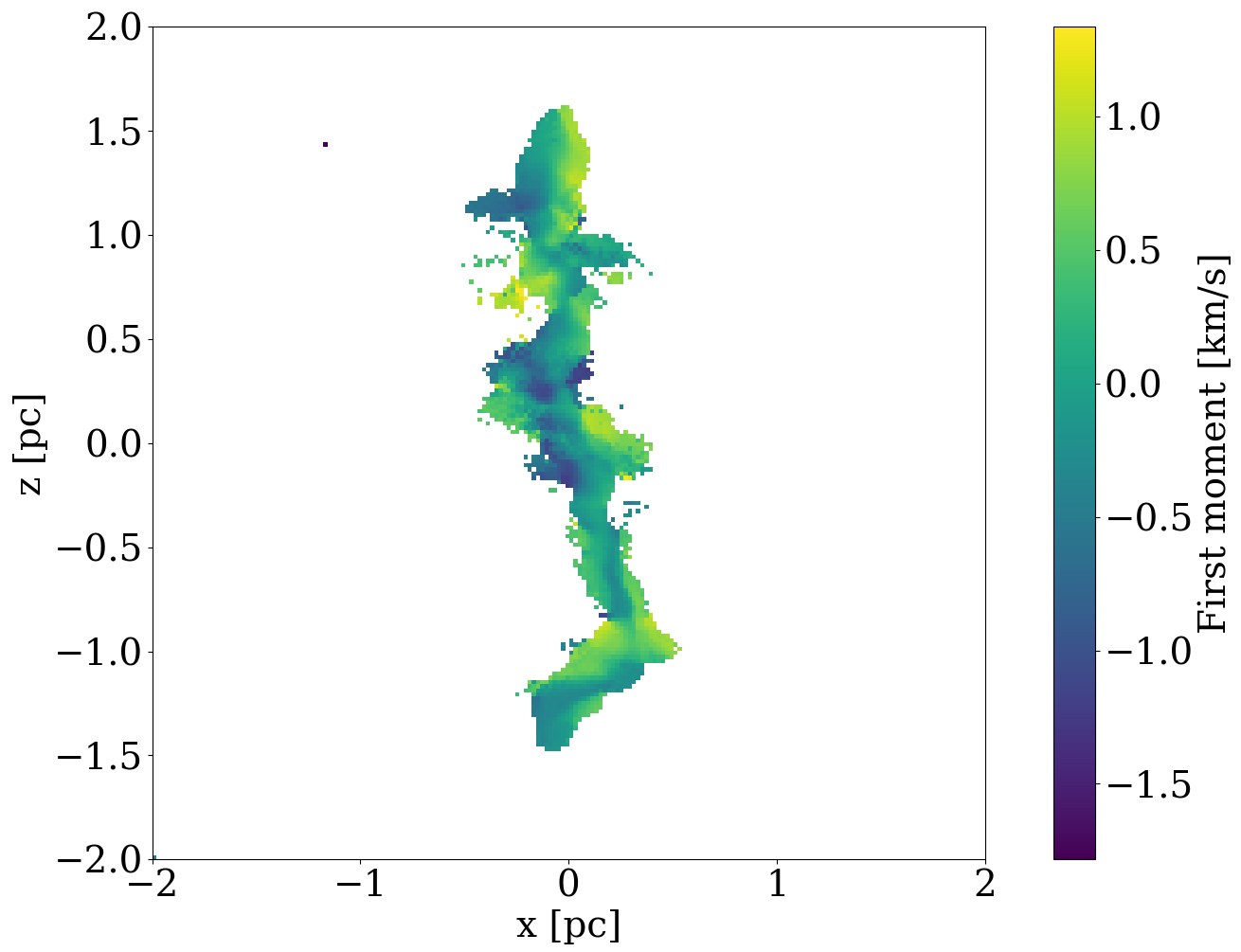}
\includegraphics[width = 0.35\linewidth]{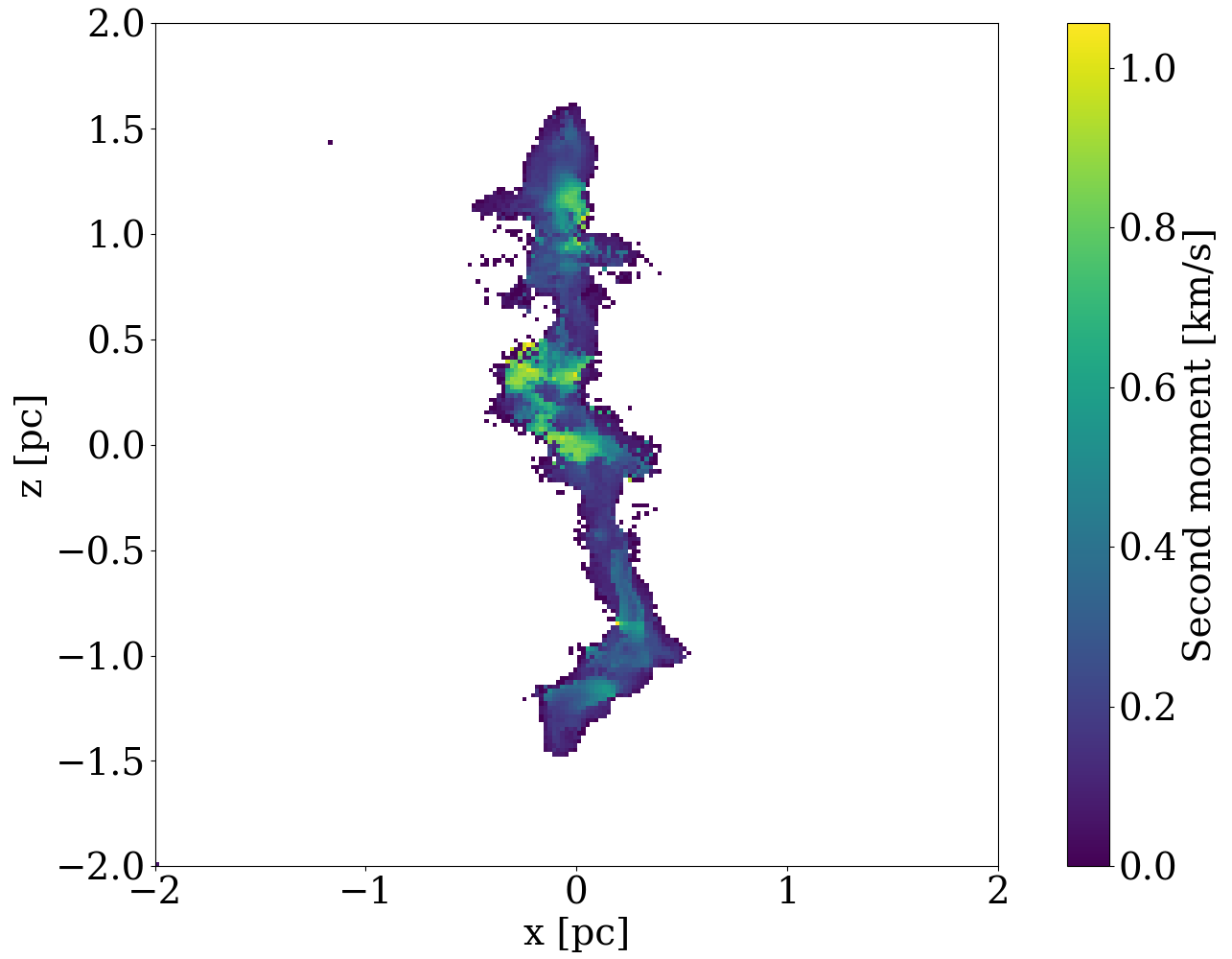}
\caption{As Fig. \ref{fig::seed1}, but for {\sc Sim}07.}
\label{fig::seed7}
\end{figure*}

\begin{figure*}
\centering
\includegraphics[width = 0.35\linewidth]{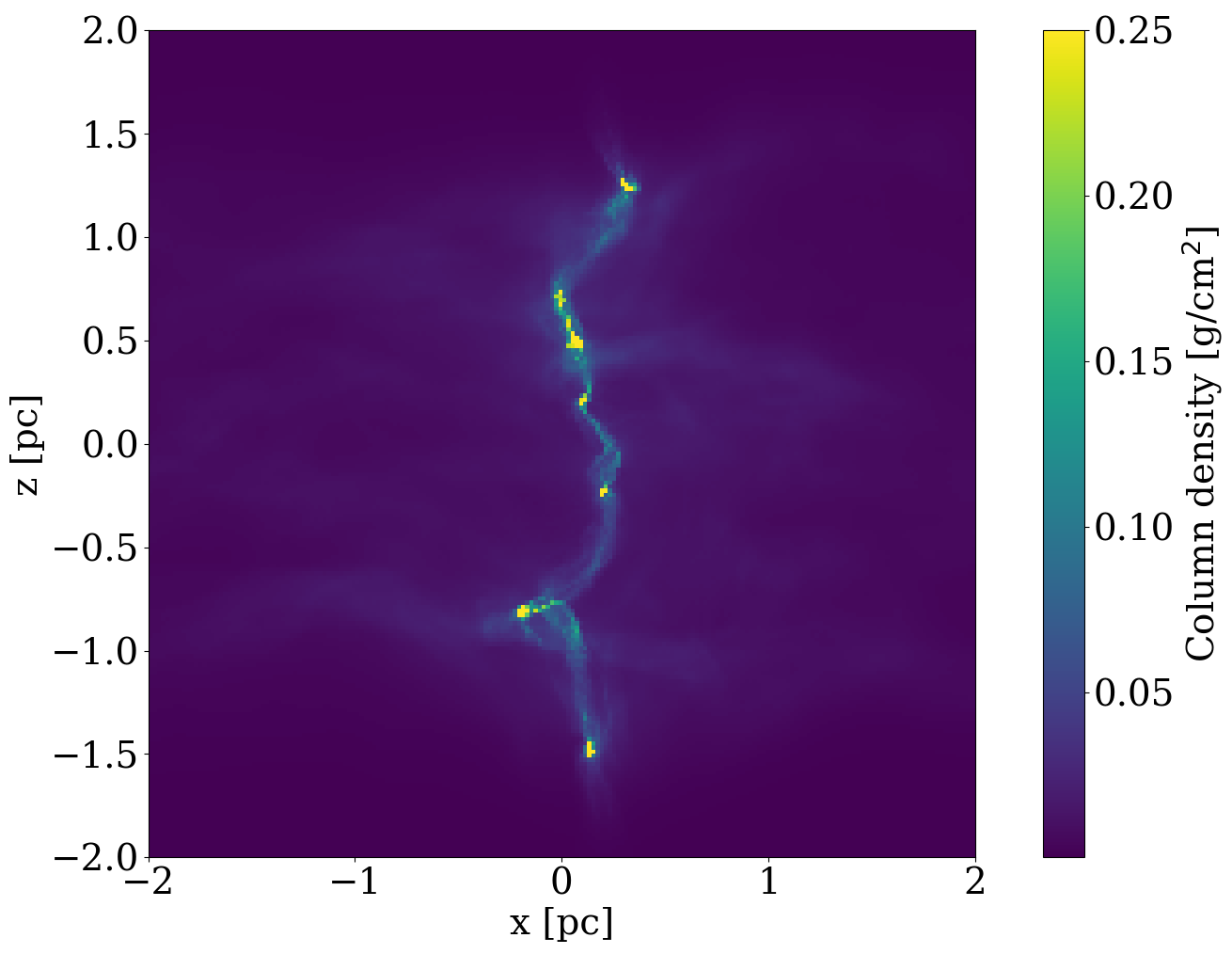}
\includegraphics[width = 0.35\linewidth]{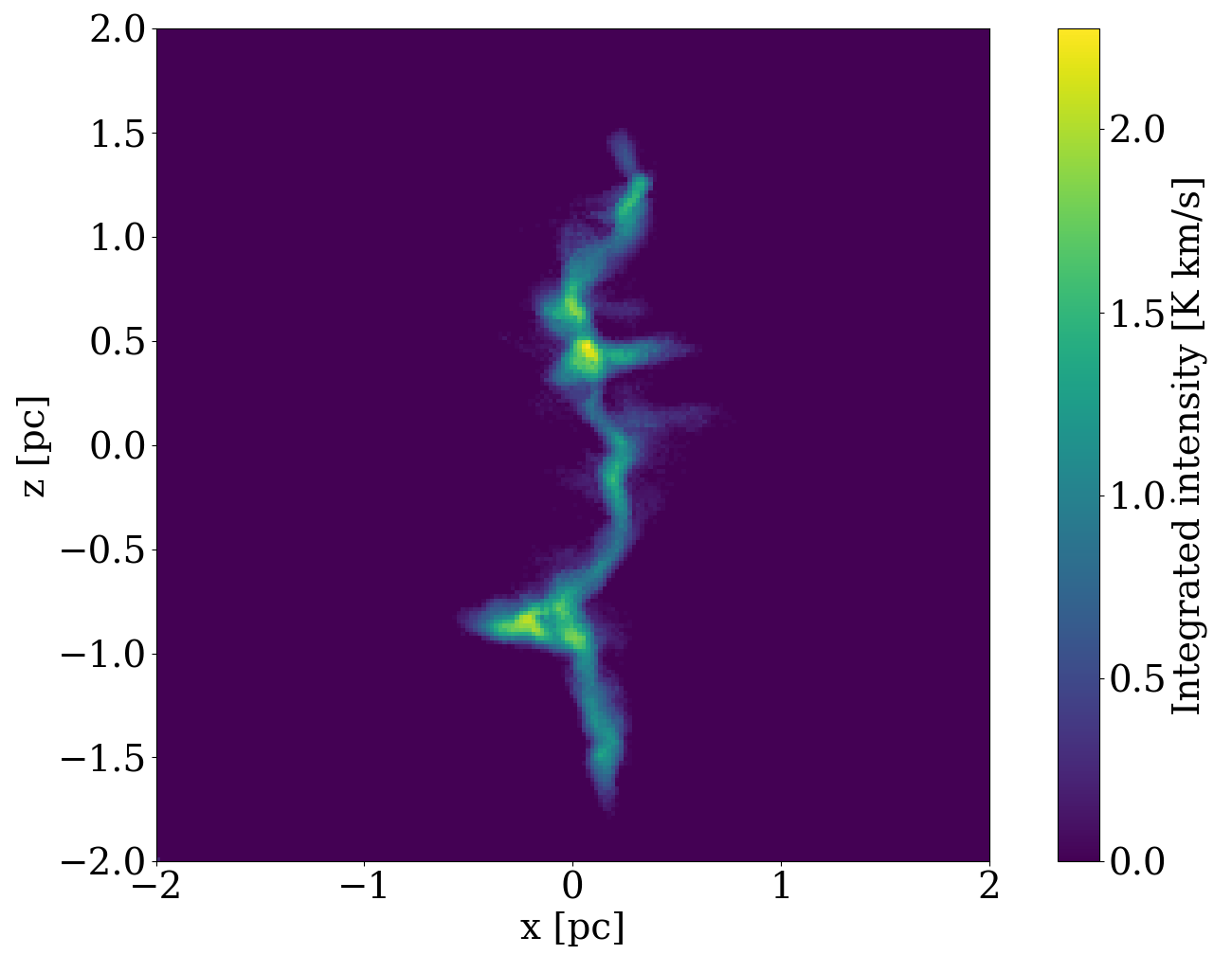}
\includegraphics[width = 0.35\linewidth]{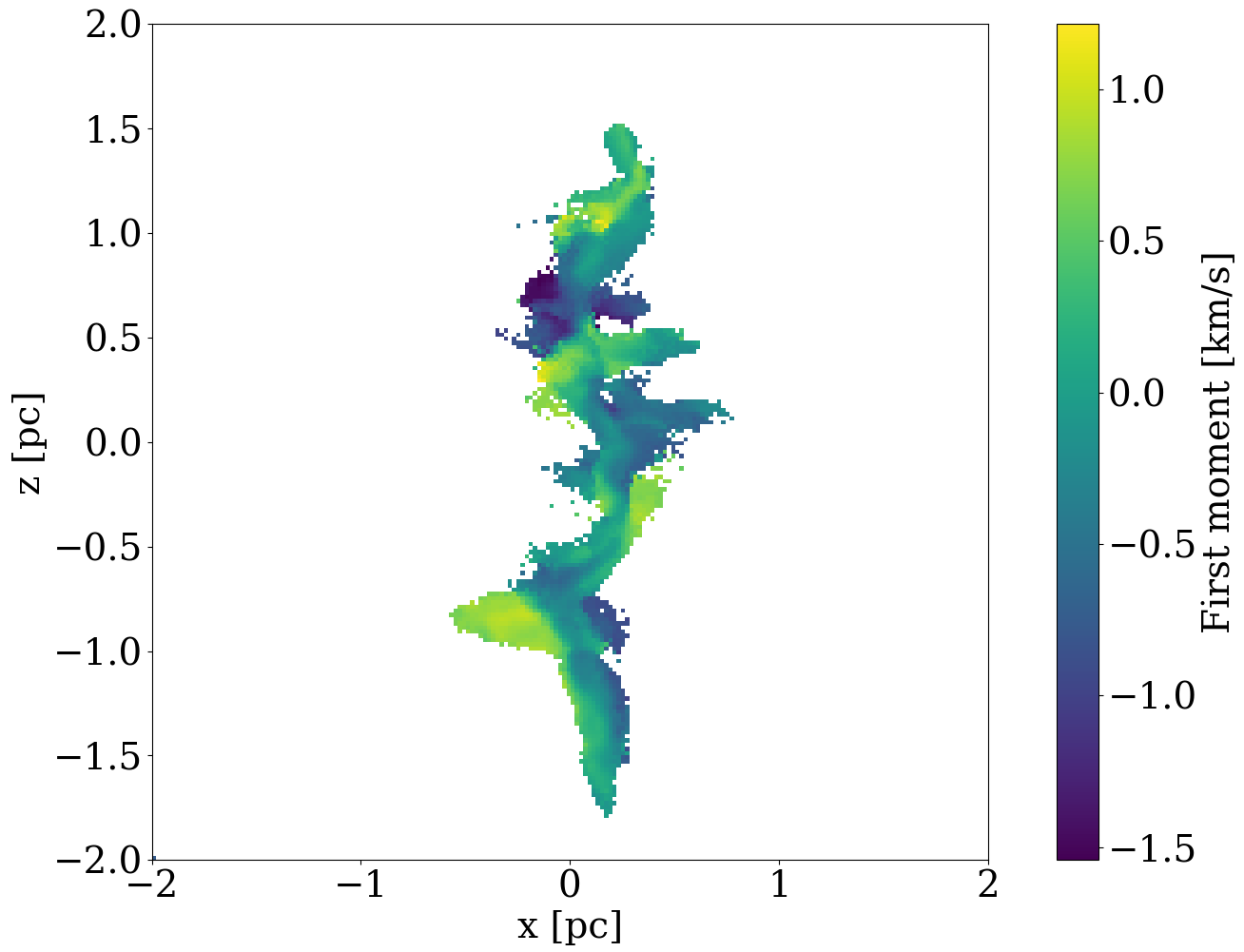}
\includegraphics[width = 0.35\linewidth]{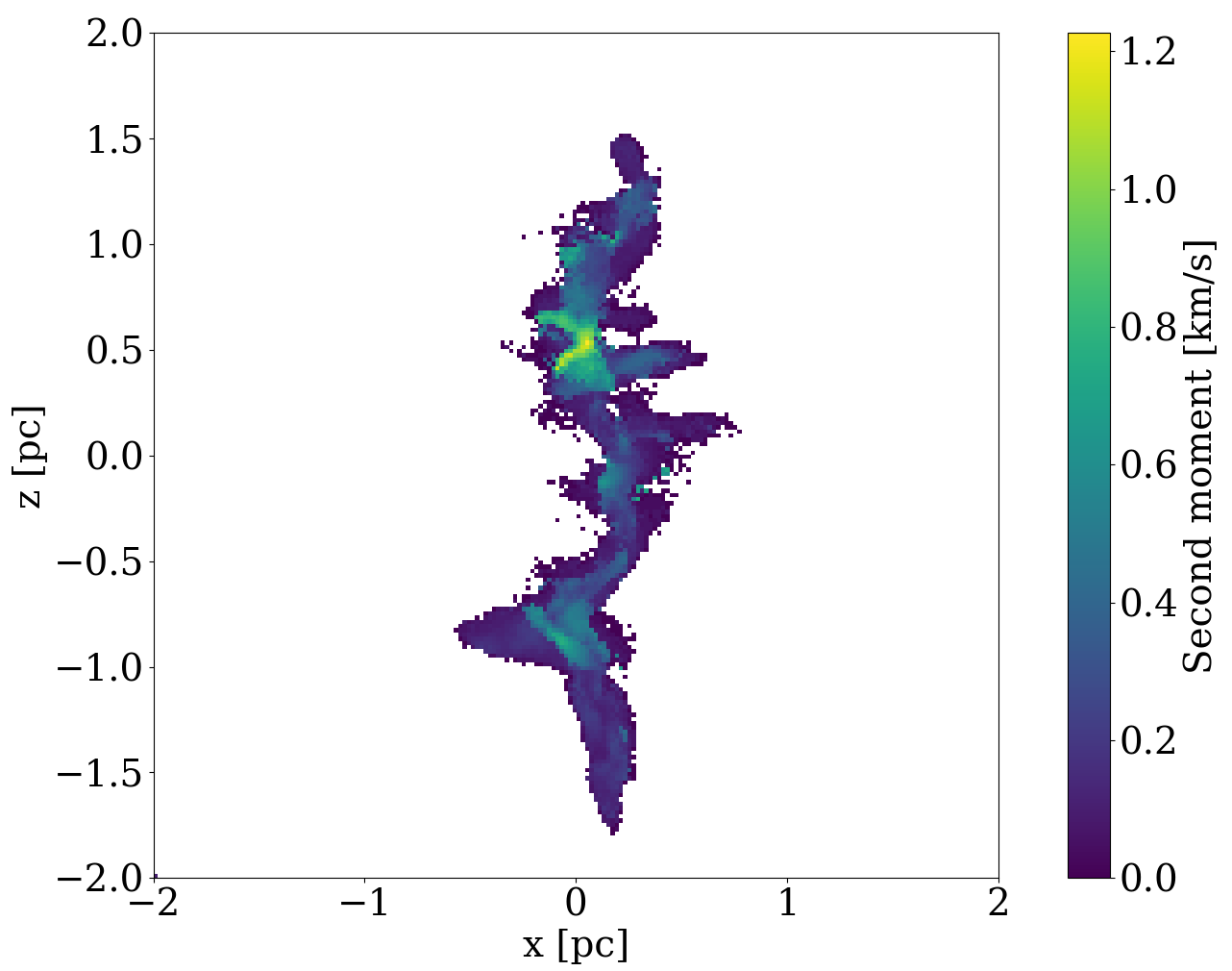}
\caption{As Fig. \ref{fig::seed1}, but for {\sc Sim}08.}
\label{fig::seed8}
\end{figure*}

\begin{figure*}
\centering
\includegraphics[width = 0.35\linewidth]{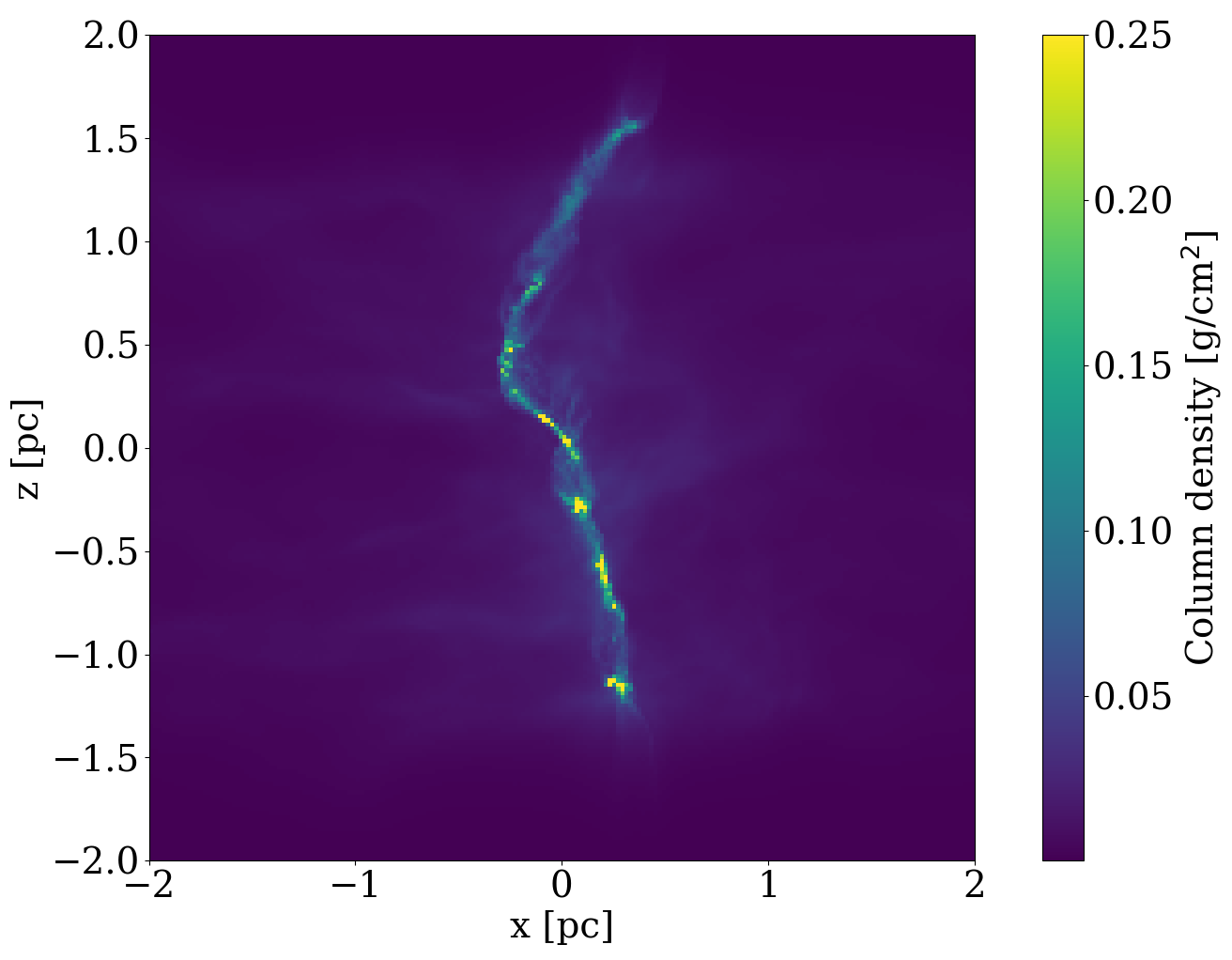}
\includegraphics[width = 0.35\linewidth]{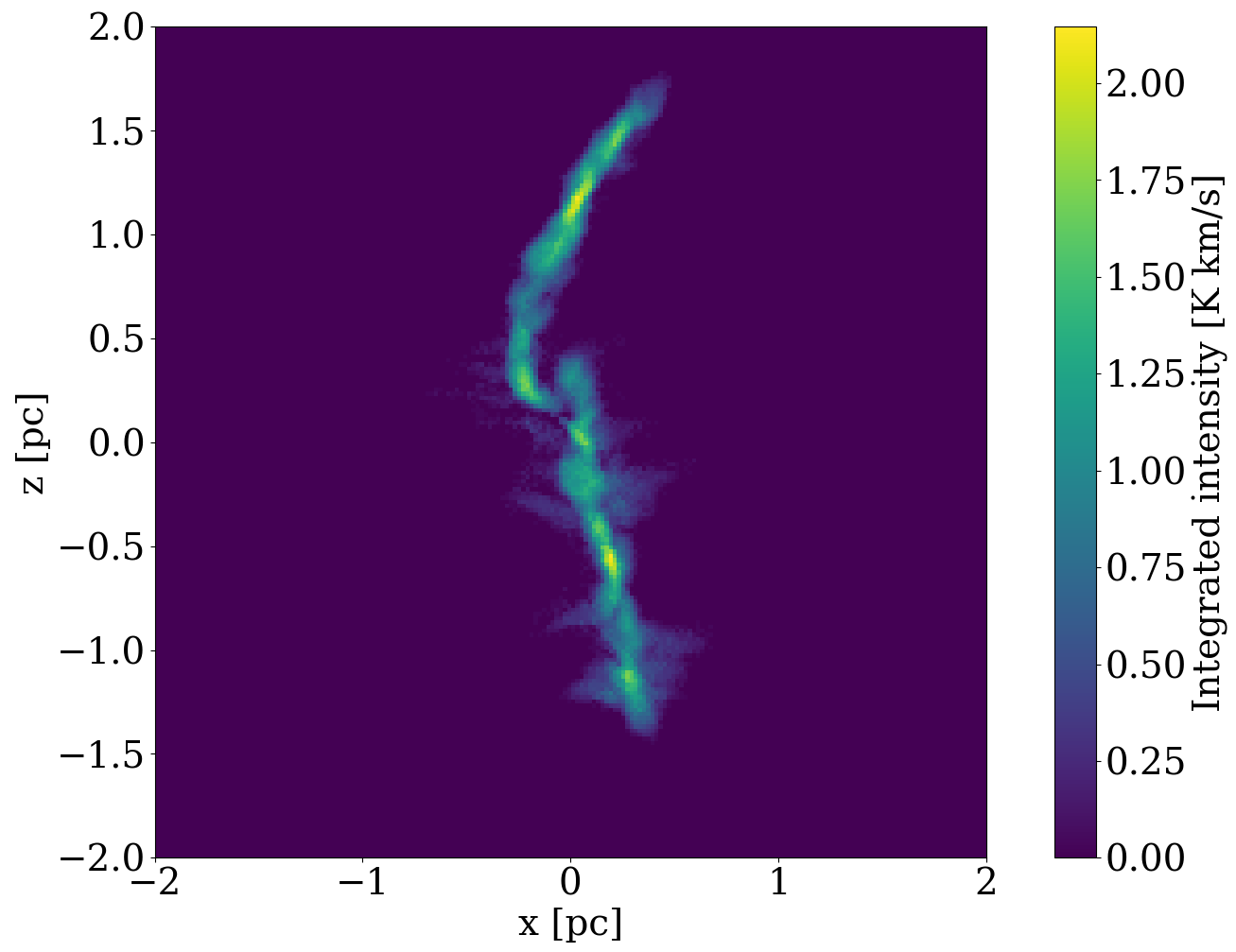}
\includegraphics[width = 0.35\linewidth]{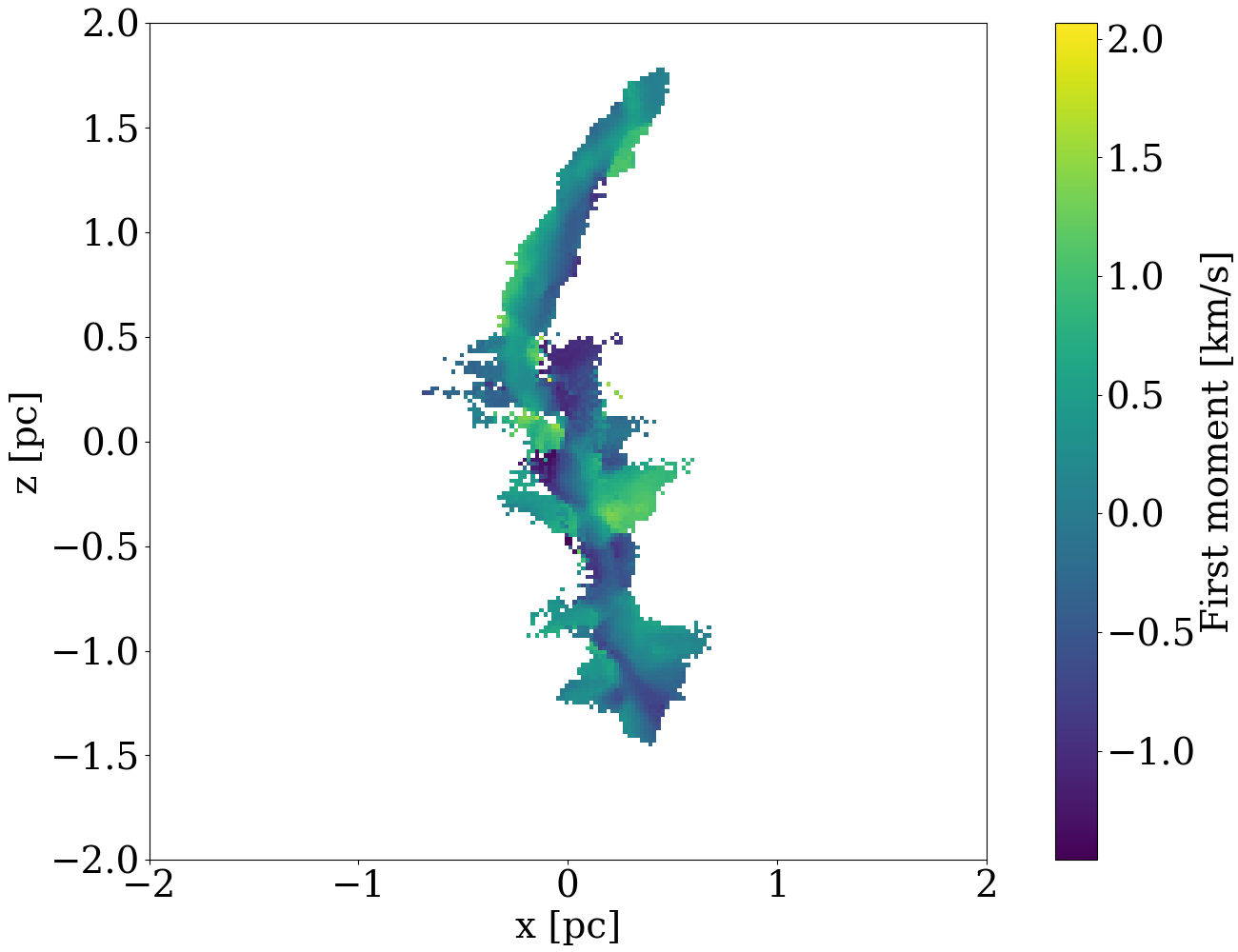}
\includegraphics[width = 0.35\linewidth]{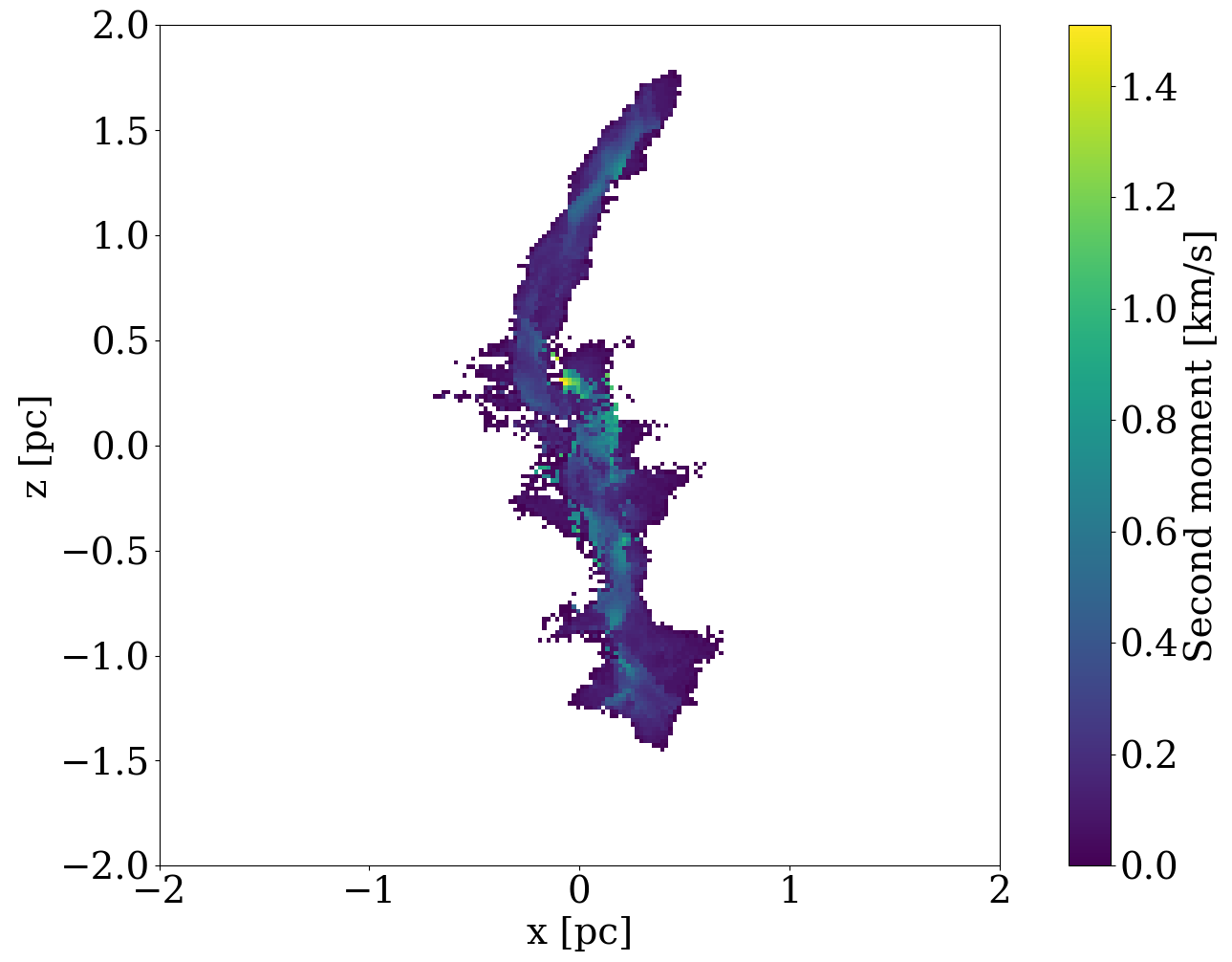}
\caption{As Fig. \ref{fig::seed1}, but for {\sc Sim}09.}
\label{fig::seed9}
\end{figure*}

\begin{figure*}
\centering
\includegraphics[width = 0.35\linewidth]{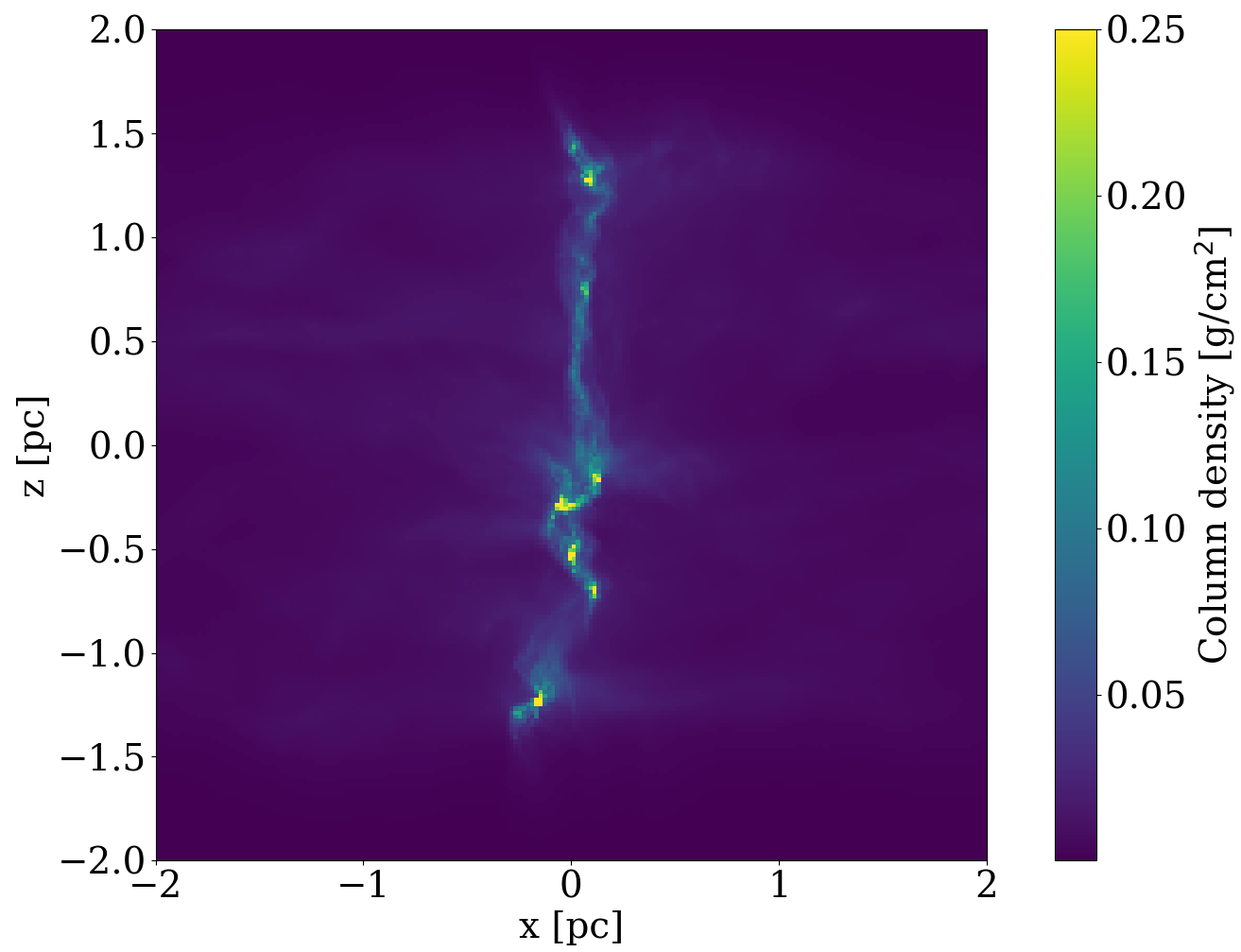}
\includegraphics[width = 0.35\linewidth]{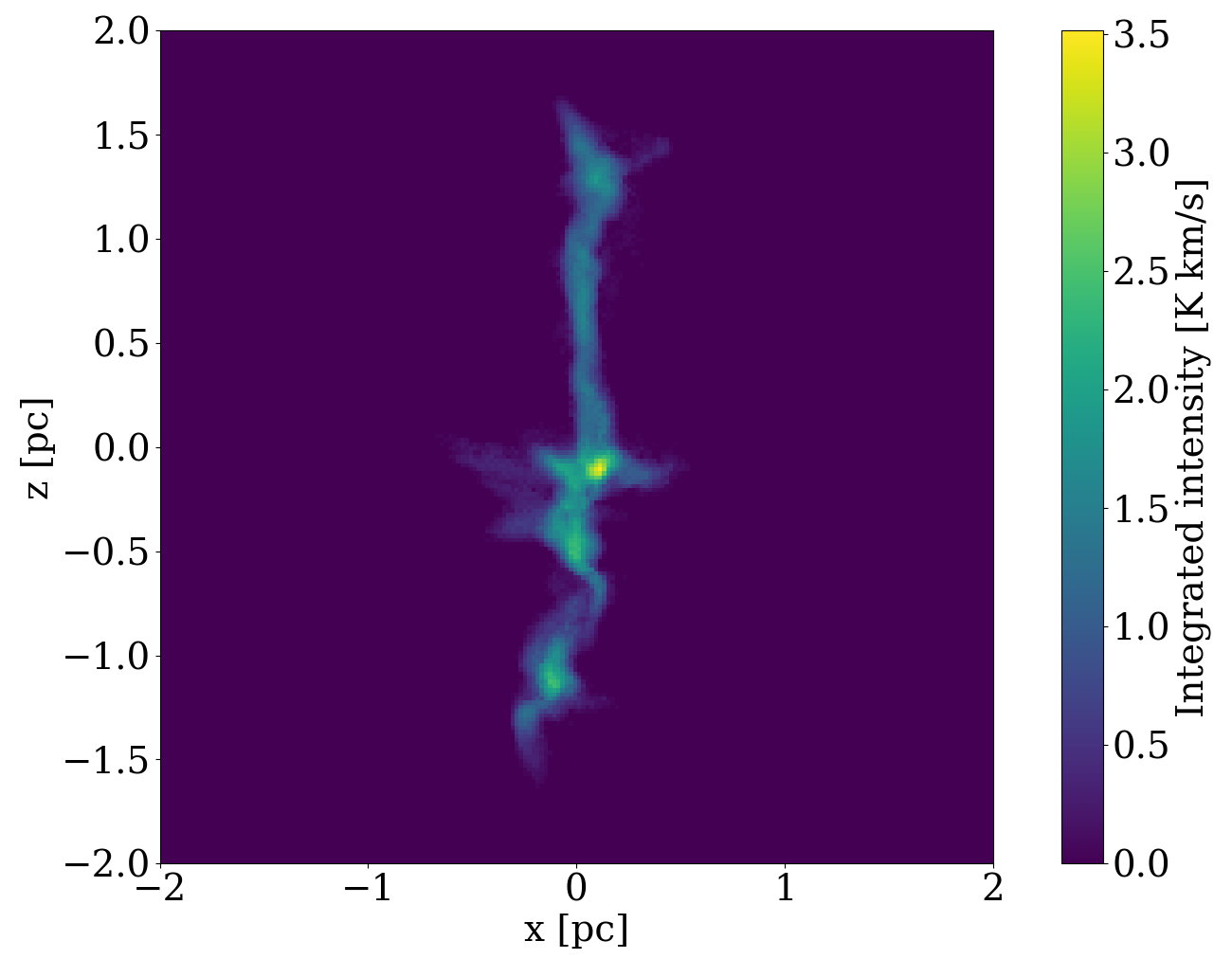}
\includegraphics[width = 0.35\linewidth]{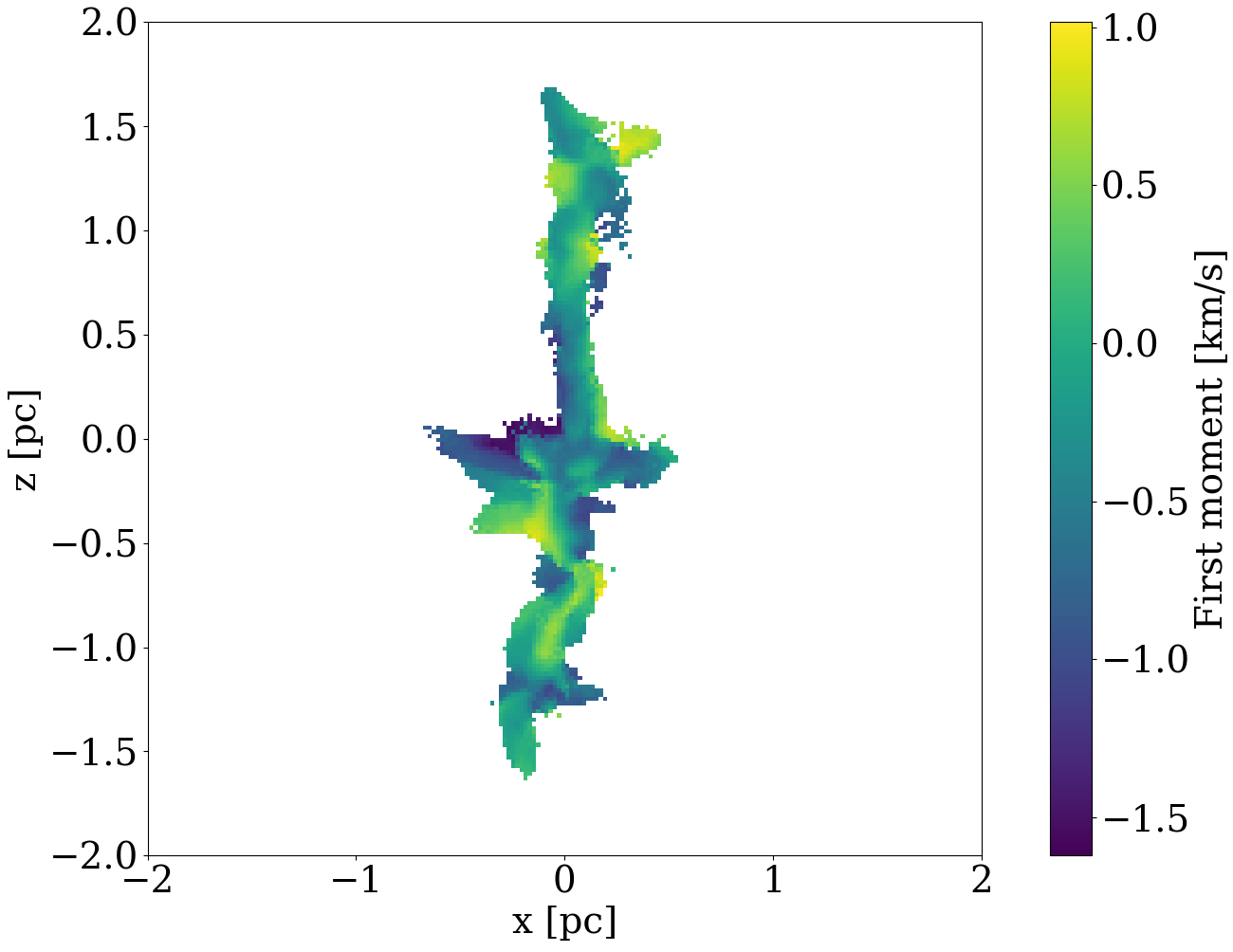}
\includegraphics[width = 0.35\linewidth]{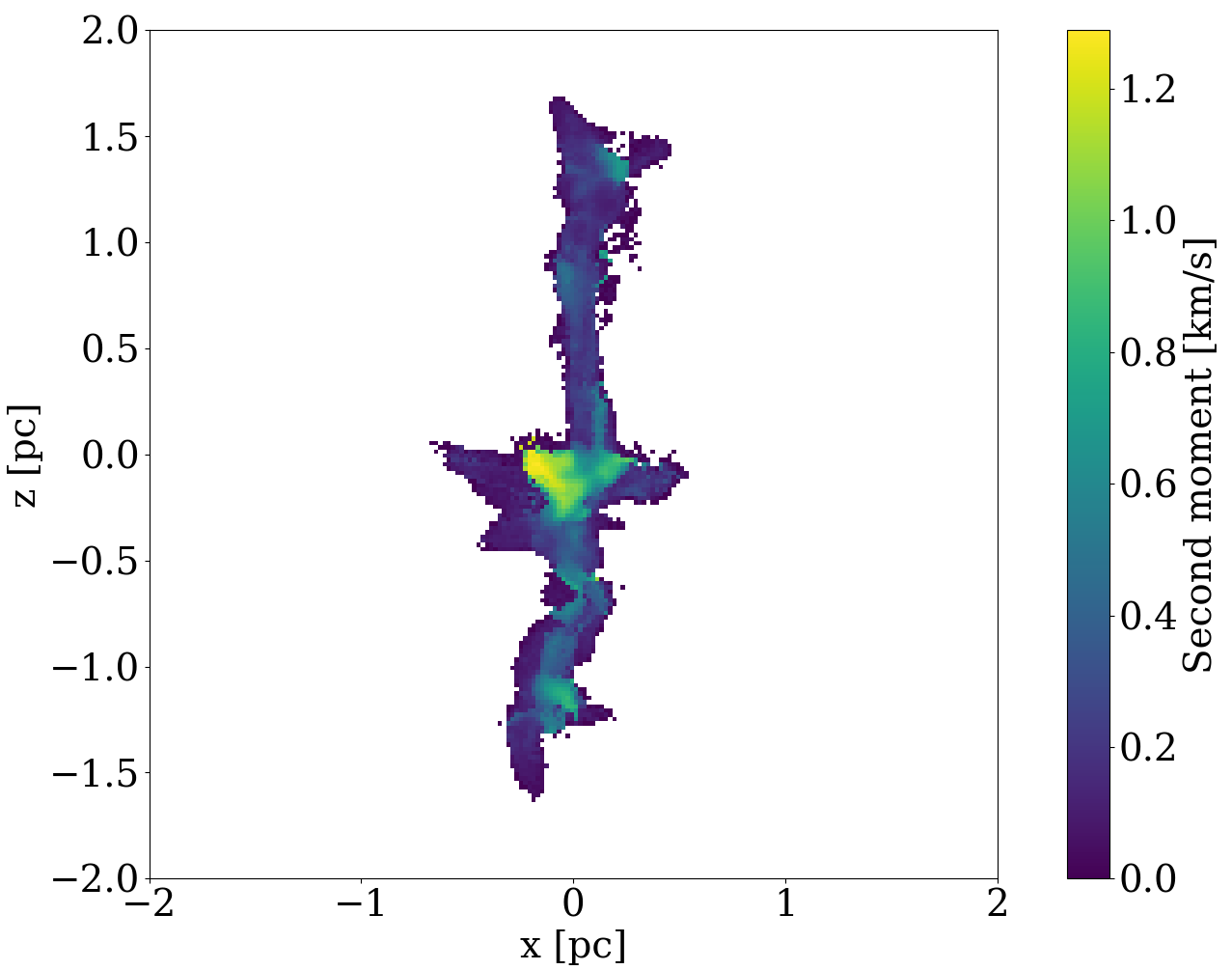}
\caption{As Fig. \ref{fig::seed1}, but for {\sc Sim}10.}
\label{fig::seed10}
\end{figure*}

Figs. \ref{fig::seed1}  through \ref{fig::seed10} show (a) column-density, (b) integrated intensity, (c) intensity-weighted velocity centroid, and (d) intensity-weighted velocity dispersion, for the synthetic C$^{18}$O observations of a single representative frame from, respectively,  {\sc Sim} 01 and {\sc Sim}03 through {\sc Sim}10. The simulation which is analysed in the main text of the paper is {\sc Sim02}, and the equivalent figure for {\sc Sim}02 is Fig. \ref{fig::column}.

\label{lastpage}

\end{document}